\documentclass[3p, final]{elsarticle} % 3p shirnks more the text and makes bigger margins

\usepackage{lineno,hyperref}
\modulolinenumbers[5]
\newtheorem{Theorem}{Theorem}[section]

\usepackage[normalem]{ulem}
\usepackage{siunitx,booktabs,}
\usepackage{amssymb,epsfig,latexsym,bm}
\usepackage{graphicx,amsmath,amssymb,latexsym,psfrag}
\usepackage{cancel}
\usepackage{amsfonts, amsmath}
\usepackage{color}
\usepackage{bm}
\usepackage{xcolor}
\usepackage{hyperref}
\usepackage{graphicx}
\usepackage{caption}
\usepackage{subcaption}
\usepackage{algorithmic}
\usepackage{comment}
\usepackage{dsfont}
\usepackage{amssymb}
\usepackage{algorithm}
\usepackage{setspace}

\newcommand{\argmax}{\mathop{\mathrm{argmax}}}

\newtheorem{Remark}{Remark}
\journal{Online Social Networks and Media (OSNEM)}

\begin{document}

\begin{frontmatter}
\title{Modeling communication asymmetry and {content} personalization in online social networks}
% to have a footnote in the title
%\title{Elsevier \LaTeX\ template\tnoteref{mytitlenote}}


\author[politoaddress]{Franco Galante\corref{correspondingauthor}}
\cortext[correspondingauthor]{Corresponding author}
\ead{franco.galante@polito.it}
% to specify a url

\author[politoaddress]{Luca Vassio}
\ead{luca.vassio@polito.it}

\author[unitoaddress]{Michele Garetto}
\ead{michele.garetto@unito.it}

\author[politoaddress]{Emilio Leonardi}
\ead{emilio.leonardi@polito.it}

\address[politoaddress]{Politecnico di Torino, Corso Duca Degli Abruzzi, 24, 10129 Torino}
\address[unitoaddress]{Università degli Studi di Torino, Corso Svizzera 185, 10149 Torino}

\begin{abstract}
{The increasing popularity of online social networks (OSNs) attracted growing interest in modeling social interactions.}
On online social platforms, a few individuals, commonly referred to as \textit{influencers}, produce the majority of content consumed by users and hegemonize the landscape of the social debate. However, classical opinion models do not capture this  communication asymmetry. We develop an opinion model inspired by observations on social media platforms {with two main objectives: first, to describe this inherent communication asymmetry in OSNs, and second, to model the effects of content personalization.}
We derive a Fokker-Planck equation for the temporal evolution of users' opinion distribution and analytically characterize the stationary system behavior. Analytical results, confirmed by Monte-Carlo simulations, show how {strict forms of} content personalization tend to radicalize user opinion, leading to the emergence of \textit{echo chambers}, and favor \textit{structurally advantaged} influencers.
As an example application, we apply our model to Facebook data during the Italian government crisis in the summer of 2019.
Our work provides a flexible framework to evaluate the impact of {content personalization} on the opinion formation process, {focusing on the interaction between \textit{influential} individuals and \textit{regular} users. This framework is interesting in the context of marketing and advertising, misinformation spreading, politics and activism.}

\end{abstract}

% MSC: mathematics subject classification at url: https://mathscinet.ams.org/mathscinet/msc/msc2020.html

\end{frontmatter}

\section{Introduction}
In recent years, the way people communicate has changed dramatically. With the advent of the Internet, new communication channels have emerged that allow people to transcend geographical and language barriers thanks to a global communication network and alternatives to text-only interaction like images and videos. Online social networks (OSNs) are probably the most notable example of such new interaction mechanisms and have greatly influenced our society by fostering discussions and disseminating information.
The amount of content produced on such social media platforms is immense. Therefore, to keep users engaged within the social network, the platform performs filtering to select the posts to offer.
{This filtering mechanism may reinforce the natural tendency of users to interact with like-minded individuals, one of the main drivers of social network building \cite{found_homophily}, known as \textit{homophily}.}
And, in turn, it can lead to the formation of \textit{echo chambers} \cite{quattrociocchi_echo_chamber},
where people who share a similar point of view interact with each other {while being}  isolated from the rest of the users. The reach of certain social media users can be extraordinary. Posts by very influential people, {often referred to as \textit{influencers} or \textit{opinion leaders},} can reach a large audience in virtually no time. Another aspect worth mentioning is the diversity of topics discussed, ranging from commentaries on the latest sporting events to debates on sensitive issues such as vaccinations. 

{From a user's perspective, there are two types of online interactions: those with acquaintances and those with influencers.  Our model focuses primarily on the latter interaction, as influencers' sway {has not yet received sufficient attention in the literature.}
	Toward the end of the manuscript, we present an extension of the model to include regular users' interactions.
}

The main objective of this work is to develop an analytical framework tailored to online interactions, incorporating the following aspects:
\begin{itemize}
	{\item The asymmetry of OSNs: a relatively small portion of \textit{well-known} users, i.e., the \textit{influencers}, % have a large following and their opinions
		can reach a vast number of far less known individuals.
		\item The closed-loop between influencers' and regular users' dynamics, triggered by the \textit{{content} personalization} mechanism applied by the social media platform in one direction and user \textit{feedback} (e.g., likes) in the other.}
\end{itemize}
{The concept of \textit{reference direction}, which is the individual's main topic of interest and expertise,
	represents a novel aspect of our approach.} 
To our knowledge, the existing literature has not yet considered a reference topic for each influencer on the opinion formation process in multi-dimensional spaces.
{Following recent work, we support the model's hypotheses with social network data \cite{das_ash_exp}\cite{Xiong_Liu_2014} from a large ensemble of Italian influencers, and compare the outcomes with emerging phenomena on two popular OSNs, i.e., Facebook and Instagram \cite{scaling_elections}\cite{quattrociocchi_echo_chamber}. The proposed model does not aim to be 
	quantitatively predictive but rather to be a tool to study qualitatively the emerging behavior in online social networks and the impact of {content} personalization, with a focus on opinion leaders.}

{For the single-topic case, we derive a Fokker-Planck equation as a second-order approximation of the opinion formation process and prove the existence and uniqueness of the stationary solution. In addition, since this approach is non-constructive we develop a less accurate first-order approximation, i.e., the fluid limit, which provides a closed-form formula expressions for the stationary solution. Then, through a Monte-Carlo approach, we  analyze the effects of the influencers' structural parameters and the interplay with content personalization in a multi-topic environment. The results highlight some of the threats associated with content filtering. We show that it favors \textit{structurally advantaged} influencers and can radicalize users' opinions, leading to the formation of \textit{echo chambers}.}

The paper is organized as follows. Section \ref{sec:related} {briefly} discusses the relevant work in the literature.
The Communication Asymmetry model is presented in Section \ref{sec:model}, along with the notation used throughout the article. Section \ref{sec:OSNobservations} presents some observations from real social networks supporting our modeling assumptions.
Section \ref{sec:analysis} is devoted to the mean-field analysis of the model as the number of users grows large. The theoretical results on the steady-state behavior of the model are proved in the Appendix, {together with the experimental analysis}. Section \ref{sec:model_vs_param} then investigates the impact of the model parameters on a reference scenario with two influencers.
Section \ref{sec:data} further validates our model with real-world data collected on Instagram and Facebook. {Finally, Section \ref{sect:conclusion} discusses limitations of the work and possible extensions and future directions.\nopagebreak}

\section{Related work}\label{sec:related}
The first steps in the study of opinion dynamics were taken in the late 1950s by a number of social psychologists \cite{Ash1955}\cite{French1956}\cite{Festinger1954}. Ash introduced the concept of \textit{social pressure} \cite{Ash1955}, a conformist tendency in individuals, while French used directed graphs to model interpersonal relationships \cite{French1956}. A landmark in the field is Festinger's work on \textit{social comparison}\cite{Festinger1954}. Individuals tend to evaluate their position by comparing it to others, and it is inversely proportional to the distance between viewpoints.
{Opinion models are continuous or discrete, according to the description of the opinion variable.  As for most models \cite{Mastroeni_IEEE_review}, the seminal work in the field is continuous.}
For example, the DeGroot model \cite{Degroot1974} considers a networked social system in which individuals interact with their neighbors. Individuals average their current opinion with the opinion of their neighbors.
Subsequently, Friedkin and Johnsen \cite{Friedkin_Johnsen_1990} {developed a linear model which} encompasses both the processes of \textit{social conformity} and \textit{social conflict} {leading } to behavior that goes beyond simple consensus.
In the early 2000s, Hegselmann and Krause \cite{HK_model} and Deffuant and Weisbuch \cite{DW_model} proposed two similar models, {introducing the idea of bounded confidence: individuals interact only with peers whose beliefs are not too different. The proposed models are nonlinear and challenging to study \cite{Lorenz_2007}.}
A great deal of attention has also been paid to discrete models. A prominent example is the voter model \cite{clifford_and_sudbury_1973}\cite{Holley_Liggett_1975} and its extensions accounting for evolving networks \cite{Holme_Newman_2006}\cite{Durrett_evolving_VM}, individuals with multiple opinions \cite{Nardini_whos_talking_1st} or spontaneous changes of opinion \cite{Granovsky_Madras_1995}.
A consistent bulk of research on opinion dynamics comes from the physics literature, among which early contributions are Ben-Naim \cite{Ben-Naim_2005} and Toscani \cite{toscani_kinetic_model}. Ben-Naim and Toscani consider two mechanisms of opinion formation: \textit{compromise}, and \textit{introspection} (in other models, e.g., \cite{Granovsky_Madras_1995},  modeled as noise), which the authors believe represents the impact of external sources of information (e.g., media).
{The Sznajd model \cite{sznajd_weon_model} is a generalization of the Ising model, which implements \textit{social validation} and for which \cite{Slanina_sznajd_analytical} derives analytical results.}
For a comprehensive review of classic opinion models, we refer to the survey by Castellano et al. \cite{Castellano_Fortunato_Loreto_2009}.

Most of the seminal literature on opinion dynamics is suited to describe the decision-making process in small groups of individuals, e.g., a board of directors, or to capture relatively \textit{regular} patterns determined by the daily personal interactions of individuals. Models such as the voter model have been studied extensively on regular lattices \cite{cox_voter} \cite{Frachebourg1996}. The structure of interactions, especially those online, is far from homogeneous. As mentioned earlier, an inherent asymmetry in communication exists in OSNs where a limited number of individuals (\textit{influencers}) monopolize the discussion. The voter model has been studied over heterogeneous networks (e.g., \cite{Suchecki2005} \cite{Sood2005}) to account for this diversity. On such networks, there can exist \textit{hubs} (strongly connected nodes) playing a role similar to \textit{influencers} in our framework. However, the authors did not explicitly make such a distinction. Other works have divided the population into classes, e.g.,~\cite{Yildiz2013} introduced \textit{stubborn} agents, and if such individuals have opposing opinions, they hinder the possibility of the population converging to consensus.
{Recent work further draws attention to online platforms by adapting classical frameworks to the specificity of online interactions.~\cite{valensise_polarization, Peralta_2021_algorithmic_bias, Perra2019}
	have developed an opinion model that embodies \textit{algorithmic personalization}. \cite{valensise_polarization} also compares model predictions and dynamics  observed on Facebook and Twitter.  
	Our work differs from \cite{valensise_polarization, Peralta_2021_algorithmic_bias, Perra2019}   as we consider distinct classes of users, precisely characterizing  \textit{influencers} and closing the interaction loop between users and the platform by a \textit{feedback} function.}

Attempts to validate opinion models are scarce due to several reasons: i) the mapping of opinions into values, ii) an adequate definition of links between agents, and iii) the change of opinion after an interaction is hardly measurable. {A recent survey \cite{data_in_opdyn_2022} examines the latest research concerning the use of data in opinion dynamics.}
Usually the approach is either through observational data \cite{scaling_elections}\cite{quattrociocchi_echo_chamber}\cite{valensise_polarization} or controlled sociological experiments\cite{pnas_opposing_views}\cite{Morgan2011}. 
The first type allows to scale to large number of users and is more related to our work. 
The political environment has classically been a florid field for opinion dynamics due to the possibility of attributing a person's opinion to the political orientation of the chosen candidate \cite{scaling_elections}. Also, a noisy voter model was used to fit data from US elections \cite{voter_for_voters}. 
The authors of \cite{Barbera_PS2015} estimate the political ideology of users on Twitter on a single axis (left-right) using  the ground truth reconstructed by roll-call votes of members of parliament and their network of followees-followers.
In \cite{De_TWEB2019}, opinions are estimated through a sentiment analysis tool applied to the text of posts on Twitter. Posts are first classified into topics according to their keywords (as we do, see Section \ref{sec:OSN_reference}), and then a continuous sentiment score is given between negative and positive.
The authors of \cite{Monti_CIKM2021} consider users' opinions in a multidimensional ideological space. Through doc2vec clustering, they identify the axes of 4 dimensions and then map users to these 4-topics opinions. Validation is made through well-known positions of famous Reddit groups. 
Other recent approaches \cite{science_diverse_news} have used shared \textit{news} on Facebook to assess the extent to which individuals are exposed to opposing views through their (online) friendship relationships, using users' self-reported ideological affiliations to infer opinion. More recent work \cite{quattrociocchi_echo_chamber} \cite{valensise_polarization} have directly employed data from online social networks, such as Gab, Facebook, Reddit, and Twitter, to observe the emergence of \textit{echo chambers} \cite{quattrociocchi_echo_chamber} and to validate a model encompassing algorithmic personalization in the process of opinion formation \cite{valensise_polarization}.

\section{The Communication Asymmetry opinion model}
\label{sec:model}

In this section, we first establish the notation used throughout the paper and then present the Communication Asymmetry model in its most general formulation, along with the limitations of the model.

\subsection{Notation }
In this work, we adopt the following vectorial notation. We denote vectors by bold characters, whereas we denote their components with normal-font characters whose subscript is the index in the vector, e.g., $\bm{a}=\{a_k \}_k$. 
Lowercase letters  denote parameters and dynamical variables associated with an individual. In general, index $i$ runs over the set of \textit{influencers} while the index  $u$ runs over  that   of regular users. For those parameters/variables that can be associated with individuals of both classes (either  influencers  or  regular users), the above indices  are indicated between superscript parentheses, e.g., $a^{(i)}, a^{(u)}$, to identify the class to which the individual belongs immediately.
If necessary, the dependence of variables on other system parameters   is made explicit by specifying the independent variables between parentheses, e.g., $\alpha(\cdot,\cdot)$. Italic capital letters denote sets, e.g., $\mathcal{I}$ is the set of all influencers in the population, while $|\mathcal{I}|$ is its cardinality. Capital letters represent outcomes of stochastic experiments whose characteristic parameters are lowercase letters: e.g., $\Omega\left(\omega(\cdot,\cdot)\right)$. The operator $\mathbb{E}[\cdot]$ represents an expected value, and a bar over a variable, e.g., $\bar{a}$, represents its average value. Whenever we need to express the probability of an event, we use the notation $Pr[\cdot]$. We employ $\mathds{1}_{\{\cdot\}}$ for the indicator function. Lastly, time is denoted by $t$ if considered continuous and by $n$ if discrete.

\subsection{Description of the model}\label{sec:model_desc}

\begin{figure*}[h]
	\centering
	\includegraphics[width=0.75\textwidth]{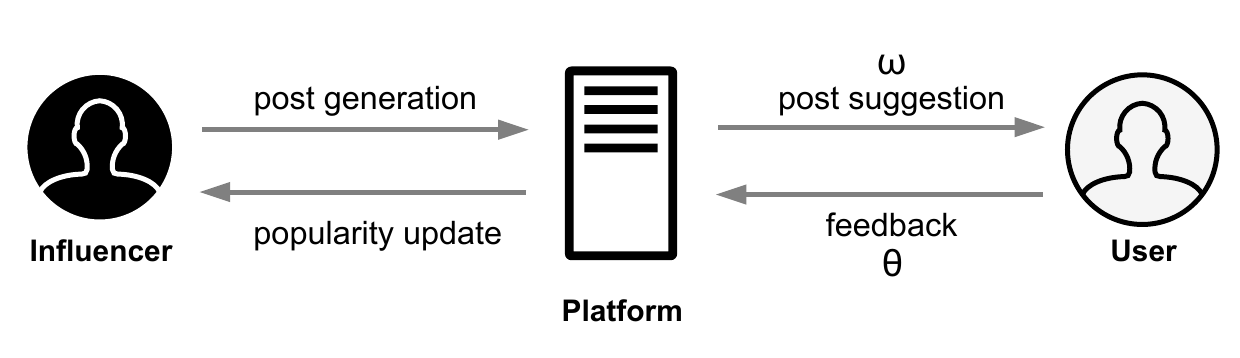}
	\caption{{Illustration of the interaction between users, influencers, and social media platforms highlighting the closed loop between user feedback $\theta$ and {content} filtering $\omega$, which will be further characterized below.}} %\lv{come segnato anche in nota 1, usiamo troppi nomi per $\omega$: post provision (in questa figura), proposition (nello pseudocodice), filtering, send, suggest, reach, matching. Forse proverei a usarne il meno possibile}}
\label{fig:rho_theta_loop}
\end{figure*}

We propose a continuous opinion model with two interacting classes of agents. Specifically, the population consists of $N_u=|\mathcal{U}|$ \textit{regular users} and $N_i=|\mathcal{I}|$ \textit{influencers}. This division mimics what happens in real social networks, where a small portion of the population, the \textit{influencers}, has a much larger number of people following their posts on the online social network. The opinion space is $\mathcal{X} \subset \mathbb{R}^d$, where each dimension represents an uncorrelated topic on which users have a belief. Hence, an opinion is a $d$-dimensional vector $\bm{x}^{(u)}(n) \in \mathbb{R}^d$, which evolves as a result of the interaction between a regular user and the influencers on every possible topic. %{\color{red}and every influencer on every possible topic.}
{Moreover, {we model the tendency toward an a priori opinion} $\bm{z}^{(u)}$, and we refer to it as the \textit{prejudice} of a user. Unless otherwise specified, we will assume that the user's initial opinion corresponds to the prejudice: $\bm{x}^{(u)}(0) = \bm{z}^{(u)}$.} We assume that the generation of new posts, i.e., messages in the OSN, is a  Poisson Point Process (PPP) with intensity  $\lambda$, where each event of the PPP corresponds to the creation of a new post from an influencer $i \in \mathcal{I}$. 
The corresponding embedded discrete time will be denoted by the integer $n\in \mathbb{N}_{+}$, $n = 1,2,\ldots$, where $n$ is the $n$-th post.
{Figure \ref{fig:rho_theta_loop} illustrates the social media platform's role in \textit{filtering} the posts sent\footnote{We use the terms \lq\lq send", \lq\lq suggest" and \lq\lq reach" interchangeably, for a post shown to a user by the platform. We will refer to \textit{regular users} simply as \textit{users}. Moreover, the terms \textit{agent} or \textit{individual} indicate a social network user of either class.} to regular users and receiving \textit{feedback} from those users.
	These two aspects implement, respectively, 
	a \textit{selective exposure} effect: namely the tendency of both the  platform and  the users to suggest/access similar content, and 
	a \textit{confirmation bias}, namely the tendency of users to value content that is close to one's point of view (see \cite{quattrociocchi_echo_chamber} and resources therein).}

Influencers are considered stubborn, i.e., $\bm{x}^{(i)}(n)=\bm{x}^{(i)}(0)=\bm{x}^{(i)}=\bm{z}^{(i)} \in \mathbb{R}^d,\quad \forall n > 0$ and $i\in \mathcal{I}$. As we will show in Section \ref{sec:OSNobservations}, each influencer has a main topic of interest on which it publishes most of its posts and typically coincides with the topic it is mainly known for on the OSN. We call it the \textit{reference direction} $r^{(i)} \in \{0,..,d-1\}$ of the influencer. Another parameter characterizing influencer $i$ is its \textit{consistency} $c^{(i)}(n)$, {which denotes the (possibly time-varying) probability of posting on the reference direction. High-consistency individuals} preferably post on their reference topic. We denote by $f^{(i)}$ the probability that a post is generated by influencer $i$ at any time instant $n$, with $\sum_{i \in \mathcal{I}} f^{(i)} = 1$.
At last, we introduce the \textit{popularity} vector  $\bm{p}(n) \triangleq \{p_i(n)\}_{i \in \mathcal{I}}$, containing the current popularity of all influencers at time $n$, before the emission of the post at time $n$.
We also introduce the normalized version of this vector $\bm{\pi}(n)= \{{\pi_i}(n)\}_{i \in \mathcal{I}}$ where the components are the normalized popularities $\pi_i = \frac{p_i}{\sum_{j \in \mathcal{I}} p_j}$.

\begin{figure*}[h]
	\centering
	\includegraphics[width=0.75\textwidth]{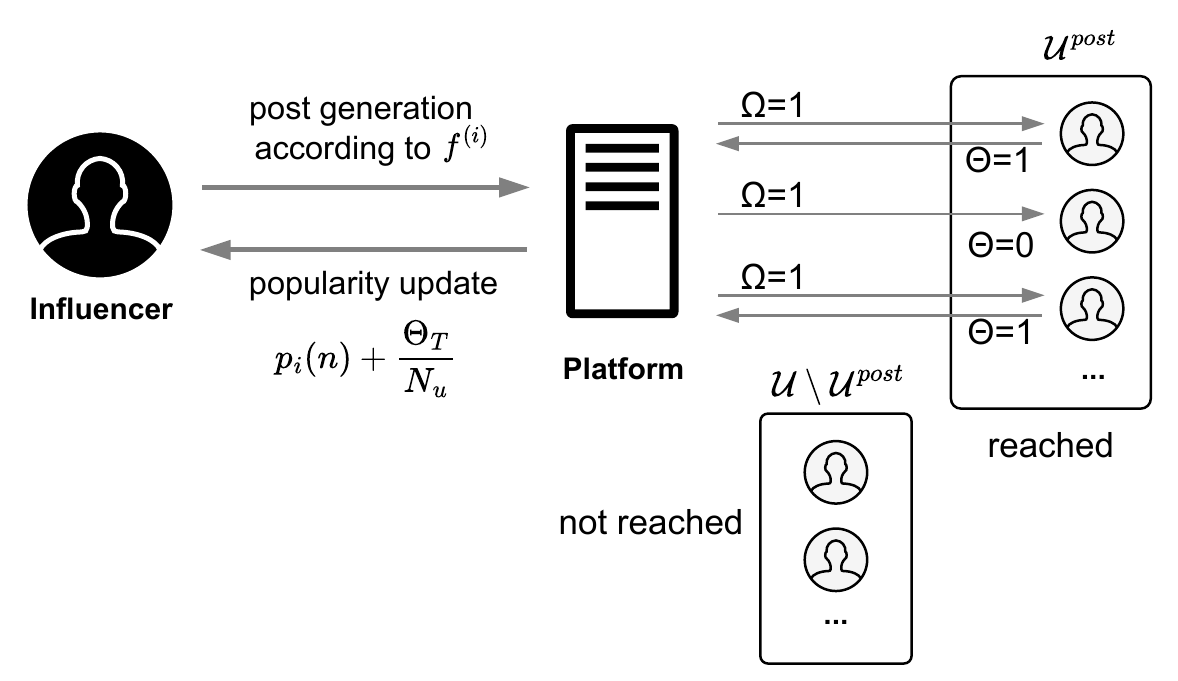}
	\caption{Schematic representation of the model dynamics. The figure highlights the proportions of users who view a particular post $\mathcal{U}^{post}$, i.e., those for which the random variable $\Omega$ equals $1$. They react with their feedback $\Theta$ (e.g., \textit{likes}), which depends on the opinion distance between them and the influencer $i$. Then the platform updates the influencer's $i$ popularity.}
	\label{fig:scheme_model}
\end{figure*}

The dynamic variables of users (i.e., their opinion $\bm{x}^{(u)}$) and influencers (i.e., their popularity ${p_i}$) are updated upon every post generation  according to Algorithm \ref{alg:model}. %It provides a detailed description of the dynamics captured by our model.
Figure \ref{fig:scheme_model} gives a schematic representation of the model.

\begin{algorithm}[H]
\caption{The Communication Asymmetry opinion model}\label{alg:model}
\begin{spacing}{1.0}        % increase the interline spacing
	\algsetup{indent=1.2em} % regulates the horizontal indentation
	\begin{algorithmic}[1]
		\REQUIRE \,\,\,\,\,$N_i$ influencers, $N_u$ users, filtering function $\omega$, feedback function $\theta$
		\ENSURE opinion of each regular user 
		$\bm{x}^{(u)}(n),\,\forall u$
		\ENSURE popularity of each influencer
		$p_i(n),\,\forall i$
		\LOOP %{\fg{over $n$?}}
		\STATE select influencer $i$ according to $f^{(i)}$
		\STATE select a posting
		direction $j$, i.e., $j=r^{(i)}$ with probability $c^{(i)}$, otherwise 
		%(with probability $1-c^{(i)}$) 
		$j$ is selected uniformly on $j\in\{0,..,d-1\}\setminus\{r^{(i)}\}$
		\STATE {$p_i(n+1) = p_i(n)$}
		\FORALL {regular user $u$ in the population}
		\STATE{$x^{(u)}_j(n+1) = x^{(u)}_j(n)$}
		\IF[post suggestion]{$\Omega\left(\omega(|x^{(i)}_{r^i} - x^{(u)}_{r^i}|, \pi_i(n))\right) = 1$}
		\STATE get feedback $\Theta\left(\theta(|x^{(u)}_j-x^{(i)}_j|)\right)$
		\IF[positive feedback]{$\Theta = 1$}
		\STATE{$x^{(u)}_j(n+1) = \alpha z^{(u)}_j + \beta x^{(u)}_j(n) + (1 - \alpha - \beta) x^{(i)}_j$}
		\STATE update popularity of $i$: $p_i(n+1) \mathrel{{+}{=}} 1/N_u$
		%\ELSE
		%\STATE{$x^{(u)}_j(n+1) = x^{(u)}_j(n)$}
		\ENDIF
		\ENDIF
		\ENDFOR
		% \STATE update influencer's popularity $p^{(i)}$
		\ENDLOOP
	\end{algorithmic}
\end{spacing}
\end{algorithm}
{At any time instant $n$, an influencer $i$, selected according to distribution $f^{(i)}$, generates a post.} The influencer $i$ posts on its reference direction $r^{(i)}=j$ with a probability equal to its consistency $c^{(i)}$. Otherwise, it posts on one of the \textit{secondary} directions $j \in \{0,1,...,d-1 \} \setminus \{r^{(i)}\}= \mathcal{N}_r$ according to a given distribution, $Pr[j=k]$ for $k$ in $\mathcal{N}_r$.
In the following, we assume this distribution to be uniform over the set of non-reference directions. 
{The emitted post carries the $j$-th component of the influencer's opinion vector $\bm{x}^{(i)}$. We suppose posts accurately reflect the influencer's true belief and no noise affects user perception.}

\begin{figure*}[h!]
\centering
\includegraphics[width=0.75\textwidth]{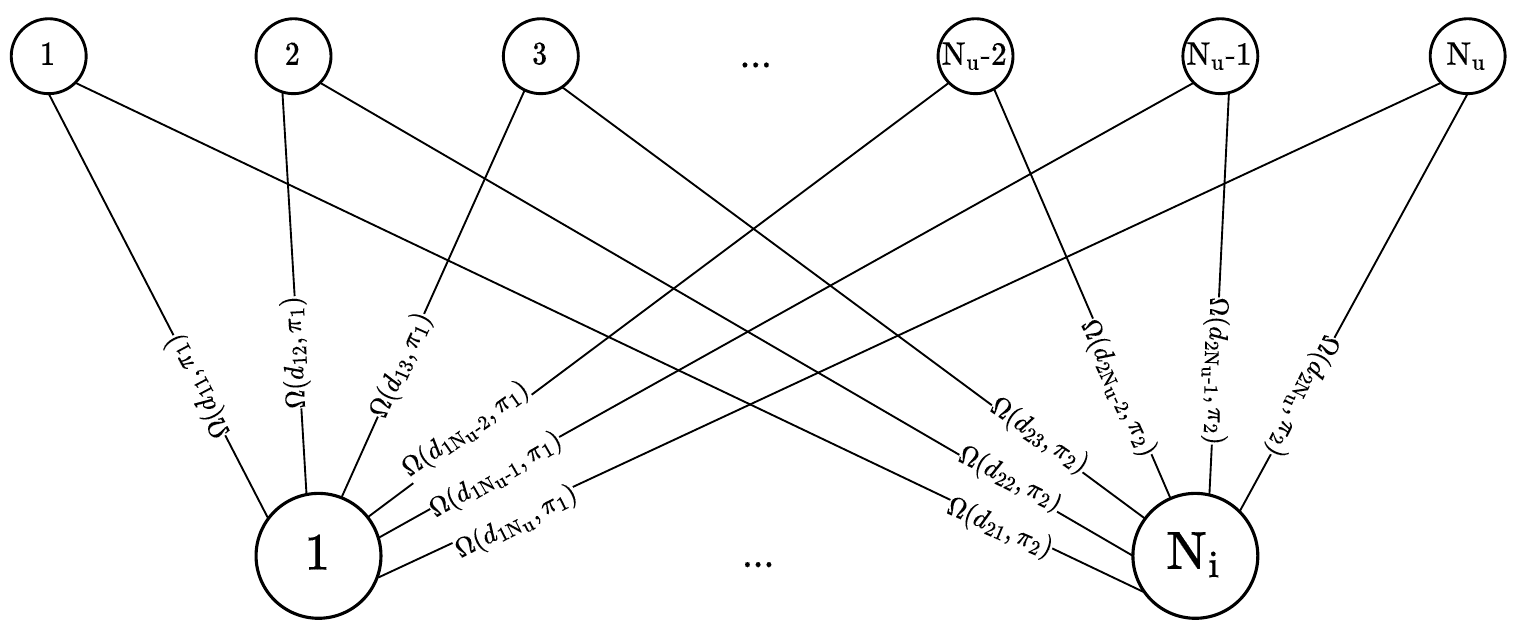}
\caption{The bipartite dynamic structure of the underlying graph between \textit{regular} users (top) and \textit{influencers} (bottom) defined by the \textit{visibility} function $\Omega\left(d_{iu}(n) = |x_r^{(u)}(n) - x_r^{(i)}(n)|, \pi_i (n) = \frac{p_i (n)}{\sum_i p_i (n)}\right)$ which depends on both the \mbox{user-influencer} opinion distance and the relative popularity of the influencer at a given time instant $n$.}
\label{fig:graph}
\end{figure*}

To decide whether the post reaches a given user (independently from other users), we extract a Bernoulli random variable $\Omega$ with parameter $\omega$. The user receives the message when $\Omega(\omega)=1$. The parameter $\omega$ can be interpreted as a \textit{visibility} function from the influencer's perspective, as it affects the subset of users reached by its posts.
{In principle, a post can reach any user as the interaction network is a dynamic complete bipartite network between the set of nodes $\mathcal{U}$ and $\mathcal{I}$ whose links are defined by $\Omega(\cdot)$ (see Figure \ref{fig:graph}).
The \textit{visibility} function $\omega$ is a decreasing function of the opinion distance on the reference direction $d_r(n) = |x_r^{(u)}(n) - x_r^{(i)}(n)|$. This dependence embeds the concept of \textit{homophily}, one of the main drivers of interaction on social networks\cite{found_homophily}: individuals with strongly divergent opinions interact less frequently than like-minded individuals. Moreover, $\omega$ is increasing in the popularity ratio $\pi_i$. The higher the relative popularity of an influencer, the more users it can reach.}
This posts-users matching process constitutes the content \textit{personalization} we consider in this paper (see Remark \ref{remark2}).
{Note that the \textit{filtering} process for selecting the subset of users who receive the post is based on the opinion distance along the reference direction between each user and the influencer who made the post.} {This because we expect that influencers mainly attract users
whose opinions are similar to their main topic.
For instance, politicians primarily attract users interested in the political landscape {and with similar orientations}.}
Adopting {this distance to perform the user selection couples the dynamics in different directions,} which would otherwise evolve independently of each other (see Remark \ref{remark1}). Users express their \textit{feedback} to a post on the platform through a Bernoulli random variable $\Theta \left(\theta(|x_j^{(u)} - x_j^{(i)}|) \right) \in\{0,1\}$ whose parameter $\theta$ depends on the difference in opinion on the \textit{actual} direction $j$ of the contribution. Only posts that receive positive feedback, i.e., $\Theta = 1$, can influence the user's opinion, reflecting the tendency to ignore unappreciated content. The social media platform collects feedback from all reached users to update the popularity $p_i$ of the posting influencer. 

The update rule for the popularity of the posting influencer $i$ reads as follows:

\begin{equation}
p_i(n+1)= p_i(n) + \frac{\Theta_T (\theta, \mathcal{U}^{post})}{N_u}  % |o_j^u(t) - o_j^i(t)|
\end{equation} \label{eq:d_model_p_update}
\vspace{-0.2cm}
\begin{equation}
\Theta_T (\theta, \mathcal{U}^{post}) = \sum_{u\in \mathcal{U}^{post}} \Theta\left(\theta(|x_j^{(u)}(n) - x_j^{(i)}(n)|)\right)
\end{equation}

where $\mathcal{U}^{post}$ is the subset of users who were made aware of the post by the platform, i.e., those for whom $\Omega(\omega)$ takes the value one. The summation in the formula gives the aggregate feedback of all users who saw the post, which is normalized by the size of the population of regular users $|\mathcal{U}| = N_u$ to update the popularity. 
This normalization is introduced only to avoid excessive popularity growth of the influencers when the number of users becomes large. It does not affect the system dynamics, which depends only on the normalized popularity $\pi_i$, which is not affected by the scaling factor $1/N_u$.

The core of the dynamics is the opinion update rule, which prescribes how the user's opinion changes on the direction $j$ of the post:
\begin{equation}\label{eq:d_model_user_update}
x_j^{(u)}(n+1) =
\begin{cases}
	\alpha z_j^{(u)} + \beta x_j^{(u)}(n) + \gamma x_j^{(i)} \;\;\; \text{if} \; \Omega \left( \omega(d_r,\pi_i) \right) = 1\,, \Theta \left(\theta(d_j)\right) = 1 \\[5pt]  % \texttt{and}
	x_j^{(u)}(n)  \quad\quad\quad\quad\quad\quad\quad\quad\;\, \text{otherwise} %\texttt{o/w}
\end{cases}
\end{equation}
%\el{il problema è  che $\Theta$  e' definita solo se $\Omega=1$, temo, forse dovremmo dire che convenzionalmente definiamo  $\Theta=0$  se $\Omega=1$ . }
{When the post reaches the user ($\Omega=1$) who likes it ($\Theta=1$), then the updated opinion is a convex combination, i.e., $\alpha + \beta + \gamma = 1$, of the current opinion $x_j^{(u)} (n)$, the prejudice $z_j^{(u)}$, and the opinion $x_j^{(i)}$ conveyed by the influencer through the post. The opinion is not updated if the user does not receive the post ($\Omega$=0) or does not like it even if it reaches them ($\Theta=0$). While it is common in the literature to express the opinion update as the convex combination in Eq. (\ref{eq:d_model_user_update}), we present an equivalent formulation that sheds more light on the meaning of the update parameters, which are practically two:}
\begin{equation}\label{eq:update_equivalence}
{x_j^{(u)}(n+1) = (1-\beta)\left[ \delta z_j^{(u)} + (1-\delta) x_j^{(i)} \right]+ \beta x_j^{(u)}(n), \quad \delta,\beta \in [0,1]}
\end{equation}
\eqref{eq:update_equivalence} can be easily derived from (\ref{eq:d_model_user_update}) by noting that $\alpha + \gamma = 1 - \beta$, and defining $\delta \triangleq \frac{\alpha}{\alpha+\gamma}$. Thus, $\beta$ represents the \textit{inertia} of users, i.e., how slowly they change, and $\delta$ their \textit{ degree of stubbornness}, i.e., the relative impact of external opinions w.r.t their prejudice.

\begin{Remark}\label{remark1}
The distance on the reference direction drives filtering action because we assume the platform is unaware of the specific topic associated with the post just created. { At the same time, homophilic connections between users and influencers primarily depend on opinion similarity on the main topic of discussion.}
Note the joint effect in the model of the distance between the user's opinion and the influencer's opinion on the reference direction and the distance along the direction defined by the post's topic. Both contribute to determining the likelihood for the user to provide positive feedback to the message.
\end{Remark}

\begin{Remark}\label{remark2}
Most OSNs have explicit subscriptions to influencers, i.e., the \textit{follow} mechanism. 
Our approach does not explicitly  represent  such {\lq\lq long-term" relationships.}
However, by applying the function $\Omega(\omega(\cdot))$,  we  dynamically determine 
the set of users reached by each influencer. Essentially, followers are regenerated at each post-emission.  The resulting  network  is  a  dynamic bipartite graph, see Figure \ref{fig:graph}, whose structure  reflects a given degree of homophily  of users' connections.
Typically homophily is one of the elements that mainly influences users' choices when they select individuals to connect to~\cite{found_homophily}. % influencers to follow.
In particular, for some  domains (e.g., product adoption, which is also a good fit for our \lq\lq competing" scenario in Section \ref{sec:model_vs_param})  \textit{homophily} has emerged as the key driver  governing the structure of the network \cite{pnas_homophily}.

At last, observe that, nowadays, most social media platforms (e.g., Facebook, Instagram, Twitter) do not only offer their users content they explicitly subscribe to, but also what they may like. %to those  users who \textit{might like}.
The selection of such users is  based on their previous activity on the platform. This mechanism  reinforces the homophilic  structure of the network and resembles what we are modeling.

\end{Remark}

\begin{Remark}
In our framework, regular users are passive, as they merely consume content produced by influencers: this constitutes a rather simplistic assumption. First, users can \textit{share} the posts they receive, which increases their reach. Secondly, users themselves write posts that reflect their opinion, influencing other users. 
The impact of \textit{active} users is  {briefly discussed in \ref{sec:extensions}.} 
\end{Remark}

\section{Observations from Online Social Networks} \label{sec:OSNobservations}
This section motivate some of our modeling choices by analyzing real-world social networks. We monitored on Facebook and Instagram the posts of 649 influencers for over 5 years. For a detailed description of the dataset used, see \ref{app:dataset}.
One of the most important features introduced in this paper is the concept of \textit{reference direction}, i.e., the main topic an influencer is interested in and on which they publish most of their posts, {which is validated here}. 
Moreover, we examine the post-generation process to justify the choice of a Poisson Point Process to describe it.

\begin{figure}[h!]
	\centering
	\begin{subfigure}{.5\textwidth}
		\centering
		\includegraphics[width=0.9\linewidth]{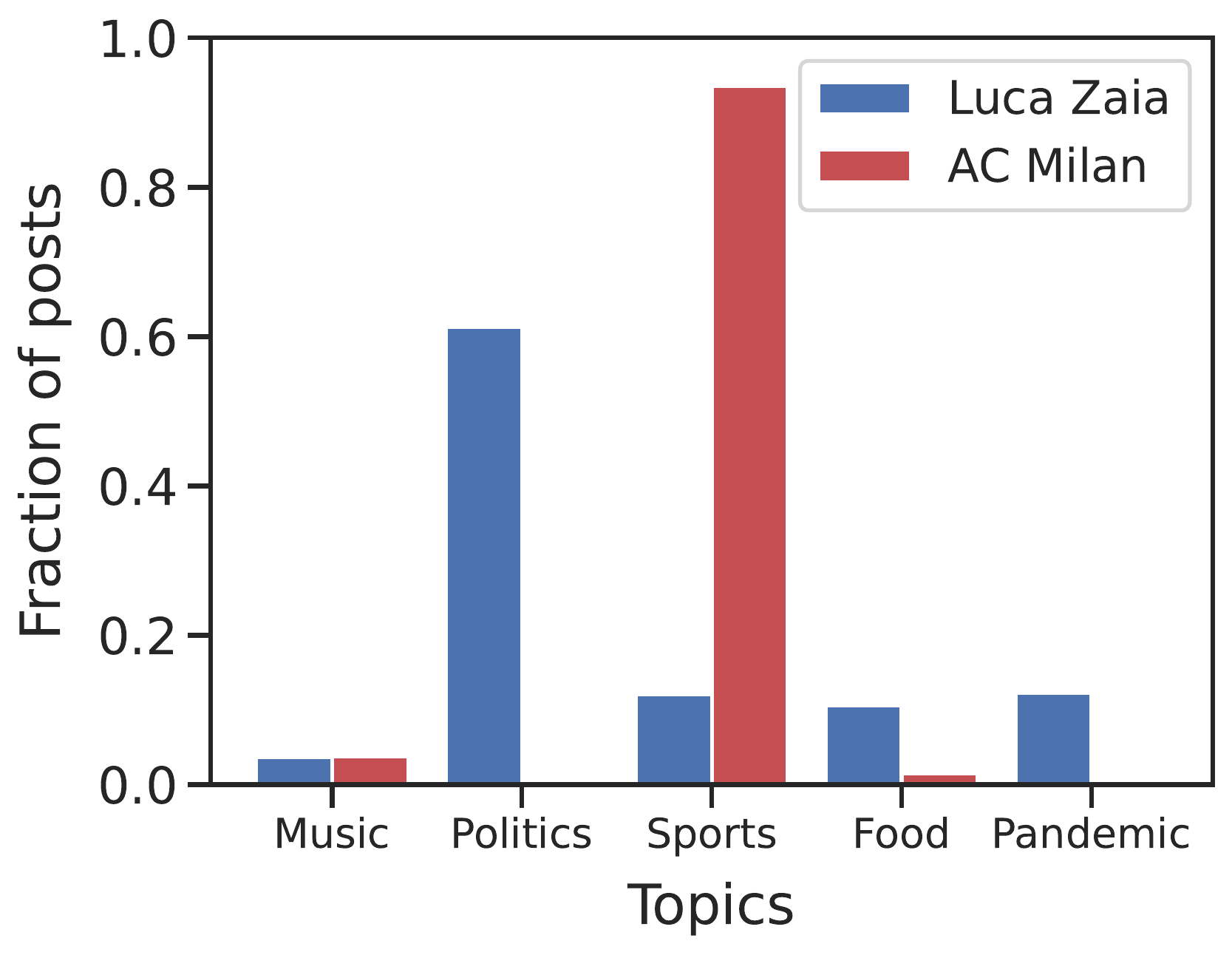}%[height=4cm]{/vect/milan_zaia.pdf}
		\caption{}
		\label{fig:milan_zaia_hist}
	\end{subfigure}%
	\begin{subfigure}{.5\textwidth}
		\centering
		\includegraphics[width=0.9\linewidth]{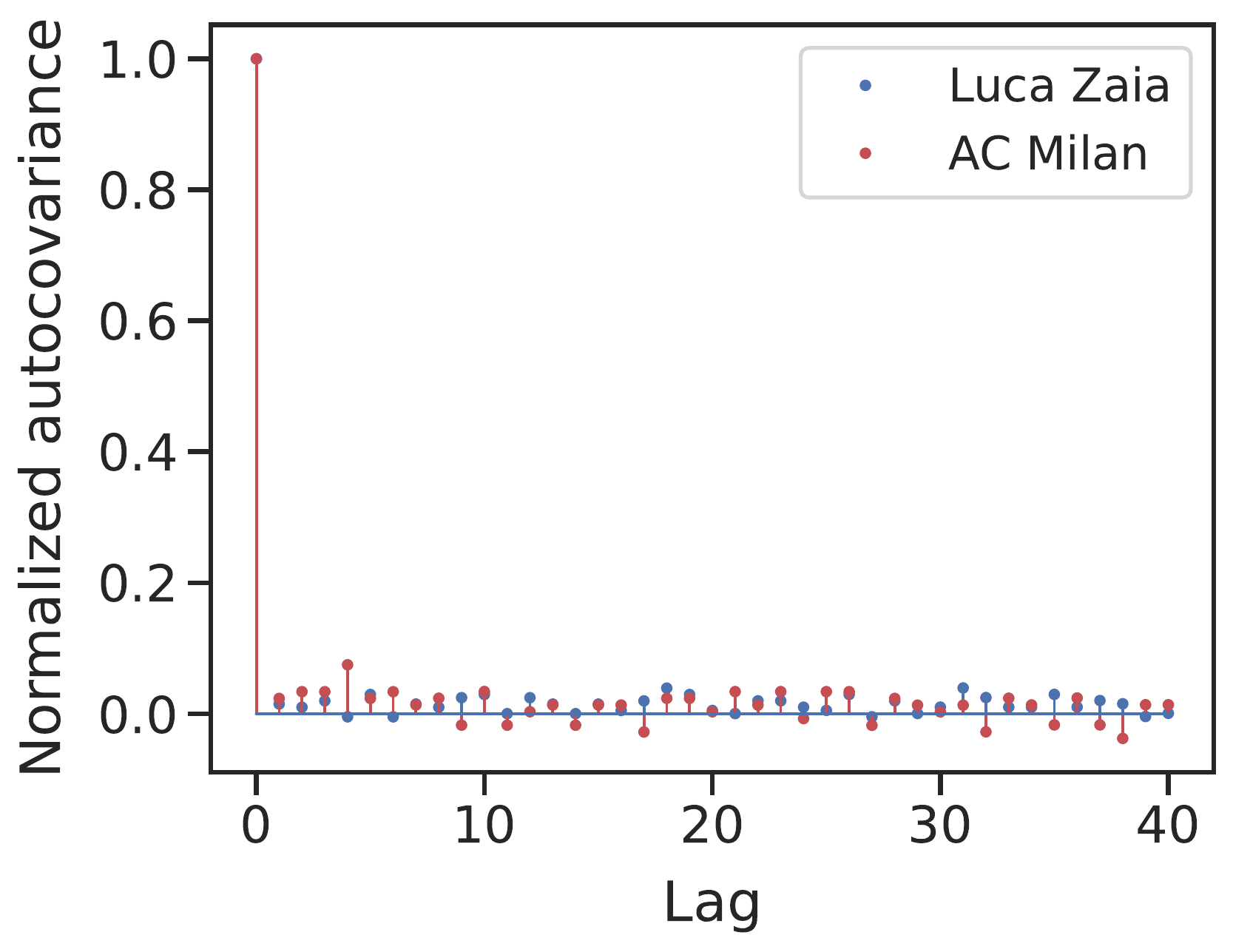}%[height=4cm]{/vect/milan_zaia_acf.pdf}
		\caption{}
		\label{fig:music_acf}
	\end{subfigure}
	\caption{(\ref{fig:milan_zaia_hist}) Percentage of labeled posts on each of the considered topics for Luca Zaia, an Italian politician, and AC Milan, an Italian football club. (\ref{fig:music_acf})Normalized autocovariance function for both influencers on a \textit{secondary} topic, i.e., music.}
	\label{fig:milan_zaia}
\end{figure}

\subsection{The reference direction}\label{sec:OSN_reference}

This section shows that influencers prefer to post about a specific topic. % rather than discuss multiple ones. 
We have developed a post classifier that flags posts based on their topic.  See \ref{app:classifier} for details on the classification and filtering process on the data. 
We should point out that classifying posts on OSNs into topics is not straightforward, and interpreting the results should be done cautiously. First, the range of possible subjects discussed in a social network is practically countless. For practical reasons, we will only focus on a subset of five topics: Sports, Politics, Food and Cooking, Music, and Pandemics.  
These can be considered popular and general enough to cover a substantial fraction of the influencers' posts taken into consideration in our dataset.

\begin{figure}[h]
	\begin{subfigure}{.5\textwidth}
		\centering
		\includegraphics[width=0.9\linewidth]{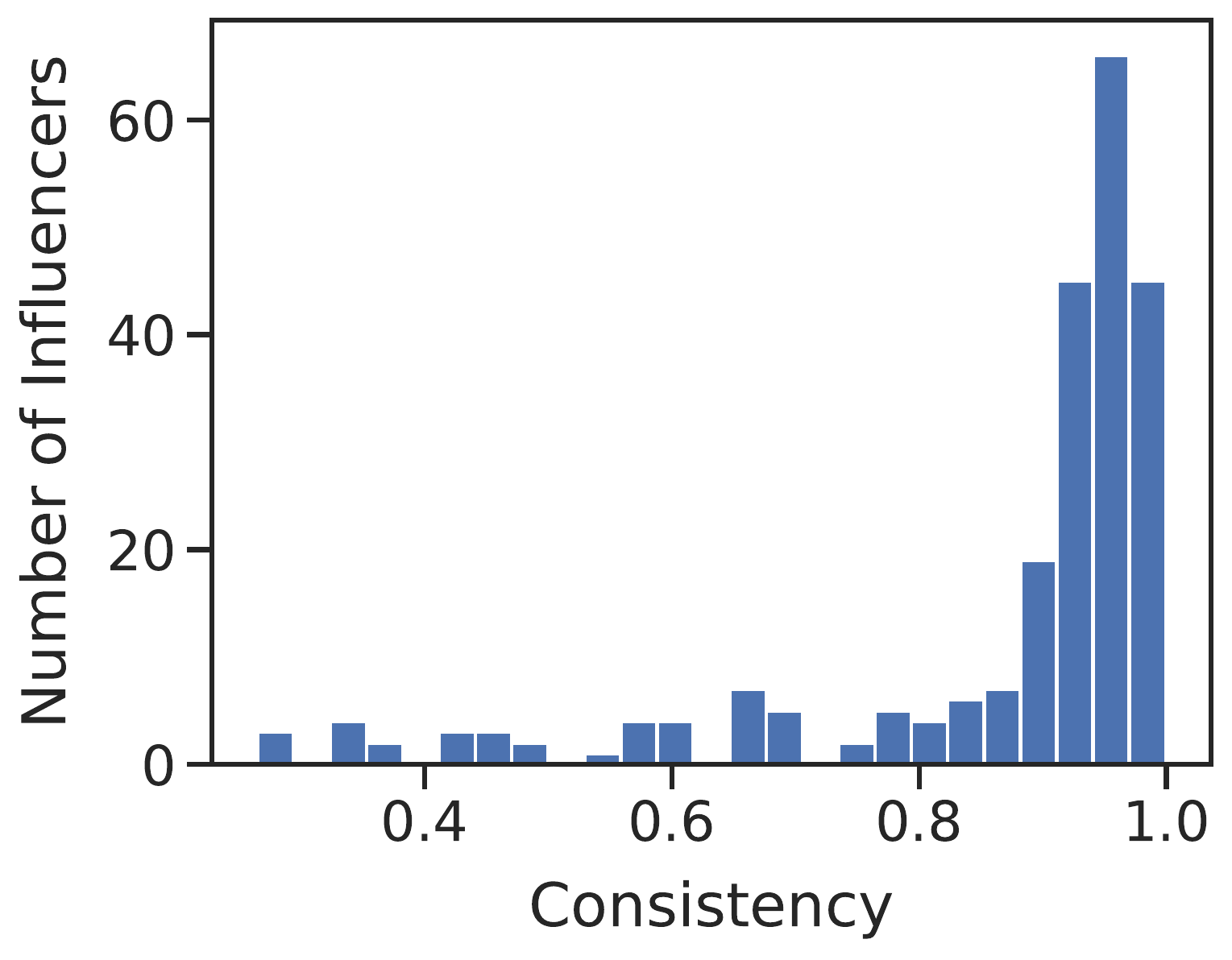}
		\caption{}
		\label{fig:cons_distrib}
	\end{subfigure}%
	\begin{subfigure}{.5\textwidth}
		\centering
		\includegraphics[width=0.9\linewidth]{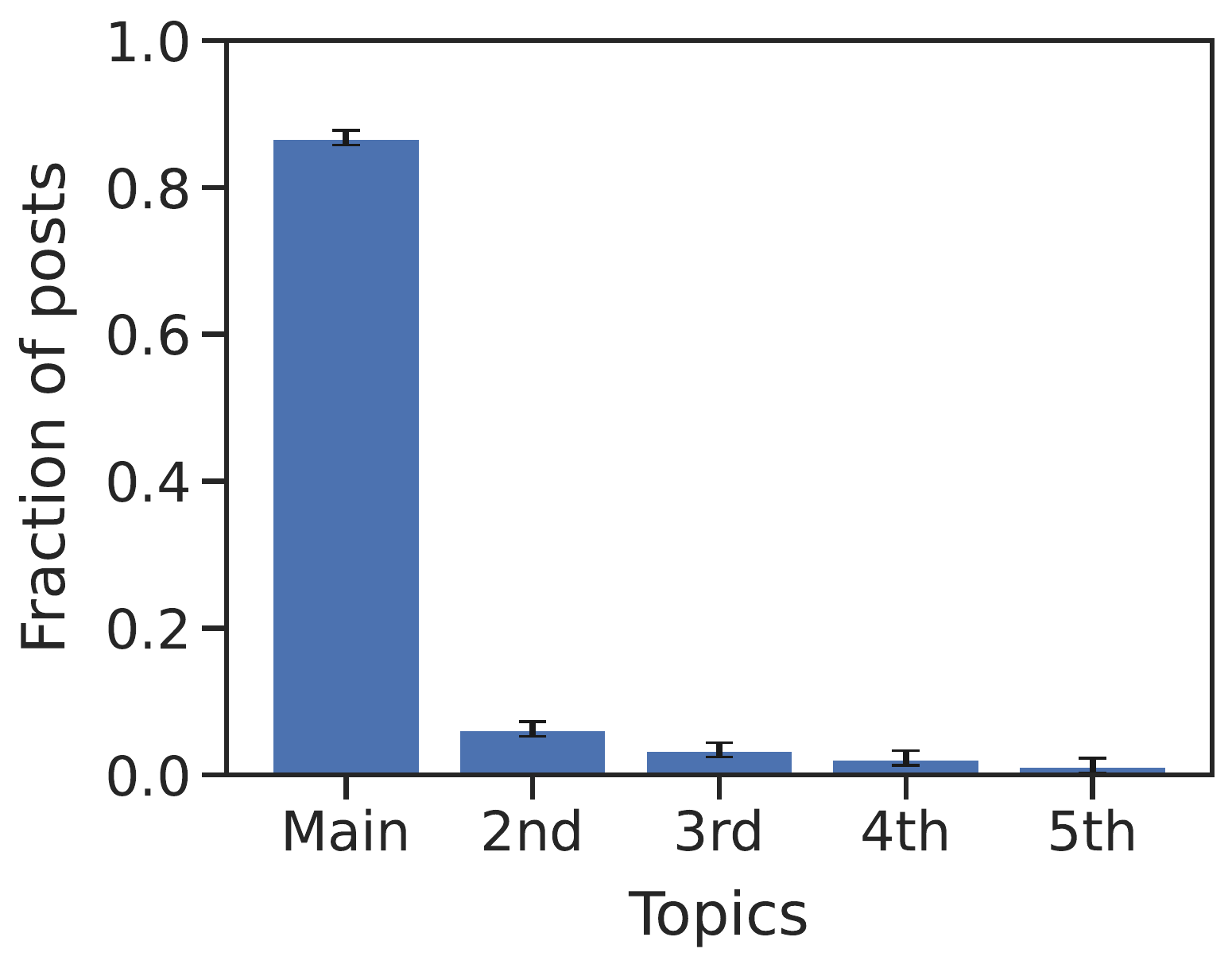}
		\caption{}
		\label{fig:topic_percentages}
	\end{subfigure}
	\caption{(\ref{fig:cons_distrib}) Distribution of the fraction of posts published on the main topic of interest by the subset of influencers considered in this experiment, i.e., their consistency.  (\ref{fig:topic_percentages}) The average percentage of labeled posts on each topic in decreasing order for all the influencers considered. The 95\% confidence interval for each average value is reported in the figure.}
\end{figure}

After classification, we examined the distribution of posts on the topics for each influencer. In Figure~\ref{fig:milan_zaia_hist}, we show two example influencers. In these two cases, the influencers have one topic on which they write most of their posts. Luca Zaia, an Italian politician, posts mainly about politics, and AC Milan, a soccer club, discuss sports predominantly. This behavior supports the existence of a reference direction for influencers. Figure \ref{fig:cons_distrib} shows the distribution of the proportion of posts dealing with the main topic of each influencer. Recall that this proportion was called \textit{consistency} in the jargon of our model.

Most influencers have a clear \textit{reference topic} on which they write more than half of their posts, i.e., with high consistency. 
Figure \ref{fig:topic_percentages} shows the average per-topic percentage of all influencers in the dataset in descending order, regardless of the specific topic. On average, almost 90\% of the posts are in the reference direction. 
We discovered that influencers with low consistency values are affected by the presence of news outlets in the considered profiles, for which the lack of a sharp main topic is sensible.

\subsection{Independence of posts' generation process on secondary directions}

Users interact in an OSN by posting content (i.e., text, images, videos) and receiving suggestions about what other users of the OSN posted, according to the filtering process set up by the social media platform.
We examine the {normalized autocovariance}\footnote{{Given  a wide-sense stationary  process  $\{X_n\}_{n}\in \mathbb{Z}$ with average $\mu$ and variance $\sigma^2$, the    normalized autocovariance is given by: $\rho_{XX}(i)=\frac{\mathbb{E}[(X_n-\mu)(X_{n+i}-\mu)]}{\sigma^2}$ }} between posts on each topic by looking at the chronological sequence of the messages of the individual influencers. {We perform this analysis only on \textit{secondary} topics, i.e., those that differ from the influencer's \textit{reference}. We do it since influencers post less frequently on these topics, and it would be easier to detect a bursty behavior pattern (which would not be well captured by the Poisson process).
	Regarding the main topic, since the consistency of the influencers is generally relatively high, we expect the covariance to be rather small, (see Figure \ref{fig:cons_distrib}). Indeed, the {covariance} on the main topic tends to zero by construction as $c^{(i)}$ approaches one. }

\begin{figure*}[h!]
	\centering
	\includegraphics[width=0.9\linewidth]{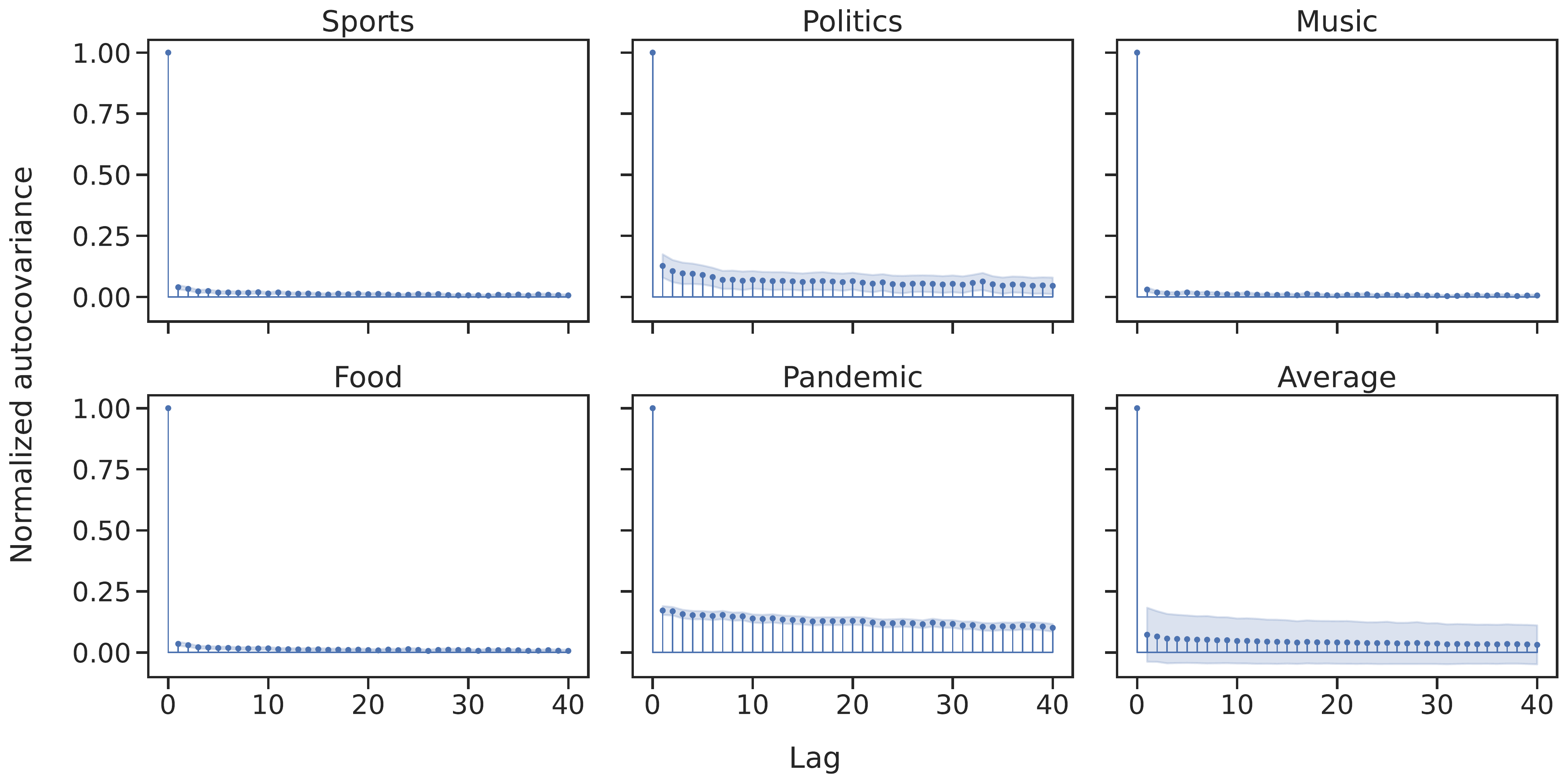}
	\caption{Mean {normalized autocovariance} values of the post-generation process for each secondary topic of all the influencers.
		The last plot represents the average value over all topics. The 95\% confidence interval is shown in each plot.}
	\label{fig:avg_acf}
\end{figure*}

In the previous section, we were able to assign a \textit{reference direction} $r^{(i)}$ to each influencer. Here we look at the time series of the Influencers' labeled posts. For each secondary direction $s_j^{(i)}$, we define an indicator function $\mathds{1}_{\{post_{label}=s_j^{(i)}\}}$ that takes the value $1$ if the post was labelled as $s_j^{(i)}$ and $0$ otherwise. For each influencer, we thus obtain four sequences (recall that we consider five topics in total) of Bernoulli random variables indicating whether a post belongs to that particular direction. We calculated the {normalized autocovariance} function $a(t)$ for these sequences. Figure \ref{fig:music_acf} shows two examples of such functions, limited to 40-time lags, for the profiles of Luca Zaia and AC Milan. The time is discretized, i.e., the actual time between postings is not considered: only the posting events matter. An {autocovariance} that equals zero everywhere except at $\tau=0$ would represent uncorrelated samples. In our case, the {autocovariance} takes moderate values in most cases ($ \ll 1$). Therefore, it is reasonable to assume that the post-generation is independent, and a Poisson Point Process is an appropriate choice. Lastly, note that the {autocovariance} function for the \textit{pandemic} topic takes larger values than for the other topics (see Figure \ref{fig:avg_acf}), suggesting that the samples are weakly correlated. This fact is due to the exceptional public interest in the topic and because the outbreak of the epidemic only interested the last part of the considered time horizon. 

\section{Asymptotic Analysis of the Model} \label{sec:analysis}

This section is devoted to the analytical study of the model. In particular, results are derived through a mean-field approach {obtained by letting the number of users} $N_u \rightarrow\infty$.
In this situation, {we will show that, under mild assumptions, the system converges to a unique steady state, independently from the initial condition.}
{Moreover, in some cases, it is possible to  analytically characterize the equilibrium value for the influencers' mean-popularity ratios $\bar{\pi}_i$ as well as users' mean opinion value $\bar{x}(z)$. %(which depend on prejudice $z$).
} 
Furthermore, transient analysis of the system can be carried out by describing the dynamics of the users through a Fokker-Plank equation. For simplicity, we restrict our investigation to the situation where the opinion space is one-dimensional. However, we remark that it is possible to extend the analysis to the more general case by following the same approach.

\subsection{Mean field approach}

When the number of users grows large, it is convenient to characterize the system state by the users' opinion {\em distribution} over the space. Moreover, from now on, we will refer to system dynamics over continuous time $t$.

Let $(X(t), Z(t))=(X(t), Z)$ be the current position (opinion) and prejudice of a randomly selected user.
We introduce the cumulative distribution function $F(x,z,t) = Pr[X(t)<x,Z<z]$. 
The corresponding probability density function is $f(x,z,t)=\frac{\partial^2}{\partial x \partial z} F(x,z,t)$. Note that, by hypothesis, there are no dynamics along the $z$-axes, thus $h(z) = \int_x  f(x,z,t) \mathrm{d} x$ does not depend on $t$ and corresponds to the initial distribution of users' prejudice.
In Section \ref{subsec:FP}, we will derive a Fokker-Plank equation for the evolution of the opinion distribution over time and space.

For what concerns the evolution of the popularity of a generic influencer $i$, recall that we distinguish between its absolute popularity value $p_i(t)$ and the normalized value $\pi_i = \frac{p_i(t)}{\sum_j p_j(t) }$. {%We remark that  i
	Influencer's popularities  concentrate around their average $\bar{p}_i(t)$ as $N_u$ grows large, as it can be easily shown.}
We can write down the equation for the evolution of the mean popularity $\bar{p}_i(t)$:

\begin{equation}\label{eqpop}
	\frac{\mathrm{d} \bar{p}_i(t)}{\mathrm{d} t }= \frac{1}{N_u} \lambda f^{(i)}
	\int_x \int_z  f(x,z,t)\,\theta\left(|x-x^{(i)}|\right) 
	\omega\left(\bar{\pi}_i, |x-x^{(i)}|\right)
	\mathrm{d}z \, \mathrm{d}x 
\end{equation} 
Indeed, the rate at which the popularity of influencer $i$ grows is proportional to its posting rate (term $\lambda f^{(i)}$) times the probability that a generic user at $(x,z)$ provides positive feedback to the post generated at  time~$t$ (integral term). 
Moreover, recall from Algorithm \ref{alg:model} that each positive feedback increases the absolute popularity of the influencer by $1/N_u$.

\subsection{Fokker-Planck equation for the opinion distribution}\label{subsec:FP}
The Fokker-Planck (FP) equation \cite{risken1996fokker} is a standard tool that describes the evolution of  an asymptotically   large population of particles moving over a given domain according to a Brownian motion, which is locally characterized by the instantaneous average velocity and the relative variance. 
In general, both average and variance may depend on the point and time at which they are evaluated.

Furthermore, the FP approach has been successfully applied in the literature to %approximately describe 
approximate the evolution of a large population of particles moving according to more general laws.
The stochastic process describing the movement of particles is approximated by a Brownian motion 
which fits the first two moments (average and variance of the velocity). For these reasons, FP description is referred to in the literature as a second-order approximation. 
Here, we essentially identify each user  opinion with a particle  moving over the space  $ \mathcal{X}^2$ 
(i.e., the set  $(x,z)$  with $x,y\in [a,b]$), and characterize by an average velocity $ v_x(x,z,t)$ along the $x$-axes,  and a variance $ \sigma^2_x(x,z,t)$, in Section~\ref{subsec:para}. 
As a consequence, the  Fokker-Plank equation for the probability density function $f(x,z,t)$ (where $x,z\in [a,b]$) is given by:
\textbf{\begin{equation}\label{FP}
		\frac{\partial f(x,z,t)}{\partial t} = - \frac{\partial v_x (x,z,t) f(x,z,t)}{\partial x} + \frac{1}{2} \frac{\partial^2 \sigma_x^2(x,z,t) f(x,z,t)}{\partial x^2}
\end{equation}}
\subsection{Identification of the parameters $ v_x(x,z,t)$ and $ \sigma^2_x(x,z,t)$}\label{subsec:para}
{To compute $ v_x(x,z,t)$}, in the continuous-time FP approximation, we assume that for the effect of a post, \lq\lq users/particles" reach their new position by moving at a constant speed during the interval $\Delta T$ equal to the average time $1/\lambda$ that elapses between the generation of two successive posts. Therefore, assuming that at time $t$ a post is generated by user $i$, the following equation describes how the opinion of a user with prejudice $z$ evolves from $t$ to $t + \Delta T$:

\begin{equation*}
	x(t+\Delta T )=\alpha z + \beta x(t) + \gamma x^{(i)}(t)
\end{equation*}
Thus, the increment is:
\begin{equation}
	\Delta x(i) = x(t+\Delta T )-x(t) = \alpha (z-x^{(i)}(t)) + (1-\beta) (x^{(i)}(t)-x(t))
\end{equation}
where we remark that $\Delta x(i)$ here represents the change in position of a user in position $x$, providing positive feedback to a post of influencer $i$.

\begin{align} \label{vel-FP}
	\mathbb{E}[  v_x(x,z,t)\mid X(t)=x,  Z=z ] &= \mathbb{E}\left[\frac{[X(t+\Delta T) -X(t)\mid X(t)=x,  Z=z ]}{\Delta T}\right] \nonumber\\
	&= \sum_i \ \lambda f^{(i)} \Delta T\,\, \theta\left(|x-x^{(i)}|\right) \omega \left(\bar{\pi}_i(t), |x-x^{(i)}|\right) \frac{\Delta x(i) }{\Delta T}\nonumber\\
	&= \sum_i  \lambda f^{(i)} \theta\left(|x-x^{(i)}|\right) \omega \left(\bar{\pi}_i(t), |x^{(i)}-x|\right) \Delta x(i) 
\end{align}
%\begin{multline} \label{vel-FP}
%	\mathbb{E}[  v_x(x,z,t)\mid X(t)=x,  Z=z ]= \mathbb{E}\left[\frac{[X(t+\Delta T) -X(t)\mid X(t)=x,  Z=z ]}{\Delta T}\right] \\
%	= \sum_i \ \lambda f^{(i)} \Delta T\,\, \theta\left(|x-x^{(i)}|\right) \omega \left(\bar{\pi}_i(t), |x-x^{(i)}|\right) \frac{\Delta x(i) }{\Delta T} \\
%	= \sum_i  \lambda f^{(i)} \theta\left(|x-x^{(i)}|\right) \omega \left(\bar{\pi}_i(t), |x^{(i)}-x|\right) \Delta x(i) 
%\end{multline}
where $\theta\left(|x-x^{(i)}|\right)$ is the probability of providing positive feedback (users move only in this case), while $\omega\left(\bar{\pi}_i(t), |x-x^{(i)}|\right)$ is the probability with which a user in $x$ is exposed to a post created by influencer $i$ at time $t$. 
Indeed, users only move if they are exposed to the post and provide positive feedback.
Note that, to avoid a cumbersome notation, we have omitted the dependency on the time of the distance term $|x-x^{(i)}|$.  

The  variance of the velocity is given by the relation:
\begin{align}
	\sigma_x^2(x,z,t) &= \sum_i \ \lambda f^{(i)} \Delta T   \theta\left(|x-x^{(i)}|\right) \omega \left(\bar{\pi}_i (t), |x-x^{(i)}|\right) \frac{(\Delta x(i)-  \mathbb{E}[  v_x(x,z,t)]\Delta T )^2}{\Delta T^2  } \nonumber\\
	&=\frac{1}{(\Delta T)^2} \sum_i  f^{(i)}{  \theta\left(|x-x^{(i)}|\right) \omega \left(\bar{\pi}_i (t), |x-x^{(i)}|\right) (\Delta x(i)- \mathbb{E}[  v_x(x,z,t)]\Delta T ) ^2} \nonumber
\end{align}

\subsection{Steady state analysis} \label{subsec:steady}
Now we direct our attention to the existence of stationary solutions for the system.
Stationary solutions of  \eqref{FP}  necessarily satisfy:
\[
\frac{\partial }{\partial x} \left(-v_x (x,z) f(x,z)+ \frac{1}{2} \frac{\partial \sigma^2_x(x,z) f(x,z)}{\partial x}\right)=0
\]
where $v_x (x,z) $ and  $\sigma^2_x(x,z)$ must be constant over time. This  requires the normalized popularities to be  static (i,e. $\omega(\cdot)$ to be constant over time).
From the previous equation, integrating both sides with respect to $x$, we get:
\begin{equation}\label{fokker-stat}
	\left(-v_x (x,z) f(x,z)+ \frac{1}{2} \frac{\partial \sigma^2_x(x,z) f(x,z)}{\partial x}\right)=c_0(z)
\end{equation}
where $c_0(z)$ is a uni-dimensional arbitrary  in $z$. Now, observe that,  for every $z$,  previous equation is a first order linear ODE in $x$, and therefore an explicitly solution for $f(x,z)$ can be obtained:
\begin{equation}\label{solution-FP}
	f(x,z)= \Big(c_1(z) \exp( A(x,z)-A(a,z))+c_0(z)\exp(-A(x,z))\int_a^x \exp (A(\theta,z))  \mathrm{d} \theta\Big) h(z)
\end{equation}
where
\[
A(x,z)= \int_a^x   \eta(u,z) \mathrm{d}  u  \qquad \eta(x,z) = - 2 \frac{v_x(x,z) - \frac{1}{2}\frac{\partial \sigma^2_x(x,z)}{\partial x} } {\sigma^2_x(x,z)}.
\] 
Function $c_0(z)$ can be obtained by imposing boundary conditions:
\[\left(
-v_x(x,z)f(x,z)+ \frac{1}{2} \frac{\partial}{\partial x}\sigma^2_x(x,z)f(x,z)\right) \Bigl\vert_{x=a,b}=0.  \qquad \forall z
\]
which leads to $c_0(z)=0$, while function $c_1(z)$ is  determined by imposing the normalization condition:
\[
\int f(x,z)\mathrm{d}x=h(z).
\]

Observe that when $\sigma_x^2(x,z)\to 0$ and $\frac{\partial\sigma_x^2(x,z)}{\partial x}\to 0$,  from \eqref{fokker-stat}, with $x_0(z)=0$, we obtain that necessarily the mass concentrates around the points for which $v_x(x,z)=0$.
Such points, improperly referred to in the following as {\it equilibrium points}, will be characterized analytically later on.

Turning our attention to popularity dynamics, recall that stationary conditions necessarily imply normalized popularities to be constant over time: 
\[
\bar{\pi}_i(t)=\bar{\pi}_i   \qquad \forall i
\]

On the other hand, absolute popularities naturally grow over time. However, the ratio between any two of them (say $i,j$) must converge to a constant value $c_{ij}$ equal to the ratio of their corresponding normalized popularities:
\begin{equation}\label{stat-cond}
	\frac{\bar{p}_i(t)}{ \bar{p}_j(t)}=c_{ij} = \frac{\bar{\pi}_i}{\bar{\pi}_j}   \qquad \forall i,j\in \mathcal{I}, i\neq j %\text{ and }
	% t\in \mathrm{R}^+
\end{equation}

Now observe that in stationary conditions the right-hand side. of \eqref{eqpop} does not depend on time. Therefore \eqref{eqpop} 
admits the following trivial solution:
\begin{equation}\label{eq:pit}
	\bar{p}_i(t)  = \left(\lambda f^{(i)}   \int_x \int_z  \theta(|x-x^{(i)}|) \omega (\bar{\pi}_i, |x-x^{(i)}|)\mathrm{d}F(x,z) \right) \frac{t}{N_u} + \bar{p}_i(0) 
\end{equation}

Therefore, %for assigned initial conditions $\{p_i(0)\}_i$, 
we meet conditions \eqref{stat-cond} for any $t\ge 0$,  iff
normalized popularities of influencers $\{ \bar{\pi}_i \}_i$
%,  $\{\widetilde p_i\}_i$ 
satisfy the following system of equations:
\begin{align}\label{pop-constraint}
	&\lambda f^{(i)}   \int_x\int_z \theta(|x-x^{(i)}|) \omega (\bar{\pi}_i , |x-x^{(i)}|)\mathrm{d}F(x,z) = c \bar{\pi}_i  \qquad 
	\forall i,   \text{ for some } c\in \mathbb{R}^+\nonumber \\
	& \qquad \text{ s.t.  } \bar{\pi}_i\ge 0 \text { and } \sum_i \bar{\pi}_i=1.
\end{align}
and the initial condition $\{p_i(0)\}_i$ satisfies \eqref{stat-cond}, i.e.,  $p_i(0) = k \bar{\pi}_i $  for some $k>0$.

Let
\begin{equation}\label{eqpi}
	k_i( \bar{\pi}_i)\triangleq \lambda f^{(i)}   \int_x\int_z 
	\theta(|x-x^{(i)}|) \omega (\bar{\pi}_i, |x-x^{(i)}|) \mathrm{d}F(x,z)
	\quad \bar{\pi}_i \in [0,1]
\end{equation}

We can show that:
\begin{Theorem} \label{theo1}
	Solutions of \eqref{pop-constraint} always exist whenever $k_i(\cdot)\in C_1[0,1]$,  $k_i(\cdot)$ is increasing, continuous  and 
	strictly concave.  
\end{Theorem}
\vspace{-1mm}
The proof is reported in \ref{app:proofs}.
\vspace{2mm}

We remark  that when $k_i(0)> 0$ $\forall i$, the solution is always unique with $\bar{\pi}_i\in  (0,1)$.
Instead when  $k_i(0)= 0$  for some $i$, the solution is not guaranteed to be unique.

Now, the problem is how to jointly solve
for stationary solutions of $\{\bar{\pi}_i\}_i$ and $F(x,z)$.
In a schematic way, on the one hand,  we have shown that 
given $\bar{\bm\pi} = \{\bar{\pi}_i\}_i$, and $h(z)$, we can uniquely determine a $F_{\bar{\bm\pi}}(x,z)=\mathcal{H}(\bar{\bm\pi})$, where   
%(where $F_{\tilde p}(x,z)=\int^x \int^z f_{\tilde p}(y,w) \mathrm{d} x $) 
$F_{\bar{\bm\pi}}(x,z)=\int_{-\infty}^x \int_{-\infty}^z f_{\bar{\bm\pi}}(y,w) \,\mathrm{d} y\, \mathrm{d} w$ is the opinion distribution of users
resulting from fixed influencers' popularities $\bar{\bm\pi}$
(by (\ref{solution-FP})).

On the other hand, under the conditions:
$k_i(\cdot)\in C_1[0,1]$,  $k_i(\cdot)$ is increasing and 
strictly concave,   $k_i(0)>0$ $\forall i$,
given $F(x,z)$,  we can obtain a  ${\bar{\bm\pi}}_{F}=\mathcal{G}(F(x,z))$ that {uniquely} corresponds to $i$ (Theorem \ref{theo1}). 
The existence of  a unique  fixed point for the joint system
of (stationary) users' opinions and influencers' popularities  
is guaranteed  under the condition that the operator $\mathcal{H}\circ\mathcal{G}(\cdot)$ is  a contraction over a complete space.

\begin{Theorem} \label{theo2}
	Under the assumption that both  $\omega(\cdot, \cdot)$ and $\theta(\cdot)$ exhibit a sufficiently weak dependence on their variables, the operator $\mathcal{H}\circ\mathcal{G}(\cdot)$ is a contraction over a complete space, and therefore a unique stationary solution exists.
\end{Theorem}
\vspace{-1mm}
The proof is reported in \ref{app:proofs}.

\subsection{Asymptotic analysis of the fluid limit} \label{sec:stoch_comp}

Previous theoretical analysis is, unfortunately, non-constructive, meaning that it does not allow for direct computation of stationary solutions of our dynamical system. 
To complement the previous analysis, in this section, we propose a methodology to numerically compute stationary solutions, even in multi-dimensional scenarios, under {stricter assumptions.}
{In particular, if the FP approach is a second-order approximation (matching the first to moments of the  instantaneous velocity),  the approach proposed in this section, and referred to as {\it fluid limit},  is a first-order approach  (matching only the first moment of the velocity and assuming the variance, as well as its spatial derivative, to be negligible).
	Therefore we expect that {\it fluid limit}  provides reasonable good predictions when $N_u\to \infty$,  $\lambda\to \infty$   $1-\beta\to 0$  (keeping $\lambda(1-\beta)$ constant).
	Previous assumptions, indeed, imply that  $\sigma_x^2(x,z)\to 0$ and $\frac{\partial\sigma_x^2(x,z)}{\partial x}\to 0$.} 

\subsubsection{Mean opinion assuming that normalized popularities converge}
As already observed in Section \ref{subsec:steady}, recall that, given $\bar{\bm\pi}=\{ \bar{\pi}_i \}_i$, the distribution of users with a given prejudice $z$ concentrates around {\it equilibrium points}, i.e., points $\bar{x}(z) $ at which $v(x,z)$,  as given in \eqref{vel-FP}, is  null  (i.e. $v(\bar{x}(z),z)=0$).  
Therefore, points $\bar{x}(z) $ must satisfy equation:
\begin{equation}\label{eq-FPequi}
	0= \sum_i  f^{(i)} \omega \left(\bar{\pi}_i, |\bar{x} - x^{(i)}|\right) \theta\left(|\bar{x}-x^{(i)}|\right) 
	\left(   \alpha (z-x^{(i)}) + (1-\beta) (x^{(i)}-\bar{x})\right)
\end{equation}

Defining for compactness $d^{i,\bar{x}} = \left|\bar{x} - x^{(i)}\right|$ and recalling $\gamma = 1-\alpha-\beta$,
from (\ref{eq-FPequi}) we get:
\begin{equation} \label{eq:stoch_e}
	\bar{x}(z) = \frac{\alpha}{1-\beta} z + \frac{\gamma}{1-\beta} \frac{\sum_{i \in \mathcal{I}} f^{(i)} \omega\left(\bar{\pi}_i, d^{i,\bar{x}} \right) \theta\left( d^{i,\bar{x}} \right) x^{(i)}}{\sum_{i \in \mathcal{I}} f^{(i)} \omega\left(\bar{\pi}_i, d^{i,\bar{x}} \right) \theta\left( d^{i,\bar{x}} \right)}
\end{equation}

{We can rewrite this relation in terms of the \textit{degree of stubbornness} of Eq. (\ref{eq:update_equivalence}) as $\bar{x} = \delta z + (1-\delta) \frac{\sum_{i \in \mathcal{I}} f^{(i)} \omega\left(\bar{\pi}_i, d^{i,\bar{x}} \right) \theta\left( d^{i,\bar{x}} \right) x^{(i)}}{\sum_{i \in \mathcal{I}} f^{(i)} \omega\left(\bar{\pi}_i, d^{i,\bar{x}} \right) \theta\left( d^{i,\bar{x}} \right)}, \, \delta \in [0,1]$. By so doing it should be clear that the assumption $\beta \rightarrow 1$ is well-founded, reinforcing the idea that $\beta$ is associated with the \textit{inertia} of the system (see the end of Section \ref{sec:model_desc}) but does not affect the equilibrium points.} This assumption is required to avoid too large oscillations of users' opinions in response to a single post generated by an influencer, which may reduce the accuracy of our mean-field approximation, {validated in \ref{sec:simulation}}.

The hypothesis is not restrictive: since $\beta$ represents the weight individuals give to their current opinion, we can reasonably assume that users do not dramatically change their opinion in response to single post events.

\subsubsection{Normalized popularities assuming opinion convergence}
Here we assume that users with prejudice $z$ are concentrated in opinion point $\bar{x}(z)$, and we look for the stationary popularity ratios $\bar{\pi}_i$.
To simplify the expressions, we introduce the quantity
$F_i(\bar{\pi}_i) \triangleq \int_z
f^{(i)}\omega(\bar{\pi}_i, d^{i,\bar{x}(z)}) \theta(d^{i,\bar{x}(z)}) h(z) 
\mathrm{d} z$.

{Observe that solutions of \eqref{pop-constraint}   are necessarily of the form:
	\begin{equation} \label{eq:tilda_p}
		\bar{\pi}_i = \frac{F_i(\bar{\pi}_i)}{\sum_{j \in \mathcal{I}}  
			F_j(\bar{\pi}_j)}  
	\end{equation}
	where $c$ appearing in \eqref{pop-constraint} is given by
	$c= \frac{1}{\sum_{j \in \mathcal{I}}  	F_j(\bar{\pi}_j )}$.  Under the assumption that  $\omega(\cdot ,\cdot)$ 
	is concave in its first argument (for any choice of the second), Theorem \ref{theo1} guarantees the existence of such solutions for every choice of  function $\bar{x}(z)$.  Moreover,  even in the more general case, i.e.,  when $\omega(\cdot ,\cdot)$  is non-concave in its first argument, solutions of \eqref{eq:tilda_p} can be found  numerically in many cases, through a fixed point iteration method.}

To conclude,  observe that a pair $(\bar{x}(z), \{\bar{\pi}_i \}_i)$  represents a stationary solution if it jointly satisfies \eqref{eq:stoch_e} and \eqref{eq:tilda_p}. The existence of such a solution  can be, again, only  verified  {\it numerically} through a fixed point approach.

At last,  note that, in the special case in which all users have the same prejudice $z$
we can rewrite (\ref{eq:stoch_e}) as:
\begin{equation}\label{eq:o_bar_comp}
	\bar{x} = \frac{\alpha}{1-\beta} z + \frac{\gamma}{1-\beta} \frac{\sum_{i \in \mathcal{I}} F_i(\bar{\pi}_i, \bar{x}) x^{(i)}}{\sum_{i \in \mathcal{I}} F_i(\bar{\pi}_i, \bar{x})} = \frac{\alpha}{1-\beta} z + \frac{\gamma}{1-\beta} \sum_{i \in \mathcal{I}} \bar{\pi}_i x^{(i)}
\end{equation}
which provides a direct formula for the mean opinion $\bar{x}$ in terms of the normalized popularities $\bar{\pi}_i$ and the influencers' opinions $x^{(i)}$.

\section{Monte-Carlo approach}\label{sec:model_vs_param}

This section presents a selection of results obtained through a Monte-Carlo approach. We vary model parameters to {explore the} impact of {content} personalization {and influencers' characteristics}. 
We focus on the two main dynamic variables of the system: the average opinion $\bar{\bm{x}}$ of regular users and the normalized popularities $\{ \bar{\pi}_i \}_i$ for influencers. {Note that the quantities shown in this section, i.e., the pair $(\bar{\bm{x}}, \bar{\bm \pi})$, are empirical averages over multiple runs, and over all the regular users, as far as $\bar{\bm{x}}$ is concerned.
	Hence, they can be regarded as empirical, finite system   approximations of the quantities defined in the previous section, which refer to the limiting case of an infinite population of users with the same prejudice $z$, and where $\lambda\to \infty$ and  $\beta$ approaches~$1$.}
Moreover, in some cases, we omit the results on average user opinion to save space as it is tightly coupled with the normalized popularities, as observed in the previous section. 

Lastly, to facilitate the interpretation of results, we restrict ourselves to the case of two \textit{\lq \lq competing"} influencers. {We are interested in {determining the conditions under which} an individual attains higher $\bar{\pi}_i$ than the other.  %$\bar{\pi}_i \rightarrow 1$
	We say that influencer  $i$ \lq\lq wins" over the opponent, {which is not anymore visible over the platform,} when  $\bar{\pi}_i \rightarrow~1 \implies \omega(\cdot,\bar{\pi}_j) \rightarrow~0 \,\, \forall j \in \mathcal{I}$.}
{Note that this scenario is by no means trivial and is relevant in various applications, for example, in marketing (e.g., two brands promoting the same product) or in election campaigns (e.g., two candidates of different parties).} We provide further details on the scenario in Section~\ref{sec:sim_scenario}.  In Section~\ref{sec:final_conf} we show final opinion distributions of the regular users in a few paradigmatic cases. Then, in Section~\ref{sec:sim_rate}, we present the behavior as  function of publication frequency $f^{(i)}$, and in Section~\ref{sec:sim_cons} as  function of consistency $c^{(i)}$.
{We provide additional details in \ref{sec:appendix_sim}.}

\subsection{Description of the scenario}\label{sec:sim_scenario}

The default parameters of our reference scenario are reported in {the Appendix} (Table \ref{tab:table_param}), unless otherwise explicitly stated. 
As mentioned earlier, we consider the case of two \lq \lq competing" influencers, i.e., $N_i=2$. 
The opinion space is supposed to be bi-dimensional.
We assume that $x_j^{(0)}=0$ and $x_j^{(1)}=1$ $\forall j\in \{1,2\}$ {hence the influencers are placed on two antipodal vertices of  the  square $[0,1]^2$. }
{We consider influencers with different reference directions $r^{(0)} \neq r^{(1)}$. Furthermore, most regular users initially take a \textit{moderate} position on both topics. More precisely, initial opinions, which coincide with prejudices,}  are distributed according to a Beta distribution, independently on each axis, with shape parameters $a=b=10$ {(Fig. \ref{fig:op_init} in the Appendix)}. 

We take as $\omega(\cdot)$ a Gaussian function similar to the \textit{trust} function in \cite{cohen_tsang_ONS}, but modulated by the normalized popularity $\bar{\pi}_i$: ${ \omega(d_r^{i,u}, \bar{\pi}_i) = e^{-\rho\frac{\left(x_{r}^{(u)}-x_{r}^{(i)}\right)^2}{\bar{\pi}_i}}}$.
Here, the coefficient $\rho$ is a parameter that controls the extent to which the social media platform filters content,
{i.e., which expresses the homophilic degree over the network in a synthetic way}. Small values of $\rho$ correspond to {\textit{smooth}} personalization, i.e., influencers can reach users whose opinion strongly differs from theirs. 
Conversely, high values of $\rho$ correspond to {\textit{sharp}} personalization: only close users (in the opinion space) are reachable with non-negligible probability.
The function $\theta(\cdot)$ is assumed to be a decreasing, linear function of the opinion difference: $\theta(d_j^{i,u}) = 1 - \left|x_j^{(i)} - x_j^{(u)}\right|$. 

\subsection{Opinion configuration considering combinations of reference directions}\label{sec:final_conf}

\begin{figure}
	\centering
	\vspace{-2.5cm}
	\begin{subfigure}{.48\textwidth}
		\centering
		\includegraphics[width=\linewidth]{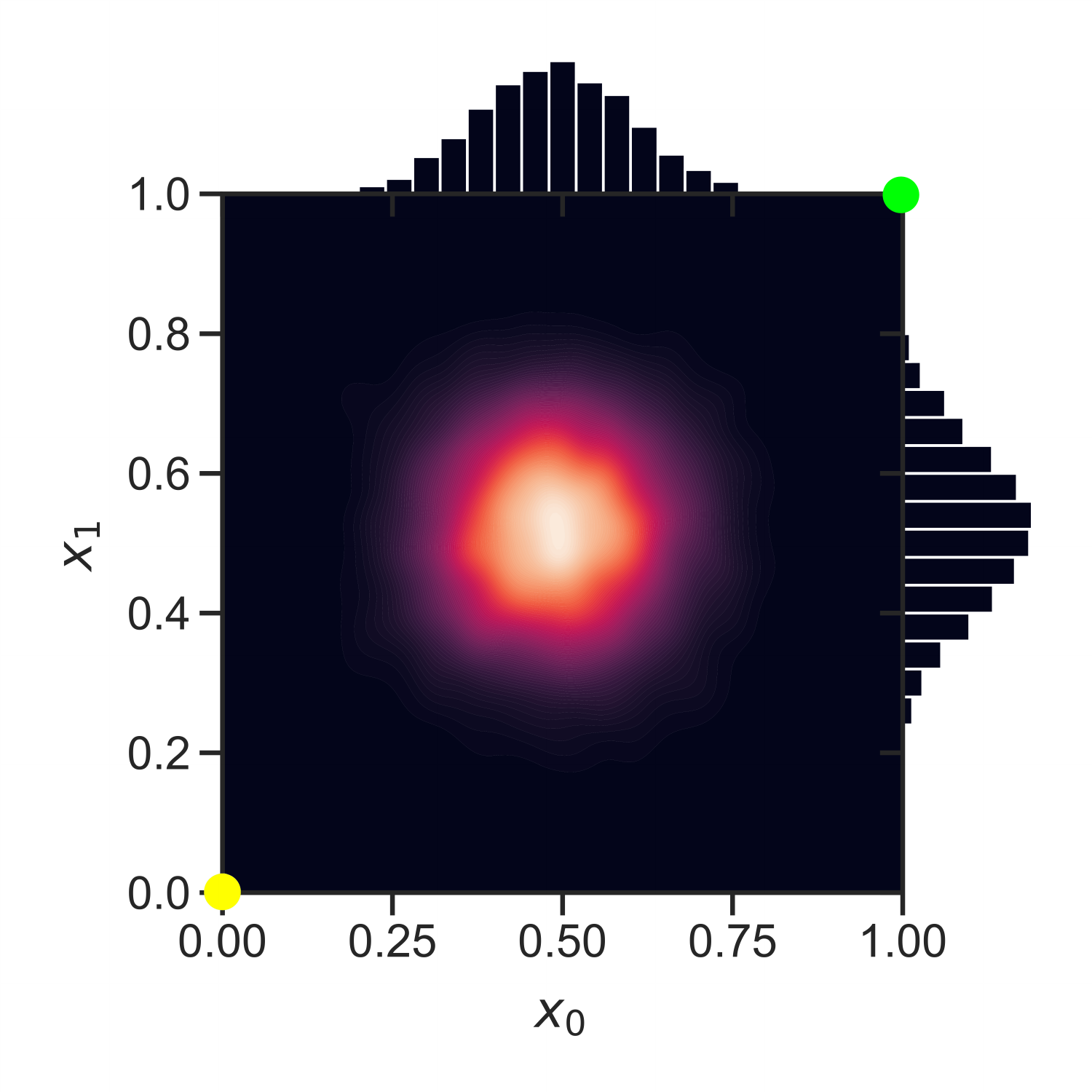}
		\centering
		\includegraphics[width=\linewidth]{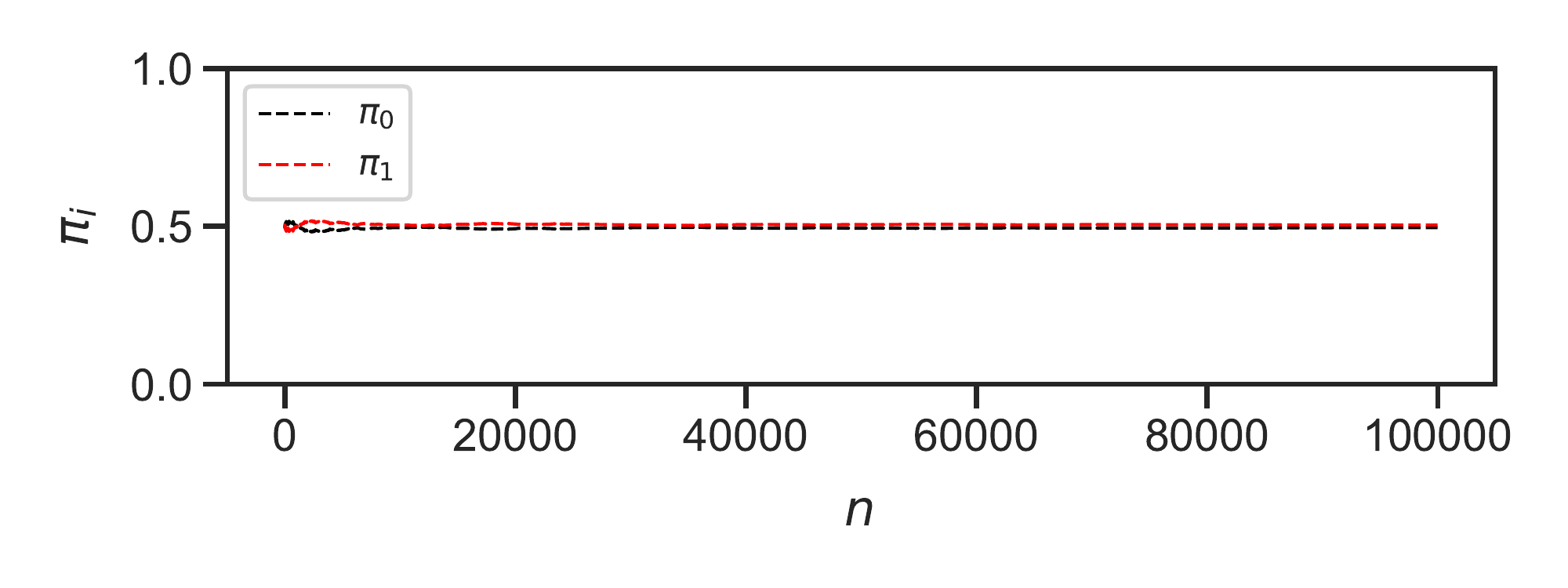}
		\caption{\parbox{0.9\linewidth}{\centering Same reference dir. $r^{(0)}=r^{(1)}=0$ \\Smooth personalization $\rho=5.0$}}
		\label{fig:same_loose}
	\end{subfigure}%
	\begin{subfigure}{.48\textwidth}
		\centering
		\includegraphics[width=\linewidth]{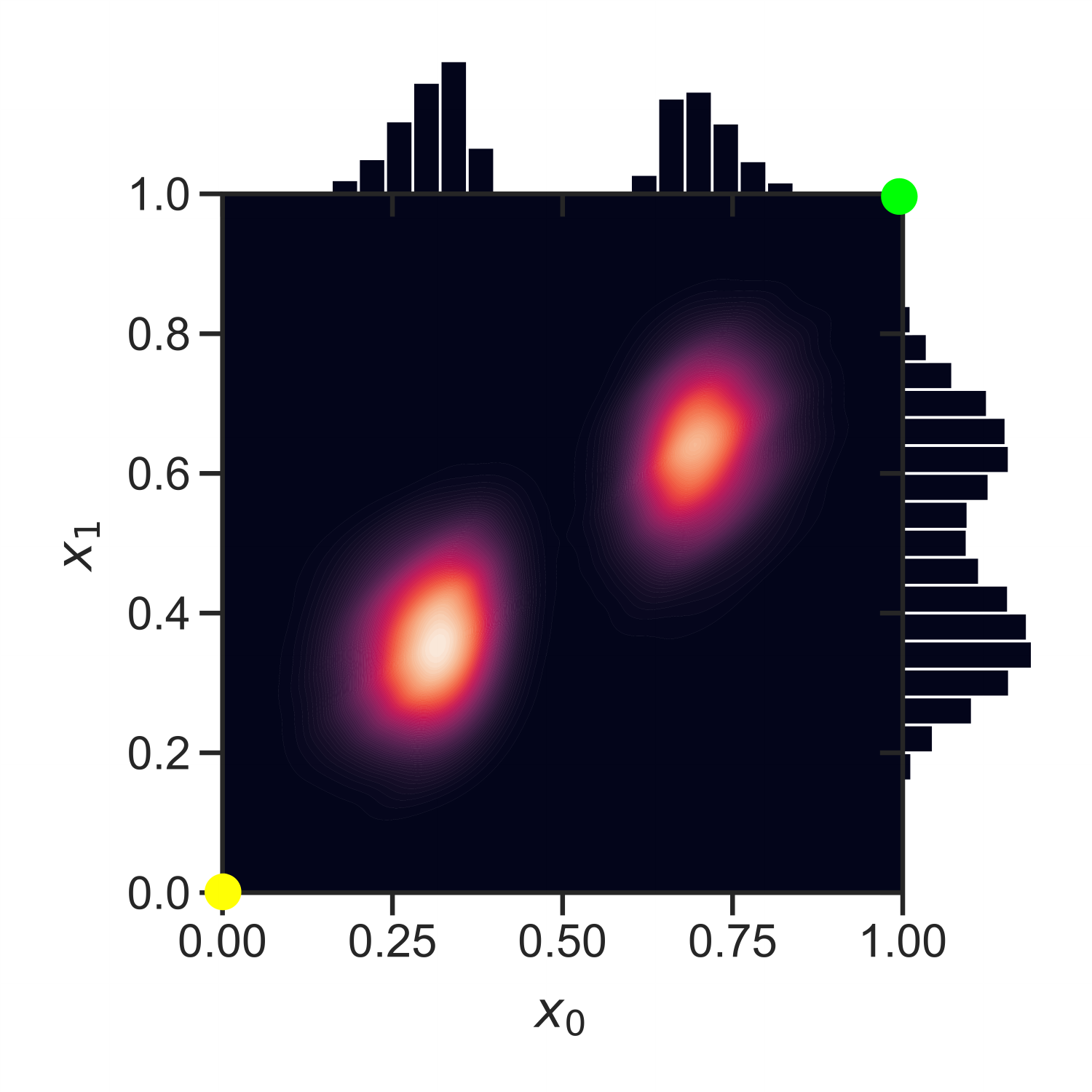}
		\centering
		\includegraphics[width=\linewidth]{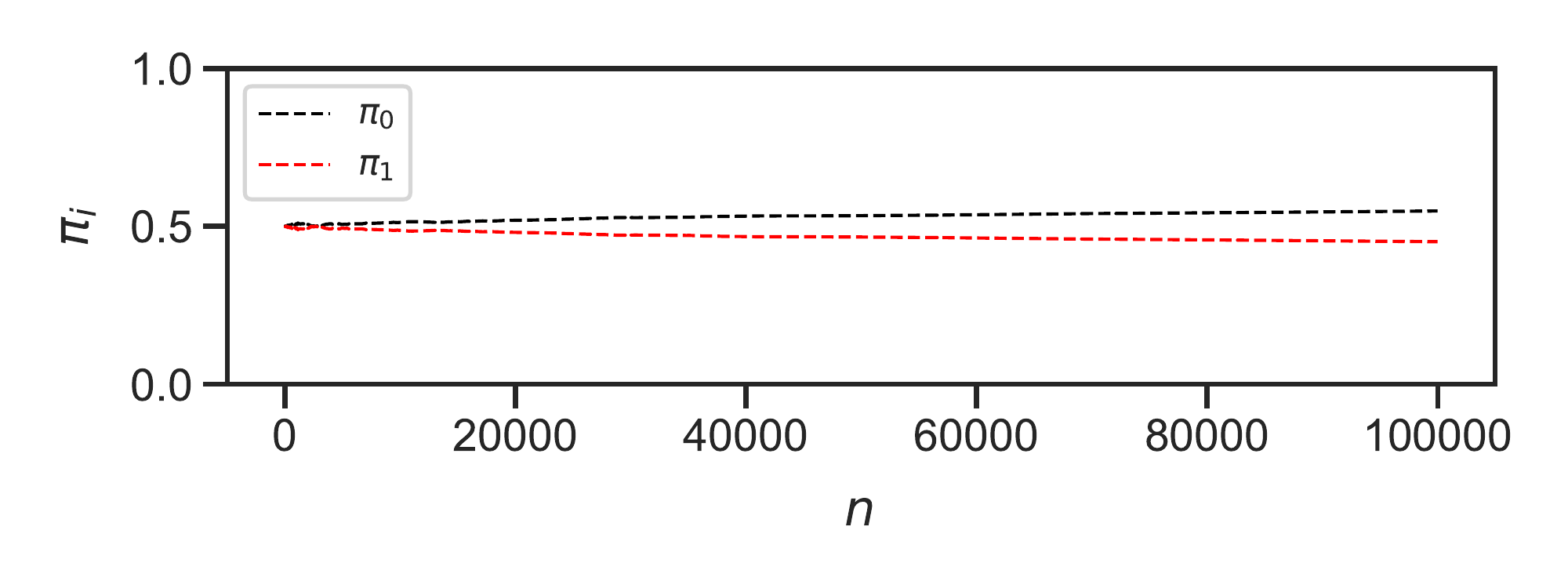}
		\caption{\parbox{0.9\linewidth}{\centering Same reference dir. $r^{(0)}=r^{(1)}=0$ \\Strict personalization $\rho=5.0$}}
		\label{fig:same_strict}
	\end{subfigure}
	\begin{subfigure}{.48\textwidth}
		\centering
		\includegraphics[width=\linewidth]{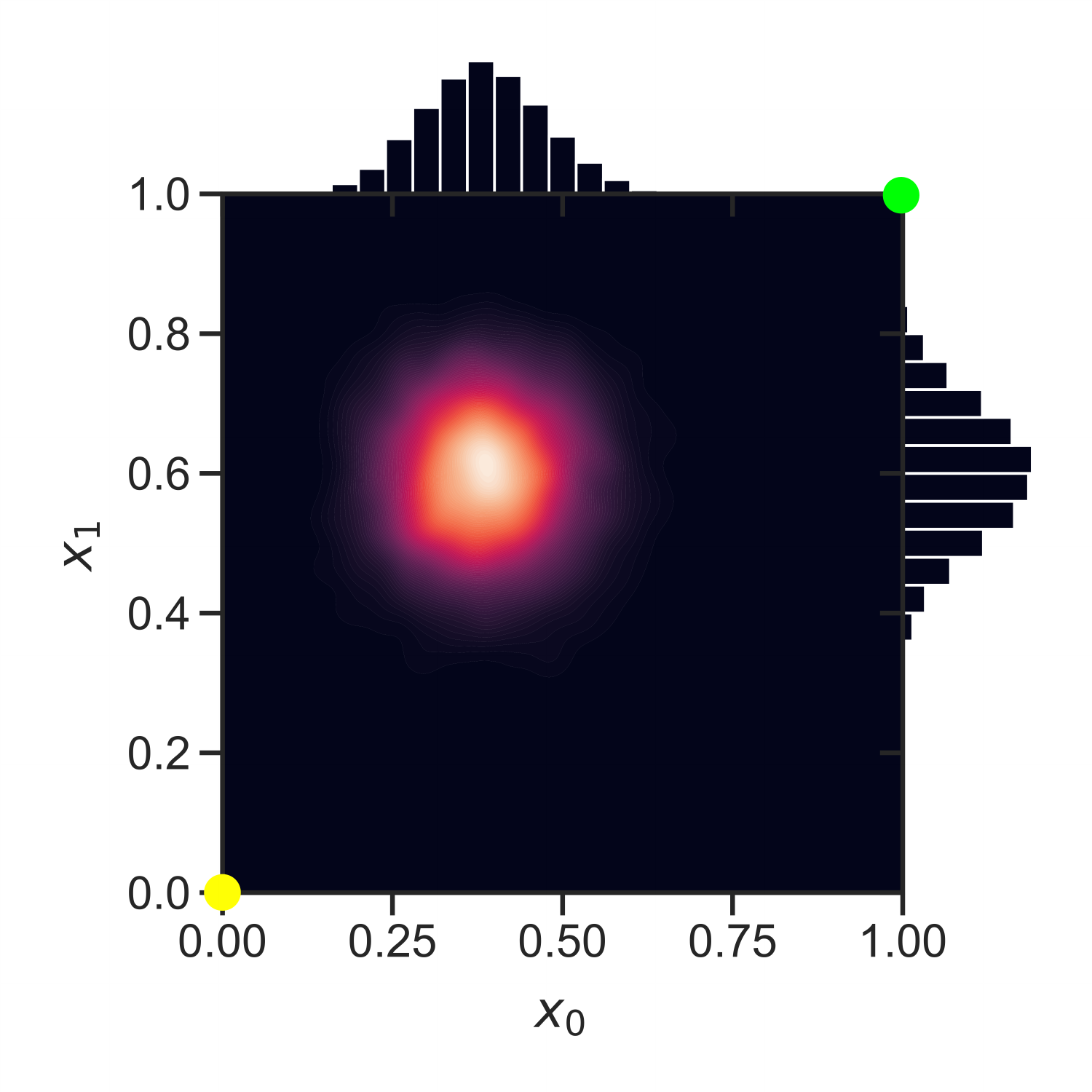}
		\centering
		\includegraphics[width=\linewidth]{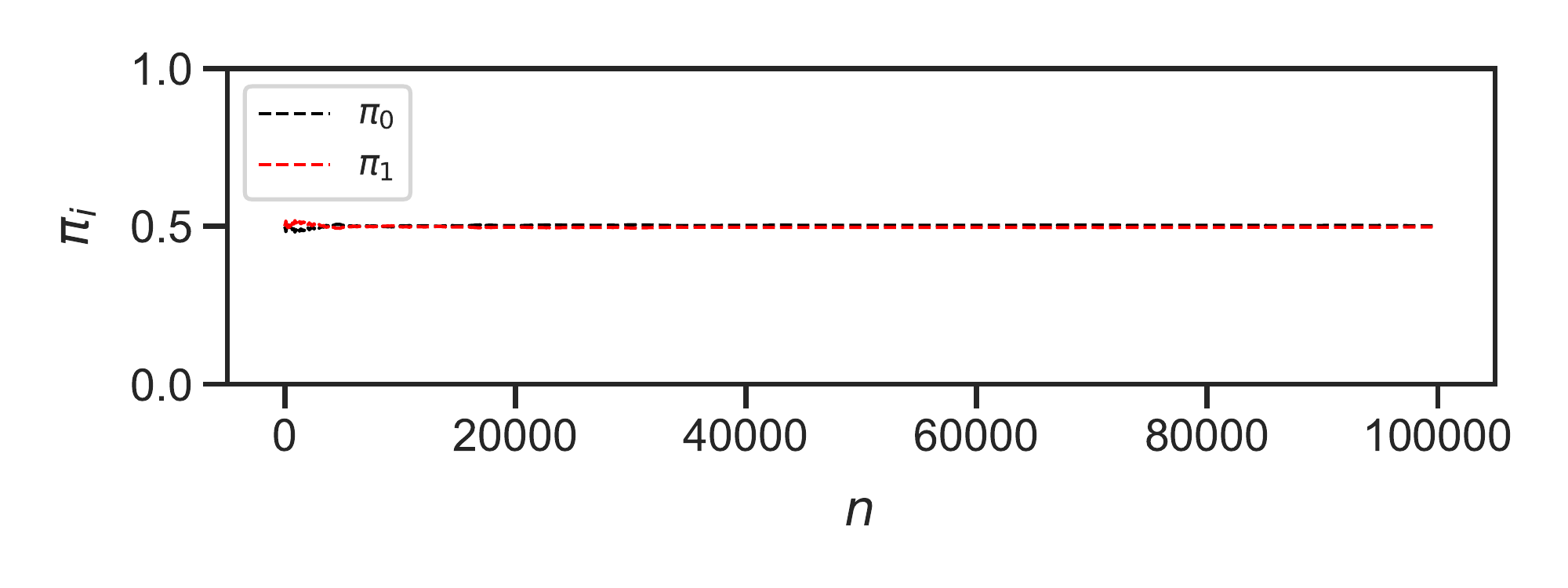}
		\caption{\parbox{0.9\linewidth}{\centering Different reference dir. $r^{(0)}=0$, $r^{(1)}=1$ \\Smooth personalization $\rho=5.0$}}
		\label{fig:diff_loose}
	\end{subfigure}
	\begin{subfigure}{.48\textwidth}
		\centering
		\includegraphics[width=\linewidth]{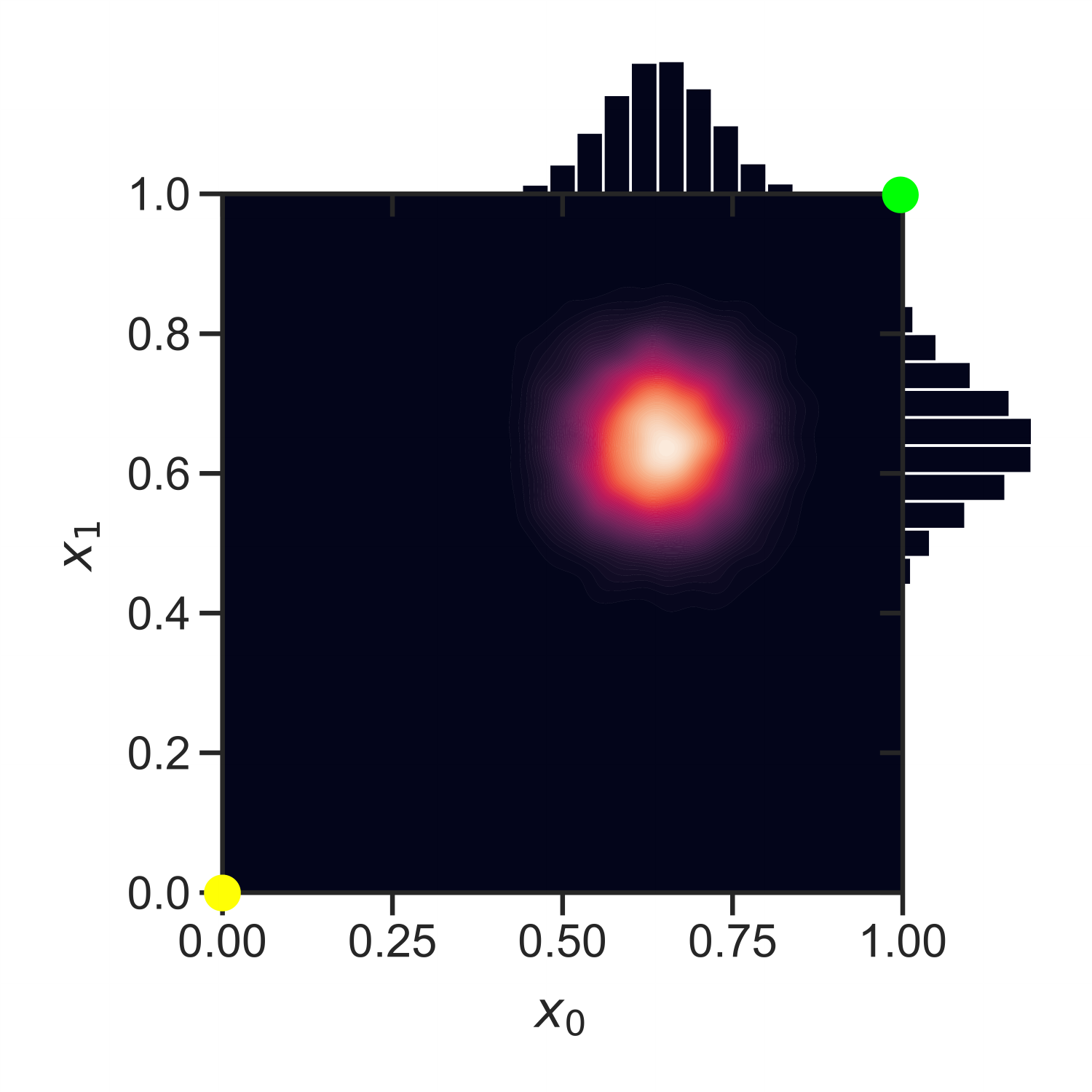}
		\centering
		\includegraphics[width=\linewidth]{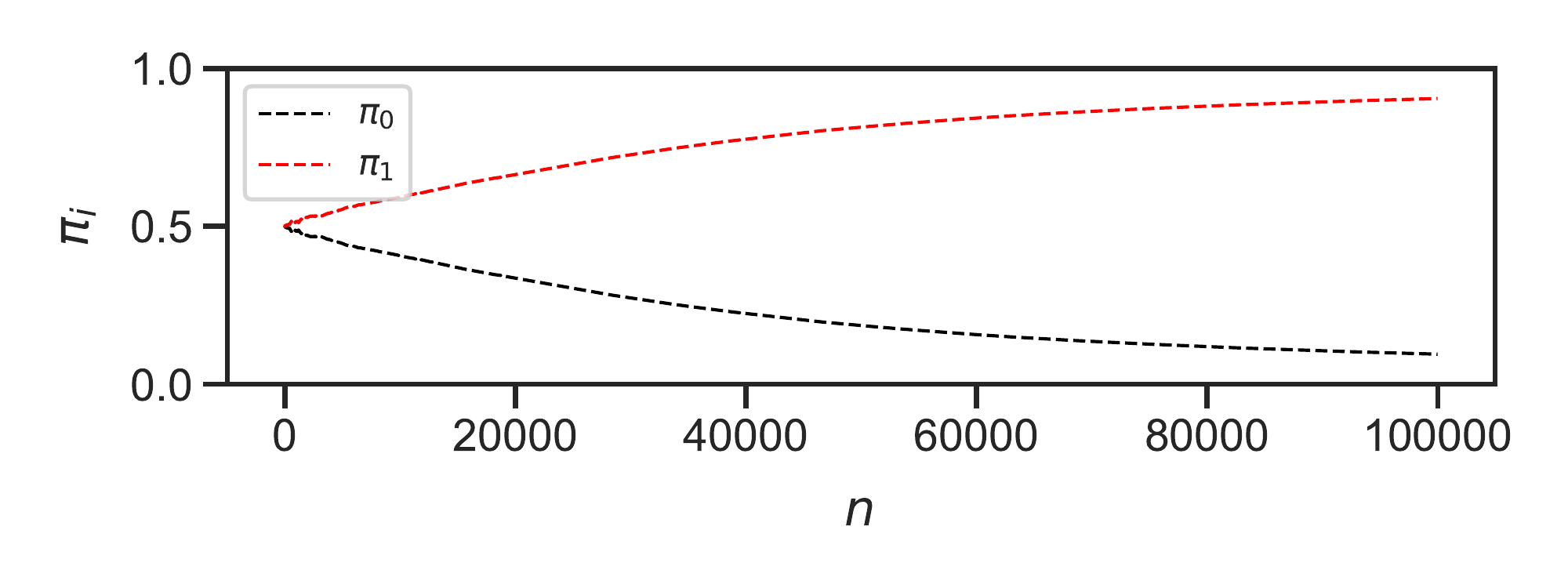}
		\caption{\parbox{0.9\linewidth}{\centering Different reference dir. $r^{(0)}=0$, $r^{(1)}=1$ \\Strict personalization $\rho=5.0$}}
		\label{fig:diff_strict}
	\end{subfigure}
	\caption{
		Users' opinion distribution and  normalized popularities. % of the two influencers.
		The two influencers have same consistency $c^{(0)}=c^{(1)}=0.8$ and same frequency $f^{(0)}=f^{(1)}=0.5$. The distributions are obtained as the time average of the opinion distribution in one realization of the process.  %It is clear that the influencers coexist except in \ref{fig:diff_strict}
	}
	\label{fig:final_distrib}
\end{figure}

{In the following sections, we will focus primarily on the influencer perspective by observing $\bar{\pi}_i$ as a function of their parameters.}
{Here, we present possible final opinion configurations of the users' population, in a symmetric scenario in terms of influencers' characteristics, i.e., frequency of publication $f^{(0)}=f^{(1)}$ and consistency $c^{(0)}=c^{(1)}$.}
As before, they hold opinions $\bm{x}^{(0)}=(0,0)$ and $\bm{x}^{(1)}=(1,1)$. We consider  {both the case of same (Fig. \ref{fig:same_loose}  and \ref{fig:same_strict}) and different (Fig. \ref{fig:diff_loose} and \ref{fig:diff_strict}) reference directions, assessing} the impact of smooth (Fig. \ref{fig:same_loose} and Fig. \ref{fig:diff_loose}) and strict content personalization (Fig. \ref{fig:same_strict} and Fig. \ref{fig:diff_strict}).

In Figure \ref{fig:same_loose}, we observe only a negligible perturbation with respect to the initial distribution (see Figure \ref{fig:op_init}). In this case, the platform practically does not filter the content, so every post reaches all users. From a regular user's perspective, individuals are  exposed to nearly identical forces, i.e., \lq\lq opposite" stimuli from the two influencers, which almost entirely cancel each other out.
In Figure \ref{fig:same_strict}, the impact of \textit{sharp} personalization is clear: the filtering effect introduced by the platform leads to the emergence of two \textit{echo chambers}, whose membership  is determined mainly by the user's prejudice.   Each user reaches an equilibrium point at which the resultant attraction  induced by the two influencers  is balanced by the attraction exerted by its prejudice.
Interestingly, users also tend to cluster in the non-reference direction ($x_1$ in Fig. \ref{fig:same_strict}) and align their opinion with the influencer associated with the echo chamber they end up in. We remark that this is a metastable condition, {i.e., the influencers have not yet reached a stable equilibrium}, as the gap in the $\pi_i$ in Fig. \ref{fig:same_strict} hints.   %By extending the time horizon,
One of the two influencers will eventually \lq\lq win" (similar to Fig. \ref{fig:diff_strict}) but in a much longer time horizon, which may be unreasonable.

Figures \ref{fig:diff_loose} and \ref{fig:diff_strict} refer to the case of different reference directions: the two influencers do not {primarily} compete on the same topic. In Figure \ref{fig:diff_loose}, it is clear that there is no competition {on their reference directions} as the two influencers are able to attract users to their \textit{reference opinion}, {i.e., $x_{r^{(0)}}=0$ for $i=0$ and $x_{r^{(1)}}=1$ for $i=1$.} %$x_0=0$ the reference opinion of $i=0$ and $x_1=0$ that of $i=1$.
This is a particularly relevant case, whose occurrence is linked indissolubly to the newly introduced concept of \textit{reference direction}. In the last scenario, Figure \ref{fig:diff_strict}, the influencer $i=1$ \lq\lq wins", i.e., $\bar{\pi}_1 \rightarrow 1$, which brings public opinion closer to its belief $\bm{x}^{(1)}$ on both issues. The final users' opinion does not coincide with $\bm{x}^{(1)}$ because users are anchored by their prejudice. {Note that here} \textit{sharp} personalization leads to a situation where only one individual monopolizes the public scene. {To better understand the dynamics, we simulated this scenario with 10 different simulator seed selections: ~5 times influencer $i=1$ won,~3 times influencer $i=0$ won, and in 2 cases, the system did not reach full convergence after $N_i=100000$ iterations. The nature of the equilibrium point $\left( \pi_0=\frac{1}{2}, \pi_1=\frac{1}{2} \right)$ appears to be \textit{unstable}. Stochastic fluctuations of the system state bring it to one of the two \textit{asymptotically stable} configurations: $\left( \pi_0=1, \pi_1=0 \right)$, where $i=0$ wins, or $\left( \pi_0=0, \pi_1=1 \right)$, where $i=1$ wins.}

\subsection{Behavior as a function of the frequency of publication}\label{sec:sim_rate}
The frequency of publication $f^{(i)}$ is one of the basic parameters that characterize influencers. The higher $f^{(i)}$, the higher the \textit{structural advantage} of the influencer because it more frequently reaches users through posts, attracting them to its own opinion. In this section, we examine the value of mean normalized popularity $\bar{\pi}_0$ as a function of $f^{(0)}$. Note that in the case of two influencers, $f^{(1)}=1-f^{(0)}$. We performed this experiment by fixing the consistency of the two influencers: $c^{(0)}=c^{(1)}=0.8$, which is approximately the average consistency observed on real-world data (Figure \ref{fig:cons_distrib}).

\begin{figure}[h!]
	\centering
	\includegraphics[width=.5\linewidth]{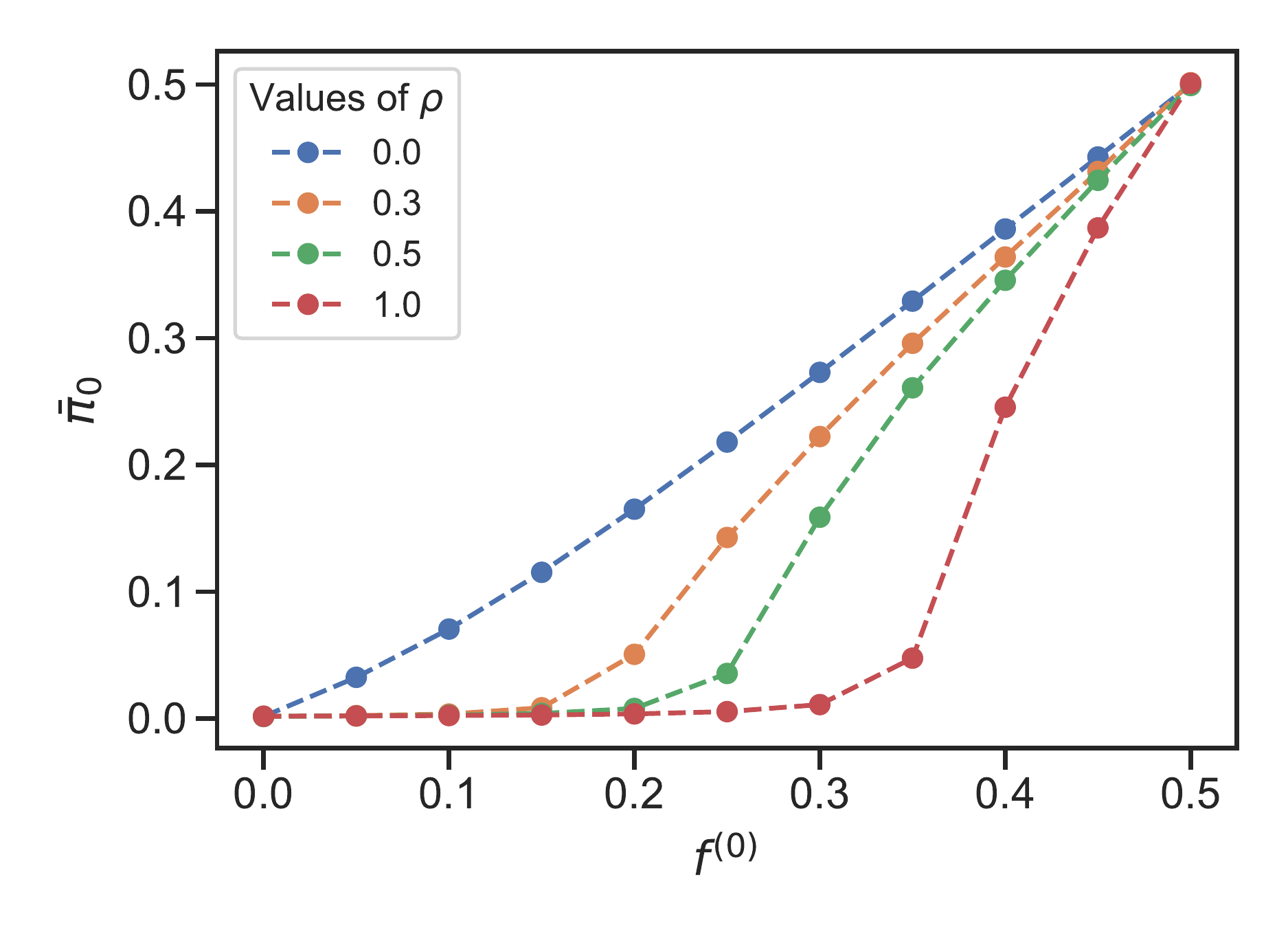}
	\caption{Popularity ratio $\bar{\pi}_0$ of influencer $0$ as function of the publication rate $f^{(0)}$. Each point is obtained by averaging over $100$ time samples and $10$ different process realizations. We consider different levels of personalization by varying the parameter $\rho$. The two influencers have the same consistency $c^{(0)}=c^{(1)}=0.8$. %Note that, in the considered scenario, the curves are symmetric for values of $f^{(0)}$ in $[0.5,1.0]$.
		{Recall that $f^{(0)}=1-f^{(1)}$ and the identity of the influencer is arbitrary. This \lq\lq symmetry" allows us to readily infer the behavior of the normalized popularity $\bar{\pi}_0$ for $f^{(0)}$ in the range of $[0.5, 1.0]$.}}
	\label{fig:metric_rate}
\end{figure}

\begin{figure*}[h!]
	\centering
	\begin{subfigure}{.5\textwidth}
		\centering
		\includegraphics[width=\linewidth]{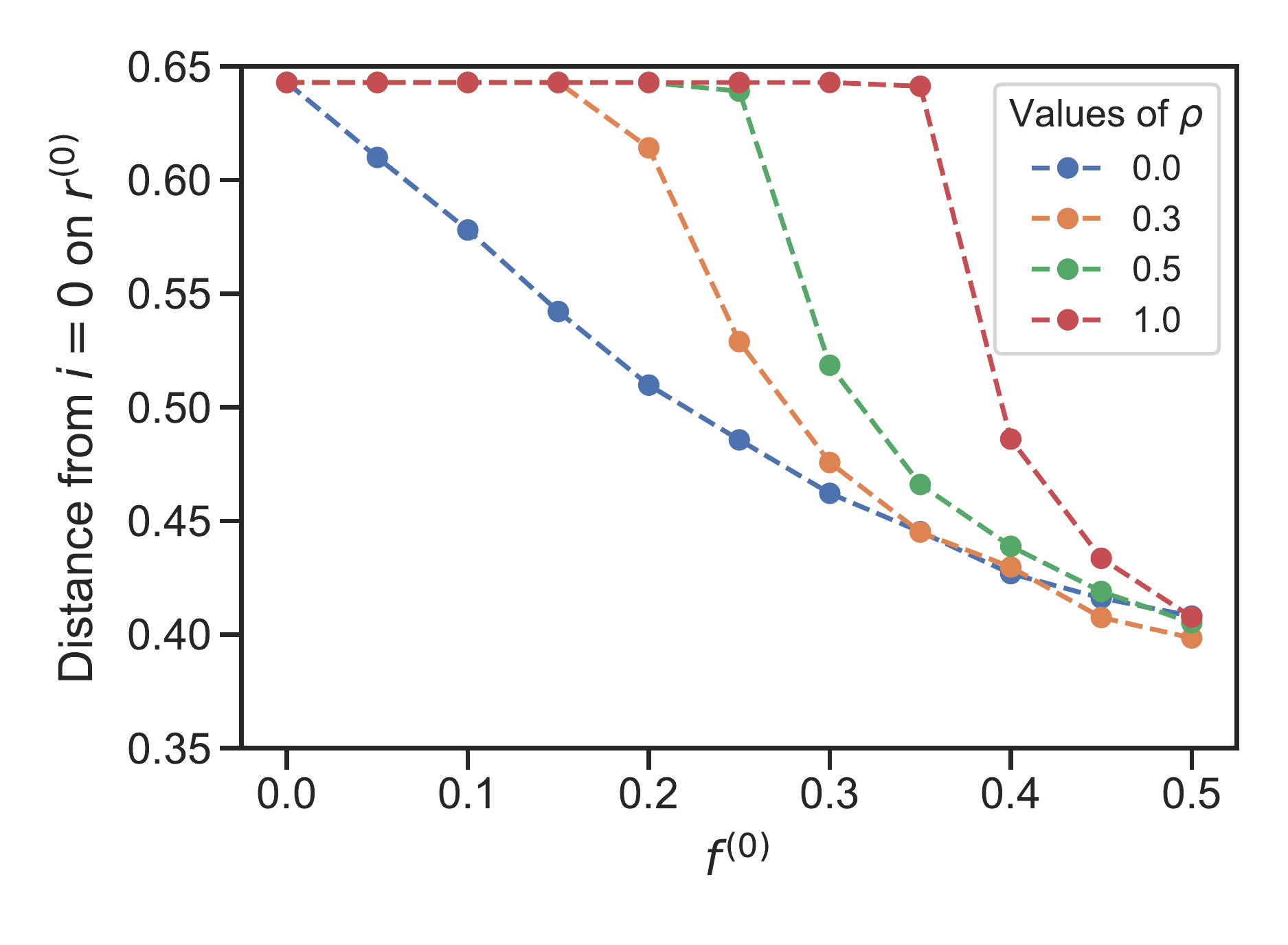}
		\caption{}
		\label{fig:rate_x0_op}
	\end{subfigure}%
	\begin{subfigure}{.5\textwidth}
		\centering
		\includegraphics[width=\linewidth]{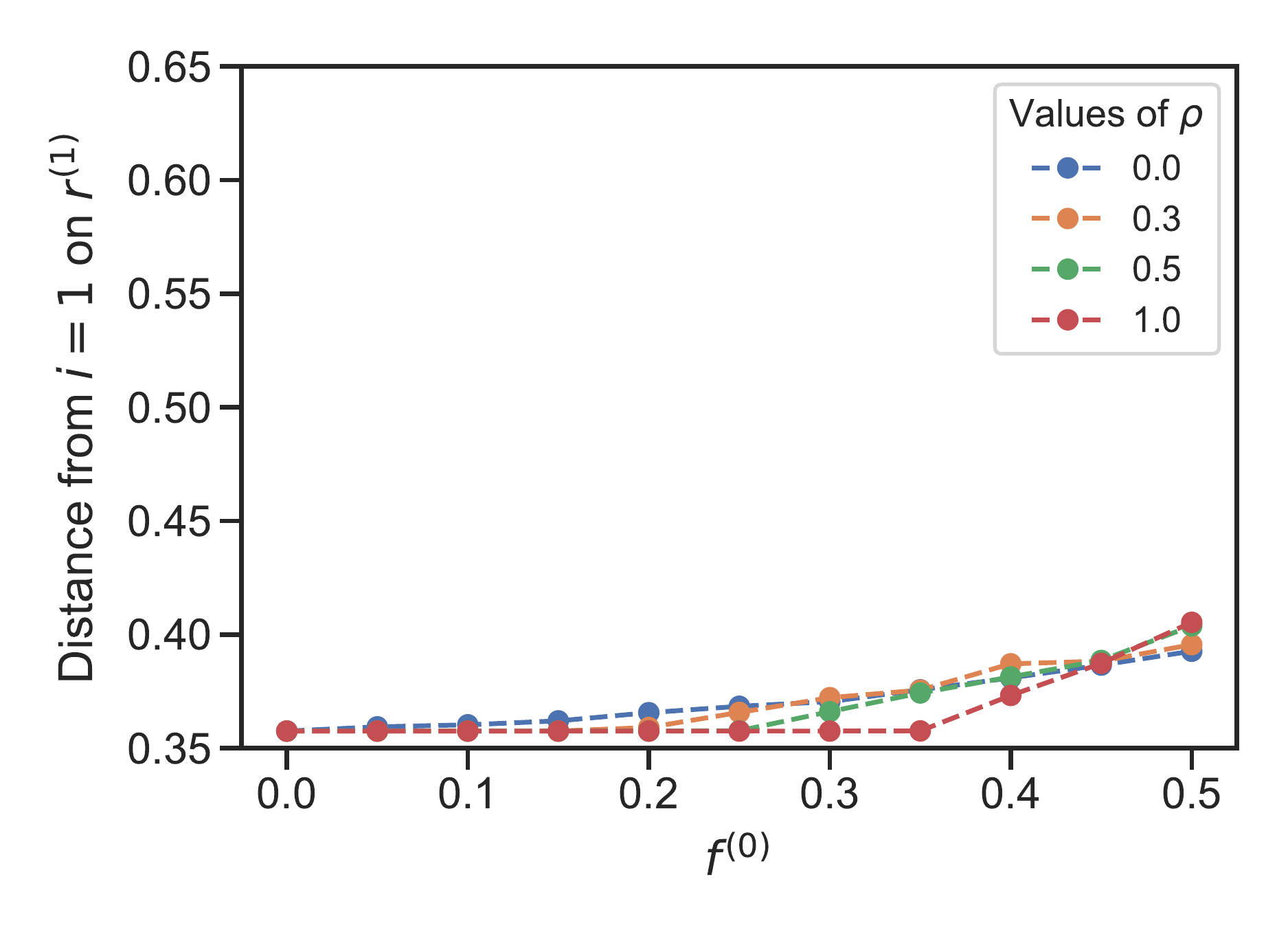}
		\caption{}
		\label{fig:rate_x1_op}
	\end{subfigure}
	\caption{{Opinion distance of the influencers' opinion on their reference direction $x^{(i)}_{r^{(i)}}$ and the average opinion of the regular users' population on the same direction $\bar{\bm{x}}_{r^{(i)}}$.} Various degrees of personalization are considered, tuning the parameter $\rho$. The setting is the same as that of Figure \ref{fig:metric_rate}.}
	\label{fig:rate_op}
\end{figure*}

In Figure \ref{fig:metric_rate}, we consider different levels of personalization by varying the parameter $\rho$ in the exponent of the visibility function $\omega$. We see that the higher the degree of personalization (i.e., the higher the value of $\rho$), the lower the normalized popularity of influencer $i=0$, for any given $f^{(0)}$. This result suggests that algorithmic personalization favors the \textit{structurally advantaged} individual, i.e., the one with higher $f^{(i)}$. This mechanism, in turn, leads to more radical positions in the population of regular users, as the platform preferentially exposes them to the belief of the \textit{advantaged} influencer. Figure \ref{fig:rate_op} clearly shows this behavior.  Note that for high values of $\rho$, the average user opinion exhibits a significant bias toward the structurally advantaged influencer.  
Such bias persists up to a critical value of posting frequency. For example, when the personalization parameter is $\rho=1.0$, the critical posting frequency value is roughly 0.35; when the personalization parameter is $\rho=0.5$, the critical posting frequency value is approximately 0.25. {Below this critical threshold the \textit{advantaged} influencers \lq\lq wins", i.e., its normalized popularity $\bar{\pi}_i$ approaches 1, completely shadowing the opposing influencer.} In Figure \ref{fig:rate_x1_op}, the opinion variation is limited since $f^{(1)} = 1-f^{(0)} > 0.5$ implies that $i=1$ exerts a strong influence over $r^{(1)}$ and suffers  little competition from $i=0$ (as $c^{(0)}=c^{(1)}=0.8$ and $r^{(0)} \neq r^{(1)}$). For instance, for $f^{(0)}=0$ the opinion values are those admitted by users' prejudice and a single \textit{winning} influencer (similar to what we discuss \ref{sec:infl_wins}). {Fig. \ref{fig:rate_x0_op} and \ref{fig:rate_x1_op}
	provide  complementary information.  Moreover  due to the aforementioned symmetry,
	the reader  can easily understand how the 		 system would evolve for  $f_0\in [0.5,1]$.}

We argue that content filtering in OSN potentially threatens opinion diversity. This premise is inextricably linked to the goal of usage maximization~\cite{Perra2019} pursued by the social media platform.
Indeed, many platforms indeed prefer to suggest just \textit{similar} content rather than exposing individuals to radically different opinions, hence often avoiding the so-called {\em serendipity}.

\subsection{Behavior as a function of the consistency}\label{sec:sim_cons}
\begin{figure*}[h!]
	\centering
	\begin{subfigure}{.48\textwidth}
		\centering
		\includegraphics[width=\linewidth]{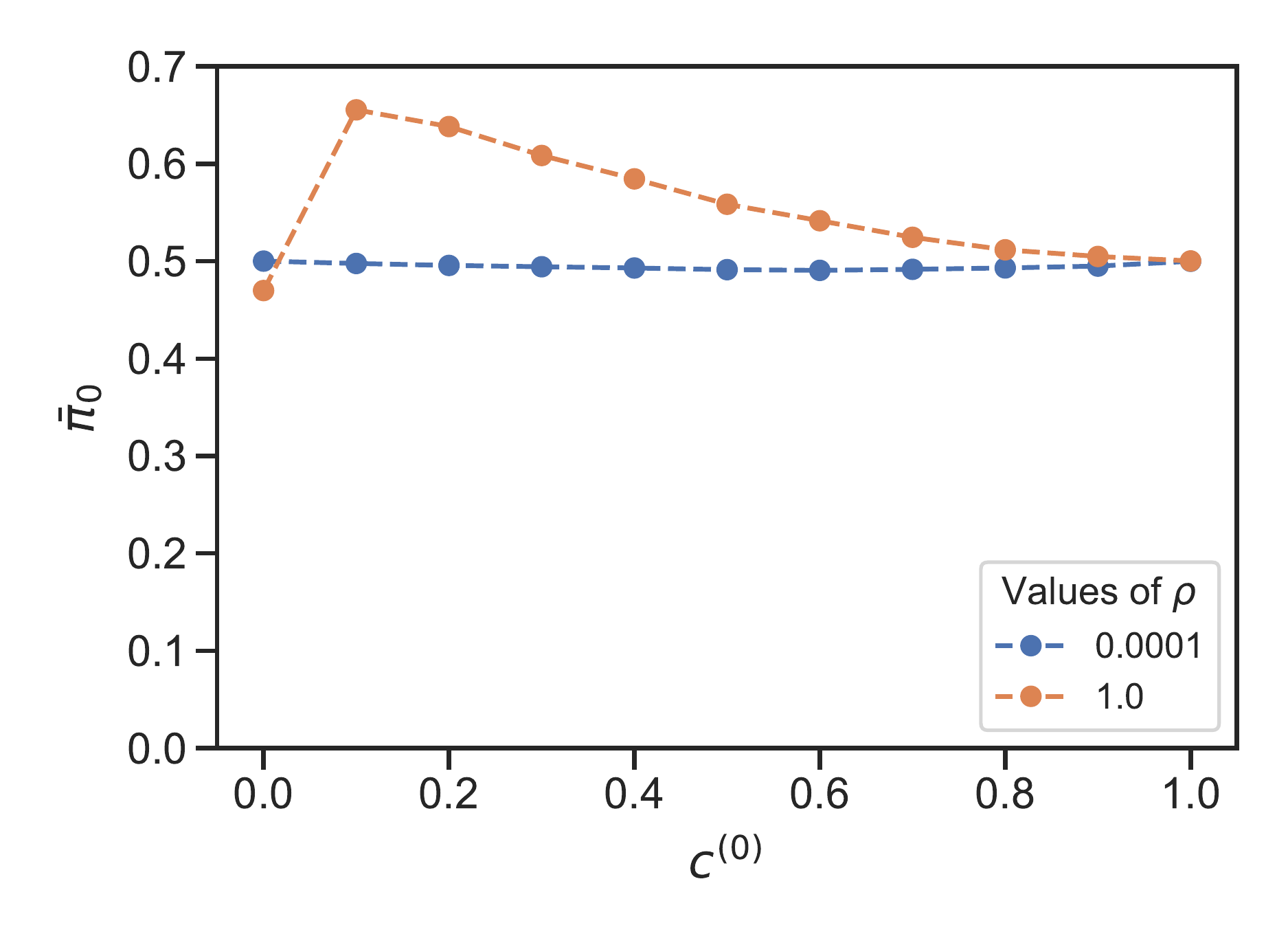}
		\caption{$c^{(1)}=1.0$}
		\label{fig:metric_cons1}
	\end{subfigure}%
	\begin{subfigure}{.48\textwidth}
		\centering
		\includegraphics[width=\linewidth]{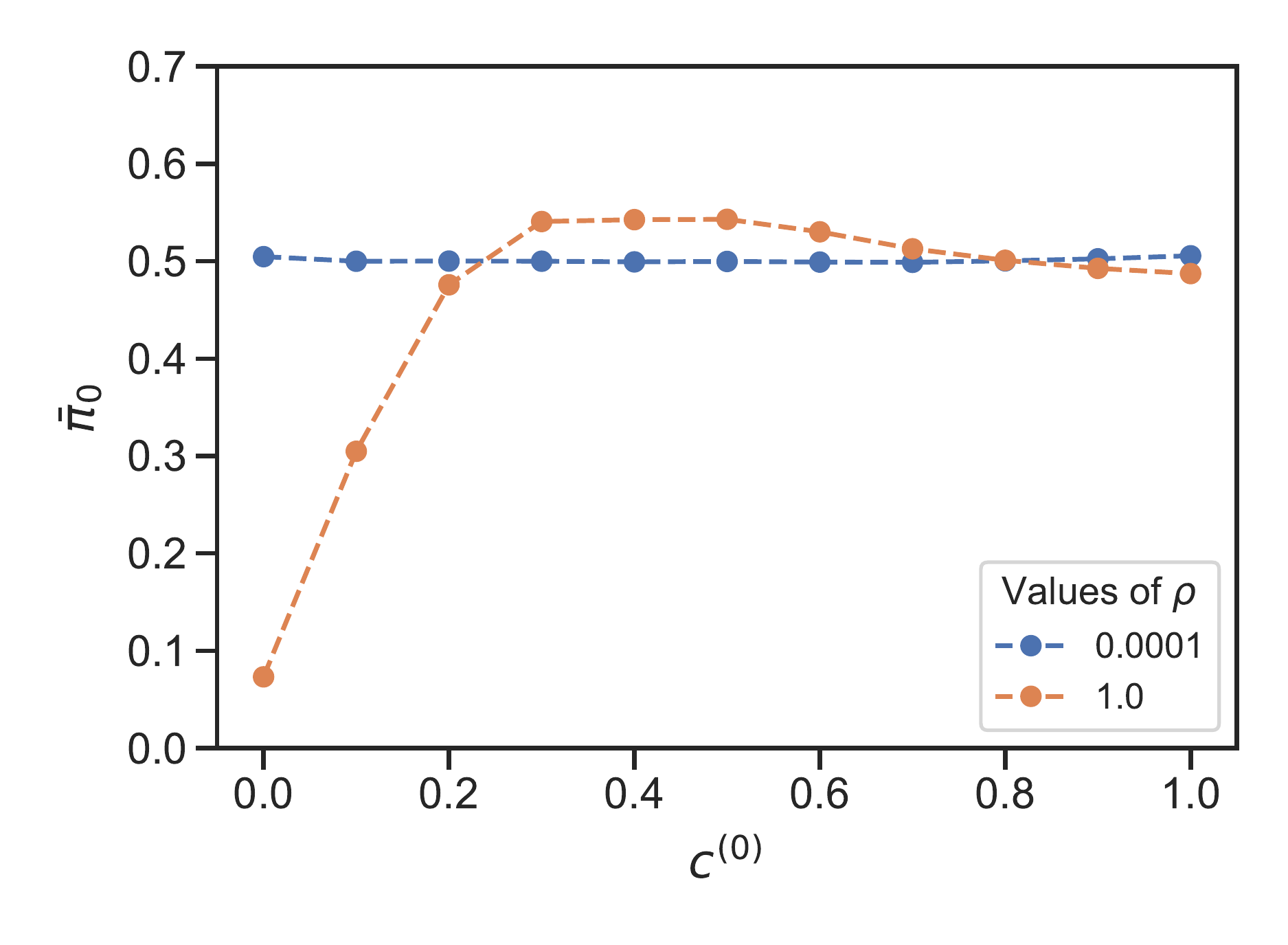}
		\caption{$c^{(1)}=0.8$}
		\label{fig:metric_cons08}
	\end{subfigure}
	
	\begin{subfigure}{.48\textwidth}
		\centering
		\includegraphics[width=\linewidth]{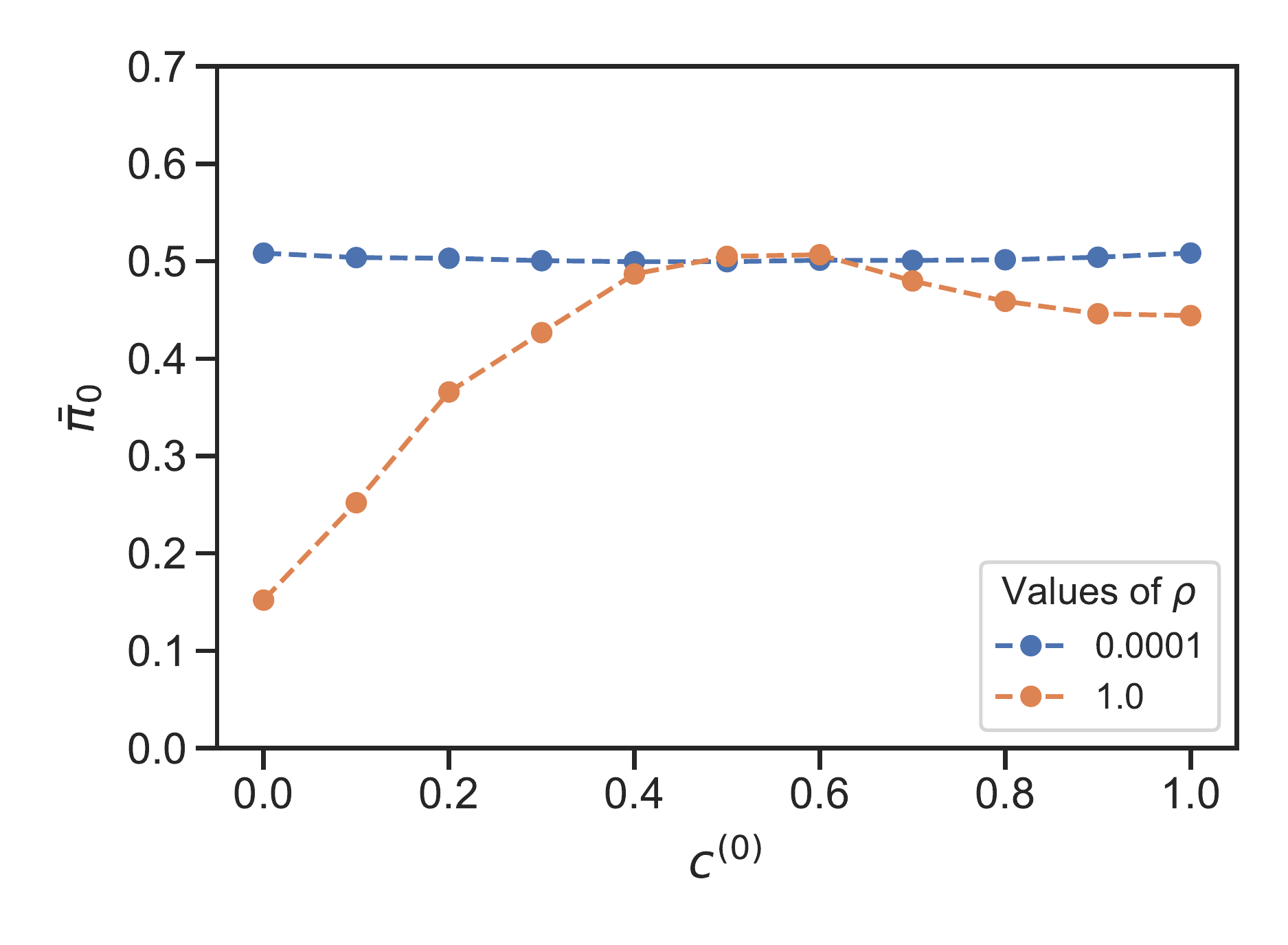}
		\caption{$c^{(1)}=0.5$}
		\label{fig:metric_cons05}
	\end{subfigure}%
	\begin{subfigure}{.48\textwidth}
		\centering
		\includegraphics[width=\linewidth]{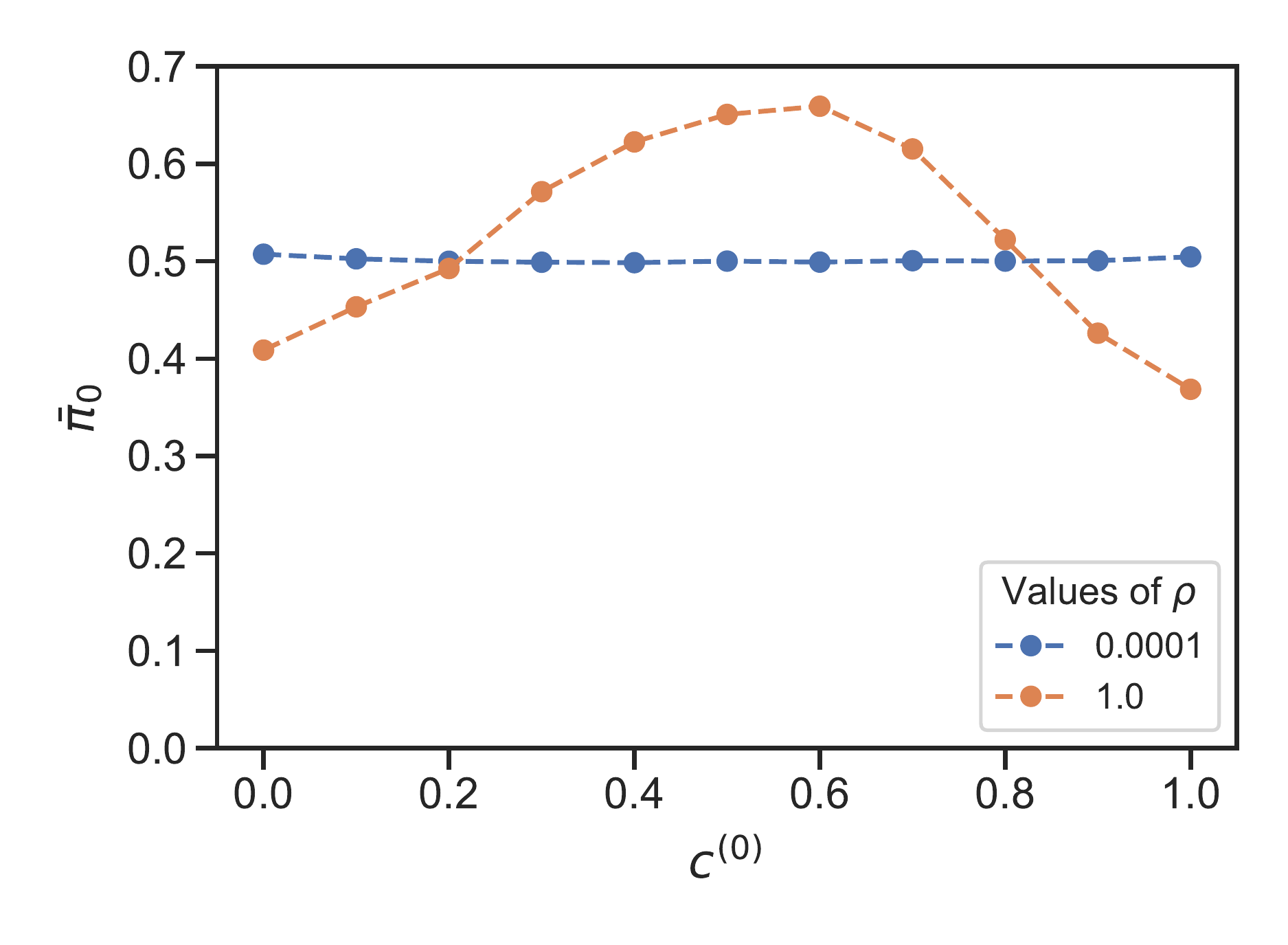}
		\caption{$c^{(1)}=0.2$}
		\label{fig:metric_cons02}
	\end{subfigure}%
	\caption{Popularity ratio $\bar{\pi}_0$ of $i_0$ as function of its consistency $c^{(0)}$, while considering $f^{(0)}=f^{(1)}=0.5$ and keeping fixed the consistency of the second influencer at (\ref{fig:metric_cons1}) $c^{(1)}=1.0$, (\ref{fig:metric_cons08}) $c^{(1)}=0.8$, (\ref{fig:metric_cons05}) $c^{(1)}=0.5$, and (\ref{fig:metric_cons02}) $c^{(1)}=0.2$. The two colors represent two different levels of personalization (i.e., smooth and sharp). Each point is obtained by averaging over $100$ time samples and $10$ different realizations of the process.} 
	\label{fig:metric_cons}
\end{figure*}

In Section \ref{sec:OSN_reference}, we showed the existence of a reference direction for real influencers. Here, we investigate the impact on dynamics of the extent to which an influencer publishes on its reference direction, i.e., its consistency $c^{(i)}$. In this experiment, we consider two influencers with the same posting frequency $f^{(0)}=f^{(1)}=0.5$, {different reference directions $r^{(0)}=0$, $r^{(1)}=1$}, and we let $c^{(0)}$ vary while keeping $c^{(1)}$ fixed. %{We then consider different choices of $c^{(1)}$ to grasp its impact on the dynamics.}
{We report plots for   a few  choices of $c^{(1)}$ since the popularity pattern as a function of $c^{(0)}$ depends on the characterization of the competing influencer $i=1$.}
Figure \ref{fig:metric_cons} shows that consistency does not significantly affect the normalized popularities when personalization is {\textit{smooth}} ($\rho = 0.0001$). In contrast, it becomes relevant when the platform applies \textit{sharp} personalization to the content ($\rho = 1$). {We consider two cases. First, that of \textit{high} $c^{(1)}$ (Figs \ref{fig:metric_cons1} and \ref{fig:metric_cons08}), which is in line with the empirical evidence of Section \ref{sec:OSN_reference}. Second, the \textit{low-consistency} scenario (Figs \ref{fig:metric_cons05} and \ref{fig:metric_cons02}) which has an interesting interpretation that goes beyond the scope of this article and is briefly {discussed in \ref{sec:strategic}.}}

In Figure \ref{fig:metric_cons1}, $c^{(1)}=1$ means that influencer $i=1$ posts exclusively in the reference direction $r^{(1)}$. The corner cases, in which both influencers post all their posts in one direction, are i) $c^{(0)} = 1-c^{(1)}$, where both post on $r^{(1)}$ but $i=1$ has a slight advantage as it is posting on its reference direction where filtering occurs, and ii) $c^{(0)} = c^{(1)}$, which is a symmetric scenario. %where the dynamics are independent on each direction.
This dichotomy is also found in Fig. \ref{fig:metric_cons08}, \ref{fig:metric_cons05} and \ref{fig:metric_cons02}. Whenever the orange curve approaches $\bar{\pi_i} = 0.5$ and $c^{(0)}=c^{(1)}$ or $c^{(0)}=1-c^{(1)}$, the influencer with the largest share of posts in the reference direction has a slight advantage. Again in Fig. \ref{fig:metric_cons1}, if $0 < c^{(0)} < 1$, the influencer $i=0$ posts in both directions and does not face competition over $r^{(0)}$. Therefore, $i=0$ attracts users towards its \lq\lq reference opinion" $x_0^{(0)}$, which in turn increases the chance of reaching users while competing with the other influencer in the non-reference direction (content filtering is performed with respect to the distance on the reference direction). In this rather extreme case, the lower the consistency $c^{(0)}$, the higher the proportion of posts on $r^{(1)}$, and the higher the final value of $\bar{\pi}_0$ as more posts \lq\lq compete" for users' attention with $i=1$. The other scenarios are not as easy to interpret. However, all \ref{fig:metric_cons1}\ref{fig:metric_cons08} \ref{fig:metric_cons05} \ref{fig:metric_cons02} are consistent in pointing out that influencer $i=0$ has a \textit{structural advantage} roughly when its consistency is $ \min \left(c^{(1)}, 1-c^{(1)} \right) < c^{(0)} < \max \left(c^{(1)}, 1-c^{(1)} \right)$.

{Figure \ref{fig:metric_cons} suggests that  a value of consistency around 0.5 is nearly optimal for any value of the opponent's $c^{(1)}$. This observation reflects the natural tendency of people to seek varied content. We evaluated $\argmax_{c^{(0)}} \bar{\pi}_0 (c^{(1)})$, and this is indeed true for $c^{(1)} < 0.8$. However, for high values of $c^{(1)}$, $i=0$ is better off reducing its consistency, i.e., post less on its reference direction $r^{(0)}$ and more on the reference topic $r^{(1)}$ of the opponent (recall $r^{(0)} \neq r^{(1)}$). This behavior points towards another potential hazard of \textit{content personalization}. If influencer $i=0$ is well-known on a platform as it deals with topic $r^{(0)}$ and starts posting massively on the other topic $r^{(1)}$ ($c^{(0)}$ drops low), it will gain an advantage over the opponent. This is because the platform filters posts considering the reference direction $r^{(0)}$ where $i=0$ faces little competition from $i=1$ (since $c^{(1)}$ is high) and therefore can attract users more easily over that topic. This results in an increase in popularity, allowing it to reach more users. This is an indication of how an influencer can leverage its \textit{importance} in the reference direction to attract users in a non-reference topic favored by content personalization.}

\section{Online social network data} \label{sec:data}
This section examines data collected from Facebook and Instagram social networks (see \ref{app:dataset}) and compares the observed behavior with some of the findings of our Communication Asymmetry model.

\subsection{Correlation between frequency of publication and popularity}

In previous sections, especially in Section \ref{sec:model_vs_param}, %and \ref{sec:comparison}, 
we discussed \textit{structural advantage} from the influencer's point of view. A key advantage parameters %, as observed across all experiments,
is the publication frequency $f^{(i)}$: the higher $f^{(i)}$, the greater the advantage (see Figure \ref{fig:metric_rate}). In this section, we attempt to validate this finding by correlating the frequency of publication of influencers with their popularity growth, using the total number of \textit{followers}, i.e., the number of people subscribed to the \textit{profile}, as a proxy for popularity. We consider temporal sequences from Instagram on a sample set of $110$ influencers.

\begin{figure}[h!]
	\centering
	\includegraphics[width=0.5\textwidth]{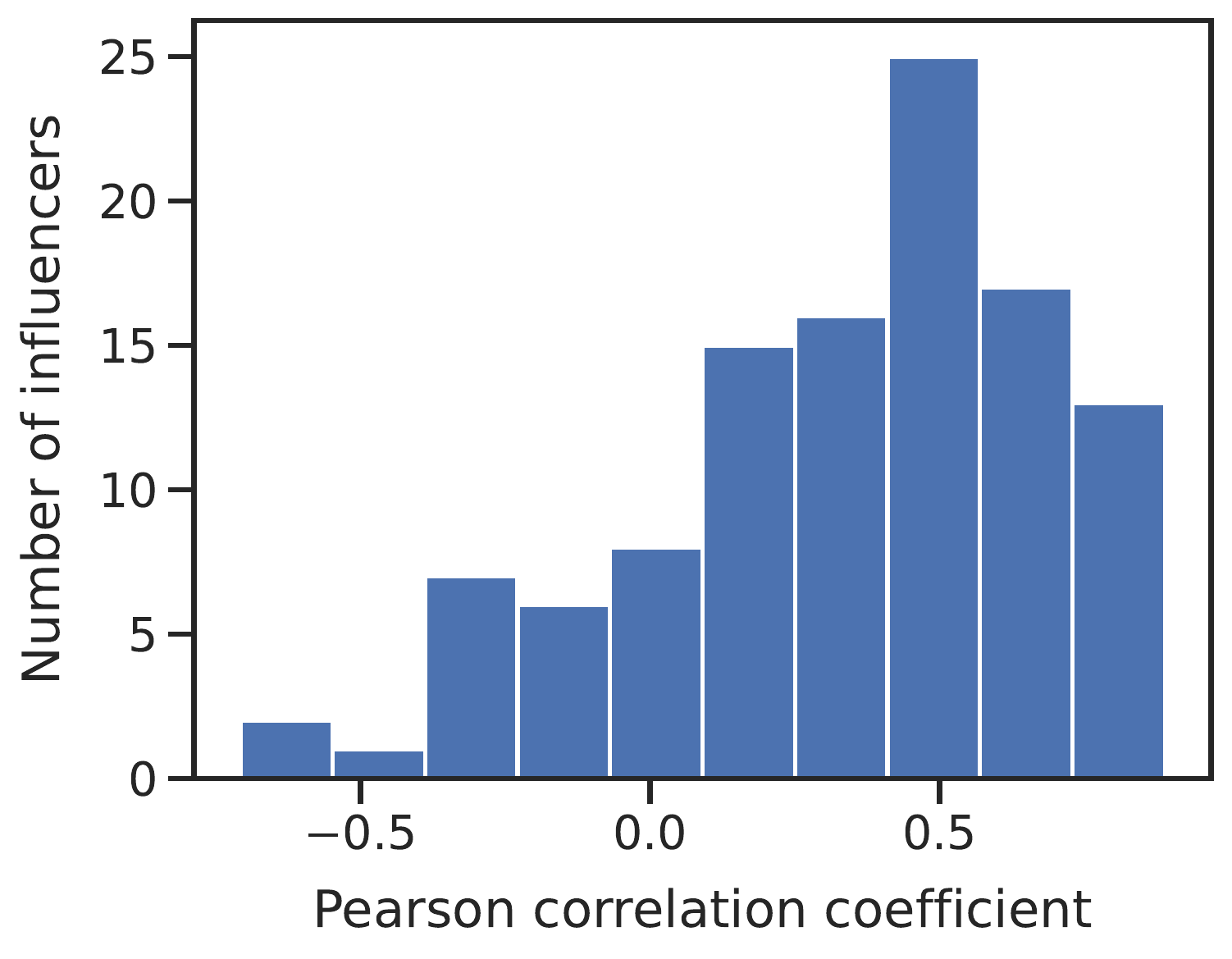}
	\caption{Distribution of the Pearson correlation coefficient between the monthly number of posts and popularity growth (in terms of number of \textit{followers}).}
	\label{fig:rate_pop}
\end{figure}

For each influencer, we considered a temporal granularity of one month, determined the number of posts during this period, and calculated the relative change in the number of \textit{followers} considering the values at the beginning and end of the interval. Then for each user, we calculated the Pearson correlation coefficient between  the number of posts and the relative variation of followers in the month. In Figure \ref{fig:rate_pop}, we show the distribution of these correlation coefficients. Results suggest that there exists, in general,  a positive correlation between the two quantities, i.e., influencers with aggressive posting habits tend (but not always) to get more followers, which likely favors them when in competition with other influencers on social media platforms.
This is consistent with the model predictions shown  in Section \ref{sec:sim_rate}.

\subsection{Case Study: Italian government crisis in August 2019}\label{sec:conte_salvini}

In June 2018, a few months after the general elections, Giuseppe Conte was appointed Italian Prime Minister. Two parties formed his supporting coalition: Movimento 5 Stelle (his party, holding the relative majority of the Italian Parliament) and Lega, whose leader was Matteo Salvini. In August 2019, Salvini decided to withdraw Lega's support to the government, starting a crisis aimed at driving Italians to new elections and gaining more votes. 
However, Movimento 5 Stelle reached an agreement with other parties to form a new government, and on September 5, 2019, Giuseppe Conte became Prime Minister for the second time, excluding Lega from the new administration.

\begin{figure}[h!]
	\centering
	\begin{subfigure}{.5\textwidth}
		\centering
		\includegraphics[width=\linewidth]{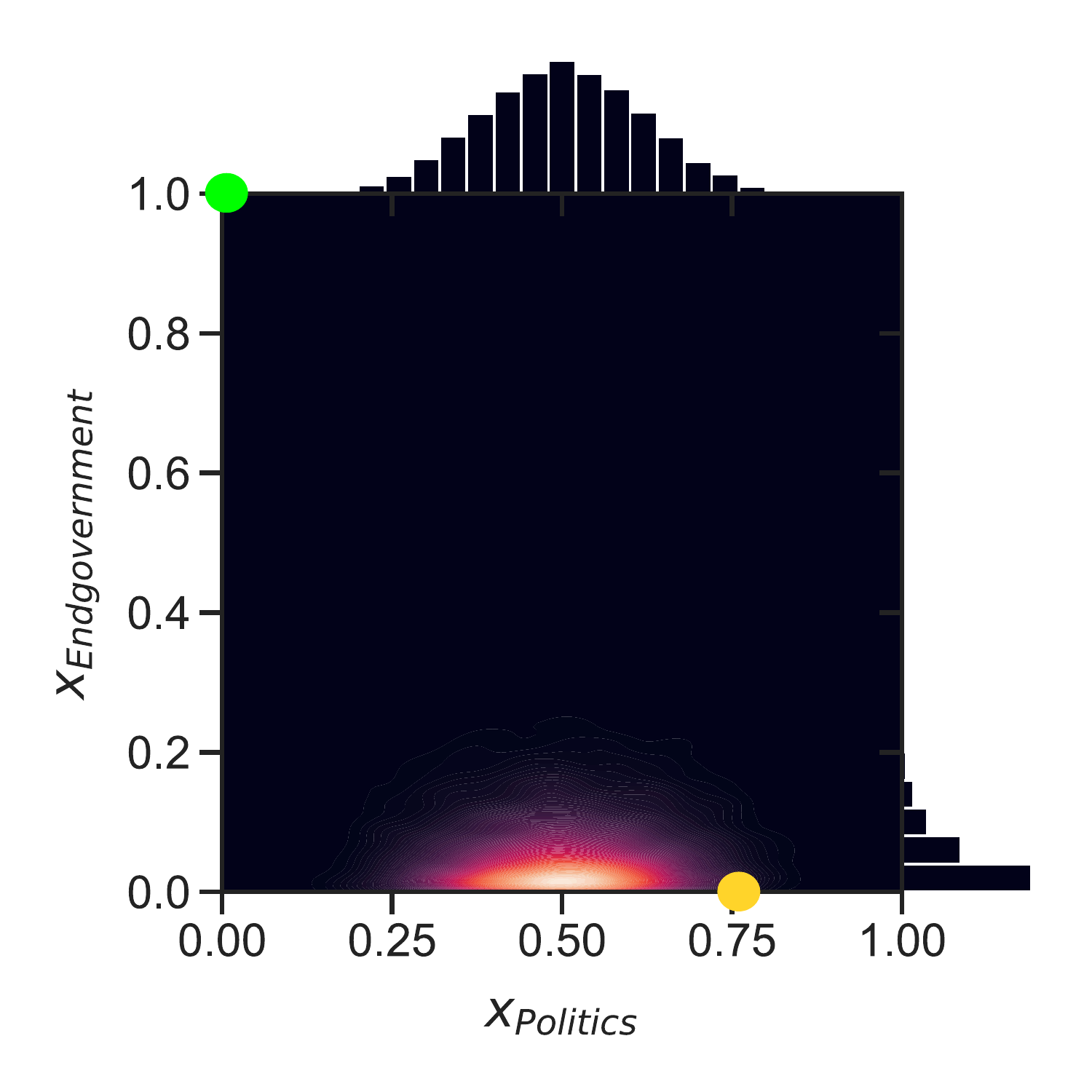}
		\caption{Initial distribution}
		\label{fig:init_cs}
	\end{subfigure}%
	\begin{subfigure}{.5\textwidth}
		\centering
		\includegraphics[width=\linewidth]{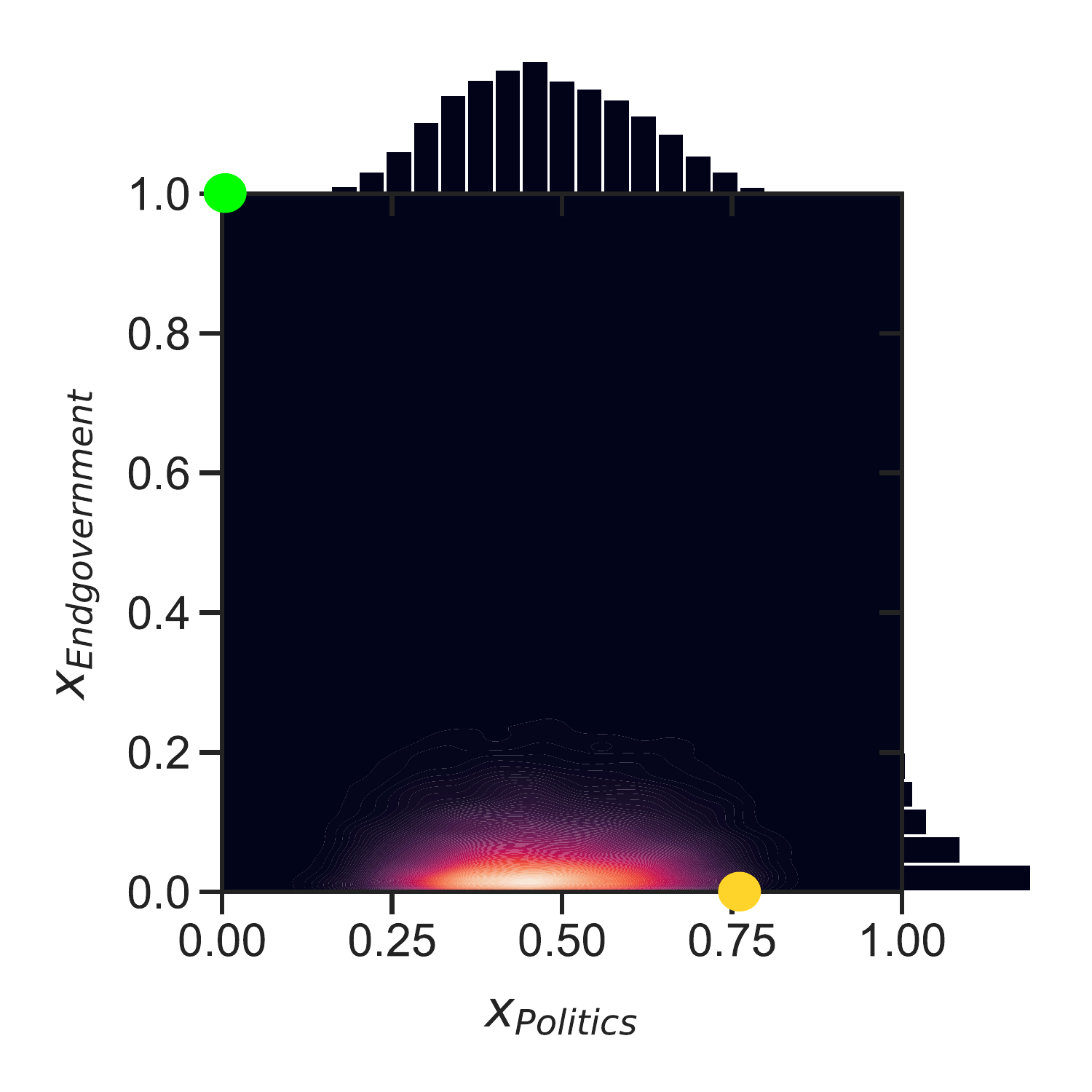}
		\caption{After the transient}
		\label{fig:before_trans_cs}
	\end{subfigure}
	\caption{(\ref{fig:init_cs}) Initial population's distribution along the \textit{Politics} and \textit{End government} direction and (\ref{fig:before_trans_cs}) {after transient, at the start of the observed period.} The opinion position of the two leaders in the space is depicted with a green (Salvini) and a yellow (Conte) point.}
	\label{fig:conte_salvini}
\end{figure}

In this section, we apply the proposed model to reproduce the sudden rise of Giuseppe Conte's popularity in social networks during this government crisis. We exploit the multidimensional capability of the model  considering two directions: \textit{Politics}, reference topic for Salvini and Conte, and attitude toward government fall, \textit{End government} (see Figure \ref{fig:conte_salvini}).

In the opinion space, we assume Salvini has a more radical political viewpoint ($x_{Politics}=0$), while Conte has an opposing and more moderate position ($x_{Politics}=0.76$). {We set these values in a somewhat discretionary manner. However, we provide a sensitivity analysis in \ref{sec:sensitivity_cs}, proving that results are robust. Conversely, it is safe to assume that the two politicians take opinions at the extreme of the spectrum on the attitude toward government fall, i.e., Salvini has $x_{End government}=1$ and Conte has $x_{End government}=0$.}
Moreover, we consider a population with a \textit{moderate} initial opinion on \textit{Politics} (centered at $x_{Politics}=0.5$, see Figure \ref{fig:init_cs}). {On \textit{End government}, we sought  a distribution that  could explain the sudden popularity leap of Conte. We found that the user population must be strongly biased} towards Conte's opinion (Figure~\ref{fig:conte_salvini}).

Some {further} simplifying assumptions are necessary to apply the model. We assume that the two politicians have a \textit{consistency} $c^{(i)}$ of exactly one (real values are often close to this value, see Figure~\ref{fig:cons_distrib}). Moreover, Giuseppe Conte and Matteo Salvini are the only influencers. Although this hypothesis is restrictive, in the scenario studied, the two influencers were the main (active and popular) protagonists during the government crisis. Moreover, we consider the simplest scenario in which personalization is not employed: $\rho=0$ and thus $\omega \equiv 1$. We consider a \textit{feedback} function of the form $\theta = e^{-8.25 (x^{(u)}-x^{(i)})^2}$ for both opinion directions. For an exhaustive list of the parameters, we refer to Table~\ref{tab:table_conte_salvini}.

A period of eleven weeks is considered, from July 7 to September 22, during which data was collected weekly from Facebook. A total of 1162 posts were published, of which 125 were by Conte. The {\textit{frequency of publication}} $f^{(i)}$ is calculated as the number of posts by an influencer relative to the total number of posts ($f^{(Conte)}=0.108$, $f^{(Salvini)}=0.892$). 

\begin{figure}[h!]
	\centering
	\includegraphics[width=0.8\linewidth]{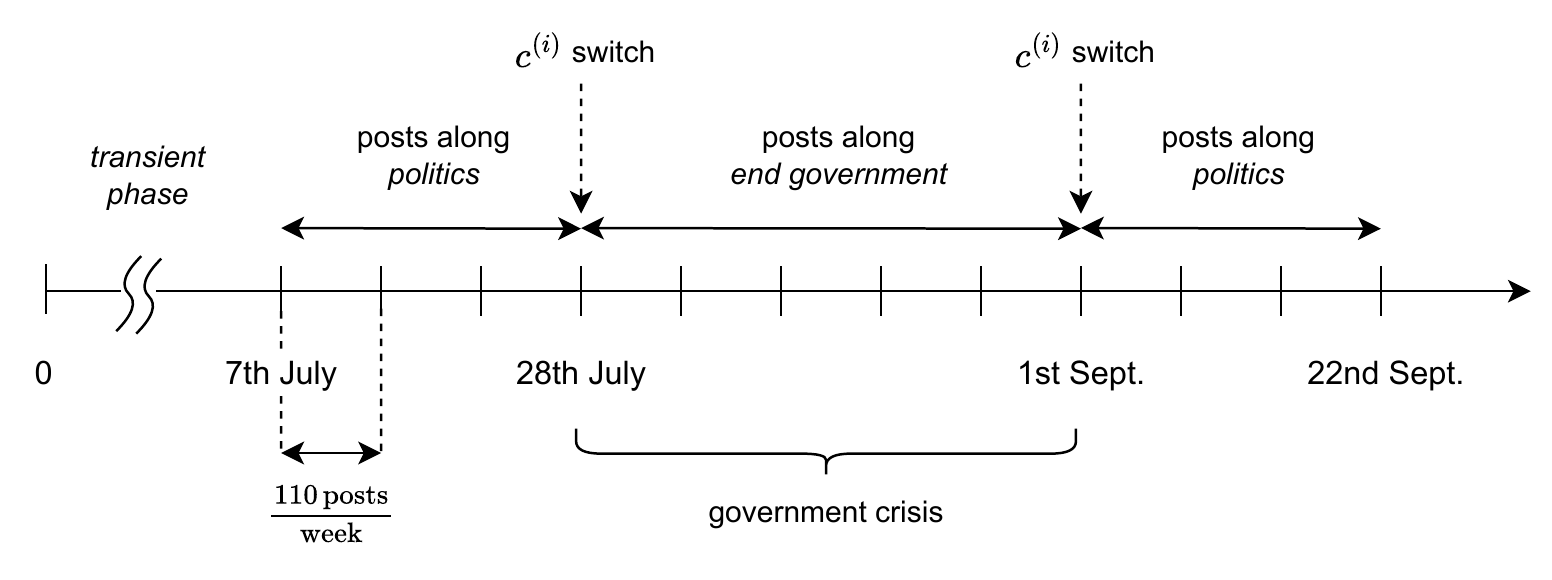}
	\caption{Timeline of modelled scenario from July 7, to September 22. From July 28 to September 1 we have a consistency switch, with posts along \textit{End government} direction.}
	\label{fig:cs_setting}
\end{figure}

Figure \ref{fig:cs_setting} shows the timeline of the experiment. The two influencers start with the same initial popularity. We consider a transient of $N_t=10000$ discrete time-units, after which the stationary normalized popularities $\pi_i$ approximately correspond to the empirical normalized popularities obtained by dividing the number of \textit{followers} of each influencer by the total number of the two. After the transient, we can see in Figure \ref{fig:before_trans_cs} that the distribution of public opinion is skewed towards Salvini, who has a higher popularity ratio due to his {much} higher publication frequency. After the transient,  the crisis starts, and both influencers post in the \textit{End goverment} direction  (i.e., we observe a consistency %shift
{switch} for both influencers) during a time window of five weeks that approximates the duration of the government crisis, after which the two 
politicians switch back to posting on the \textit{Politics} direction.

\begin{figure}[h!]
	\centering
	\includegraphics[width=.8\linewidth]{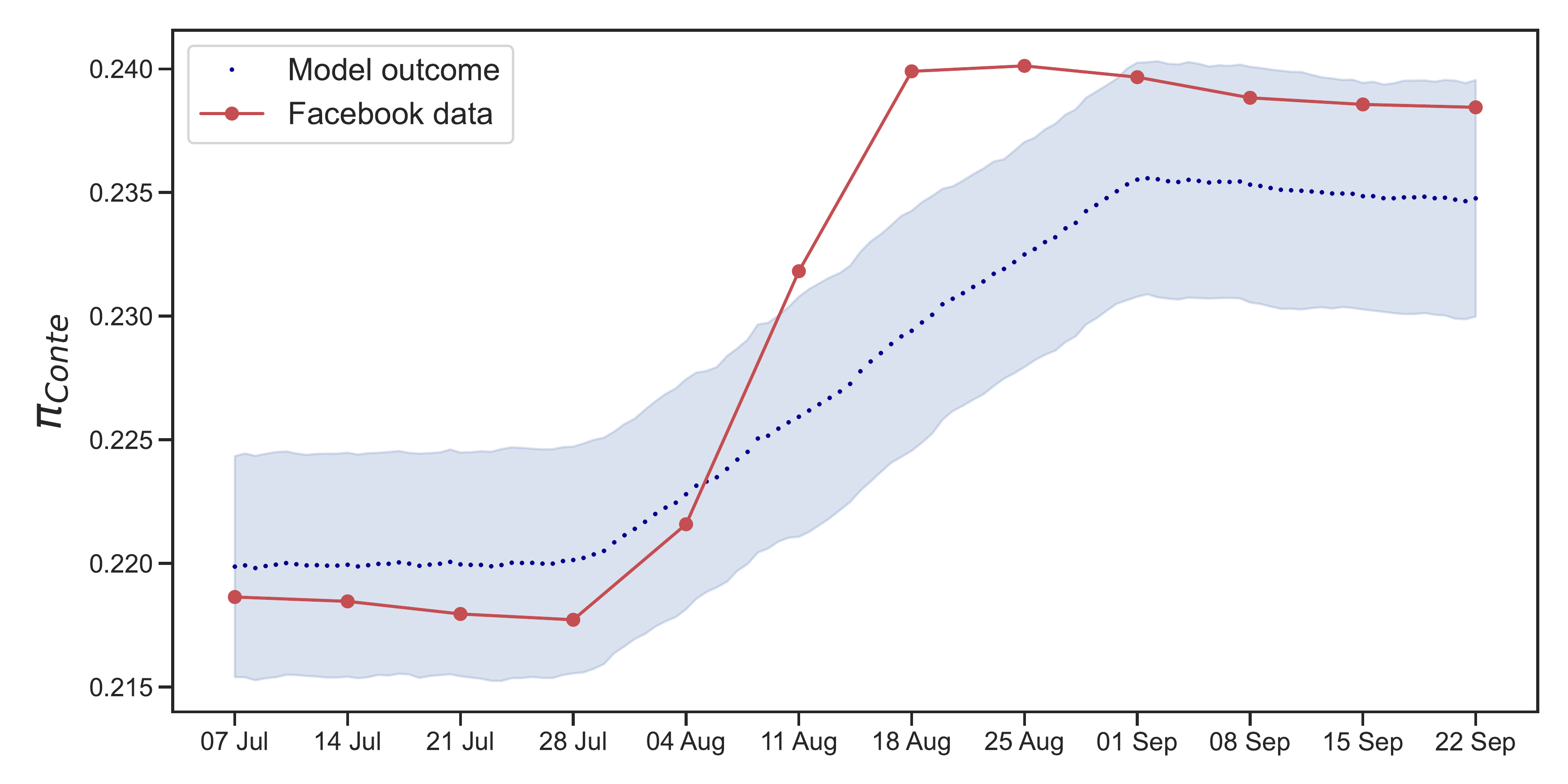}
	\caption{Popularity ratio $\pi_{Conte}$ for Conte, from Facebbok data (red) and from the model (blue) along with its 95\% confidence interval, computed over 10 realizations of the process. One can see how the model follows the increase in popularity during August~2019.}
	\label{fig:conte_salvini_res}
\end{figure}

Even with {these simplifications}, %these limitations, 
it is possible to reproduce the observed social behavior as a whole: it corresponds to a situation where an influencer is in stark contrast to the opinions of its user base and loses ground with respect to the other influencer. 
In the model, only a very unbalanced population distribution towards Conte's opinion (against  the government fall) can explain the sudden increase in Conte's popularity, despite the remarkable differences in popularity ratios in favor of Salvini. Figure \ref{fig:conte_salvini_res} compares the simulation results of the described setting and Facebook's measurements.

Clearly, our model does not precisely fit  empirical observations but provides qualitative insights into the possible causes of the rather sudden popularity shift that was observed. Many of the model's parameters are unknown, such as the opinion distribution, the weights of the updating rule, or the \textit{feedback} function. {However, by measuring some parameters, e.g., $f^{(0)}$, {the phases' duration, $\pi_{Conte}(\text{7 July})$}, and making reasonable assumptions about the others, i.e., $\alpha$, $\beta$, $\theta$, $\bm{x}^{(i)}$, the emerging behavior is consistent for several choices of the parameters as demonstrated in \ref{sec:sensitivity_cs}.}
We conclude that the observed popularity trends can be explained mainly by considering the fear of political instability in the user base.

\section{Conclusion} \label{sect:conclusion}

Online social interactions have recently played an increasingly important role in opinion formation. Understanding the mechanisms underlying this modern communication paradigm requires the development of new, flexible frameworks suitable for describing interactions on social media platforms.
In this work, we developed an opinion model tailored to online interactions, with particular attention to the interplay between \textit{regular users} and \textit{influencers}. In addition, we characterize influential individuals on online platforms by grounding our design decisions in data from real-world online social networks. Similar to other work in the recent literature, we integrated \textit{content personalization} in a flexible and tunable manner. We have shown how content personalization reinforces inequality by favoring the \textit{structurally advantaged} influencer and, in most cases, preventing the \lq\lq competing" influencer from remaining visible to the population, potentially hindering diversity of opinion among the population. Furthermore, even under structurally balanced conditions, personalization can lead to the emergence of \textit{echo chambers} in which users' opinions become radicalized by the influencer's point of view, even on topics that are not the influencer's \textit{reference} one. {Or to unstable situations in which a single individual hegemonizes the online scene.}

The proposed model is {one of the first attempts} to describe the complexity of online interactions faithfully and comes with some limitations. In our model, users are passive entities, and influencers are stubborn agents. Moreover, \textit{homophily} is the primary driver of user interaction, as no other relationship structure was considered. Nonetheless, despite the simplifying assumptions, the emergent behavior of the model proved rich enough to show the impact of content personalization and shed light on the dynamics of influencer popularity.

Our work points to several research directions, such as considering users as active agents capable of publishing their own posts and forwarding (i.e., sharing) posts from influencers, as we discuss in \ref{sec:user_interactions}.
This may pose significant challenges in terms of analytical tractability. Another promising direction might be to consider influencers as \lq\lq strategic" actors who aim to maximize their popularity on the platform by exploiting the internal mechanisms of the platform itself, e.g., content filtering, as we briefly discuss in \ref{sec:strategic}.

\vspace{0.4cm}
\textbf{Declaration of Competing Interest}\\
The authors declare that they have no known competing financial interests or personal relationships that could have appeared to influence the work reported in this paper.

\appendix
\section{Description of the dataset} \label{app:dataset}

We collected data from real online social networks to support the hypotheses of our model and compare emergent behaviors. We focus on two popular social networks: Facebook (FB) and Instagram (IG). Facebook has long been the most popular social media application, while Instagram has undergone a surge in popularity in recent years.

In Facebook and Instagram, a \emph{profile}, i.e., a social network user, can be followed by other profiles, i.e., its \emph{followers}. A profile with a large number of followers is also called an \emph{influencer} - we consider profiles with more than ten thousand followers as influencers. Influencers post content (i.e., \emph{posts}) that profile's followers and anyone registered on the platform can see, {\it like} and comment. Note that when we use the term influencer, we do not only mean individuals but also groups, soccer teams, newspapers, or companies.

To get the list of such popular profiles, we exploited the online analytics platform \url{hypeauditor.com} for IG, and \url{www.socialbakers.com} and \url{www.pubblicodelirio.it} for FB. We restricted the analysis to influencers with at least $10,000$ followers on June 1, 2021. The obtained $649$ influencers are the same of our previous paper \cite{VassioAsonam} and are publicly available.\footnote{\url{https://mplanestore.polito.it:5001/sharing/P4WnRClQn}}
In this work, we are interested in the posts of influencers and their temporal sequence.
For each monitored influencer, we downloaded all the data related to the posts published between January 1, 2016, and June 1, 2021, using the CrowdTangle tool and its API\footnote{\url{https://github.com/CrowdTangle/API}}. CrowdTangle is a content discovery and social analytics tool owned by Meta and available to researchers and analysts worldwide to support research, subject to a partnership agreement.  
Finally, we stored the data, which takes around 110\,GB of disk space, on a Hadoop-based cluster, and we used PySpark for scalable processing.

\section{Details on post classification} \label{app:classifier}

We developed a classifier that can categorize posts according to a particular set of subjects, similarly to what we have done in our previous work \cite{SocialCovid}. First, we arbitrarily identified a subset of topics that sufficiently characterize the discussions on the monitored profiles. Specifically, these topics are \textit{sports}, \textit{politics}, \textit{food and cooking}, \textit{music}, and \textit{pandemics}, which are intentionally loose and relatively uncorrelated to each other.
We developed a keyword classifier to classify the posts. For each topic, we manually defined a list of representative keywords. For example, if we consider \emph{pandemic}, we search for words like COVID, pandemic, and coronavirus in Italian (and commonly used terms in other languages).
%We use stemming to reduce inflected words.
We search for the topic-specific terms in the text corpus of the post, and if we find a match, we mark the post as belonging to the topic. Notice that since keywords of various topics may be present in the same corpus, we can flag a message as discussing \textit{multiple} topics. In this work, we discard posts marked as multiple and only consider posts associated with a single topic.

We are not interested in classifying all posts by an influencer, first because our list of topics does not cover all possible ones, and second because we only need a large enough subsample of posts to make some statistical considerations. Conversely, it is of utmost importance that the accuracy of the classifier is high since misclassified posts could lead to wrong conclusions about the distribution among the available topics. Therefore, we manually validate the accuracy of our methodology for topic detection, as described in the following paragraph.

\subsection{Classifier Precision Evaluation}

We empirically evaluated the accuracy of the classifier by taking a random subsample of the labeled posts, i.e., $100$ posts for each topic for a total of $500$ messages, and manually classifying them. To this end, we defined a lower and upper bound for accuracy. Indeed, even for a human being, it is challenging to univocally classify posts based on their content. Therefore, we defined three possible states for each classification decision: \lq\lq t" correct classification, \lq\lq f" incorrect classification, and \lq\lq ncc" standing for not completely correct (indicating that the assigned topic is related to the post but may not be the main topic of the post or the classification of the post is difficult). Given this states subdivision, the precision bounds are as follows:
\vspace{-0.15cm}
\begin{gather}\label{eq:upper_lower_precision}
	P_L = \frac{N_t}{N_t+N_f+N_{ncc}} \\
	P_U = \frac{N_t+N_{ncc}}{N_t+N_f+N_{ncc}}
\end{gather}

We refined our term selection for each topic to improve precision based on this analysis .\footnote{We make the final list of terms available at \url{https://mplanestore.polito.it:5001/sharing/0wD5oU6xr}.} % Then, we randomly pick a new set of $100$ posts per topic and repeat the manual labelling process to validate the performance. 
The classifier's precision is subject-dependent but was consistently above $80\%$ considering the upper bound defined in \eqref{eq:upper_lower_precision}. The classification is particularly effective in the case of \emph{politics} and \emph{pandemic}, where the precision goes above $90\%$. Table \ref{tab:precision} summarises the bounds on precision achieved by the procedure described above. These results are sufficient to use the classification to support our modelling assumptions.

\begin{table}[h!]
	\begin{center}
		\footnotesize
		\caption{Per-topic Precision}
		\label{tab:precision}
		\begin{tabular}{l|
				S[table-number-alignment=right]
				S[table-number-alignment=right]}
			\toprule % <-- Toprule here
			\textbf{Topic} & \textbf{Precision l.b.} & \textbf{Precision u.b.}\\
			\midrule % <-- Midrule here
			Sports   & 76.9 & 83.2\\
			Politics & 87.0 & 94.4\\
			Music    & 53.4 & 84.5\\
			Food     & 65.5 & 82.4\\
			Pandemic & 76.6 & 93.1\\
			\bottomrule % <-- Bottomrule here
		\end{tabular}
	\end{center}
\end{table}
The average percentage of messages classified is $27.8\%$ for all influencers in the dataset. Considering the final classifier and the analysed dataset, we automatically flagged about one million posts\footnote{1167963 posts were tagged with at least one label.} with at least one topic. Of these, only $6.7\%$ of the posts were flagged with multiple labels, indicating the message dealt with more than one topic.
We decided to consider in the rest of the work only influencers for whom it was possible to classify more than a thousand posts in the observed period. At the end of this filtering process, we could keep $237$ influencers for whom the average posts' classification percentage is $53.2\%$.

The dataset used contains a subset of Italian politicians. To check the correctness of the labelling procedure, we checked whether the derived reference topic for all politicians was \textit{politics}. It turned out that two politicians did not have \textit{politics} as reference: Vincenzo De Luca had \textit{pandemic}, and Renata Briano had \textit{food}. However, this is entirely understandable as the latter runs a food blog and the former was known for his firm and frequent statements on the pandemic situation during the COVID -19 pandemic.

\section{Proofs of Theorems \eqref{theo1} and \eqref{theo2}}\label{app:proofs}

\subsection{ Proof of Theorem \eqref{theo1}}

Let us start  assuming   $k_i(0)> 0$ $\forall i$. In such a case we denote with $i_0=\arg\max_i k_i(1)$ and with $K:=k_{i_0}(1)$.
First we show that the  problem:
\vspace{-2mm}
\begin{equation} \label{relaxed-pop-constraint}
	k_i(  y_i)-c y_i =0,  \quad\text{with } y_i\in[0,1]\;\; \forall i 
	\vspace{-2mm}
\end{equation}
admits a solution for any $c\ge K$.  Indeed  by 
choosing $c\ge K$  we have that necessarily  $k_{i}(1)\le k_{i_0}(1)\le c\cdot 1\; \forall i$ while 
$ k_{i}(0)> c\cdot 0=0$; therefore a zero $z_i(c)$ must exist for every $i$.  This zero is unique  as a consequence of the concavity of $k_i(\cdot)$. 
The set of zeros  ${z_i(c)}_i$  provides a solution of \eqref{relaxed-pop-constraint}. Now to get a solution of the original problem  \eqref{pop-constraint}
we need to show that  there exist a $c$  such that $\{z_i(c)\}_i$ are normalized. Observe that for $c=K$ by construction $z_{i_0}(K)=1$ while 
$0<z_i(K)\le 1$ for $i\neq i_0$, therefore $\sum_i z_i(K)>1$. Now, due to the monotonicity and concavity of $k_i(\cdot)$,  $z_i(c)$  is by construction decreasing with respect to
$c$, moreover  $z_i(c)\to 0$ as $c\to \infty$ $\forall i$, therefore since $\sum_i z_i(\cdot)$ is a continuous function of its argument, there will necessarily be a $c_0$ in correspondence of which  
$ \sum_i z_i(c_0)=1$.
In the case in which   $k_i(0)= 0$,  observe that $0$ is a solution of  \eqref{relaxed-pop-constraint} for any $c$, i.e. $z_i(c)=0$.  Moreover for any $c\ge K$ a second zero may exist. For example, by construction, $z_{i_0}(K)=\{0,1 \}$.
Therefore for $c=K$,  as before, we can always choose as set of zeros $\{z_i(K)\}_i$, such that
$z_i(K)=0$ if $k_i(0)=0$, and $i\neq i_0$, $z_{i_0}(K)=1$. By construction  $\sum_i z_i(K)\ge 1$. In particular 
$\sum_i z_i(K)> 1$ is there exists a $i$ such that $k_i(0)>0$. In this latter case, by increasing $c$ all the non null zeros decrease, therefore, as before, there will necessarily be a $c_0$ in correspondence of which  
$ \sum_i z_i(c_0)=1$. $\square$\\

\vspace{-5mm}
\subsection{ Proof of Theorem \eqref{theo2}}

We first show that  
$||\bar{\pi}^{(1)} -\bar{\pi}^{(2)} ||_{L_{\infty}}=\max_i |\bar{\pi}_i^{(1)} -\bar{\pi}_i^{(2)}|=
|| \mathcal{G}({F_1}(x,z))-\mathcal{G} ({F_2}(x,z)) ||_{L_\infty}\le M||F_1(x)-F_2(x) ||_{L_\infty}$;
then we show that we can always enforce:
$||F_1(x,z)-F_2(x,z) ||_{L_\infty}= || \mathcal{H}({\bar{\pi}^{(1)}})-\mathcal{H} ({\bar{\pi}^{(2)}}) ||\le 1/(2M)||\bar{\pi}^{(1)} - \bar{\pi}^{(2)} ||_{L_\infty}$ 	
by properly choosing  	$\omega(\cdot,  \cdot)$ and $\theta(\cdot)$. Therefore, we can
conclude that $|| \mathcal{H}\circ \mathcal{G}( F_1(x,z)) -\mathcal{H}\circ \mathcal{G}(F_2(x,z))||\le 1/(2M)|| (\mathcal{G}( F_1(x)) -(\mathcal{G}(F_2(x))|| \le M/(2M)|| F_1(x,z) - F_2(x,z)||=1/2  || F_1(x,z) - F_2(x,z)||$.
%||\bar{\pi}^{(1)} - p^{(2)} ||

First note that $||F_1(x,z)-F_2(x,z) ||_{L_\infty}= \sup_x|F_1(x,z)-F_2(x,z)| $ coincides with the Kolmogorov distance between
the two distributions.

Let us denote with
\[
k_i(y,F_1(x,z))=\lambda f^{(i)} \int \int  \theta(|x-x_i|) \rho (\bar{\pi}_i, |x-x_i|)\mathrm{d}F_1(x,z) ,
\]
and similarly  for $k_i(y,F_2(x,z))$ we assume that: 
\[
\sup_{y\in[0,1],i} |k_i(y,F_1(x,z))-k_i(y,F_2(x,z))|:=\Delta K(F_1,F_2)  \le a||F_1(x,z)-F_2(x,z)  ||_{L_\infty} \qquad a\in \mathbb{R}^+
\]
and 
\[
\frac{\mathrm dk_i(y, F_1(x,z))}{\mathrm dy} \mid _{y=0}< \max_i  k_i(1, F_1(x,z)) \qquad  \frac{\mathrm dk_i(y, F_2(x,z))}{\mathrm dy} \mid _{y=0}<\max_i  k_i(1, F_2(x,z))  \quad \forall i .
\]
Without lack of generality we assume $\max_i  k_i(1, F_1(x)) \ge \max_i  k_i(1, F_2(x)) $.
Let the pair  $(\bar{\pi}^{(1)}= \{\bar{\pi}^{(1)}_i\}_i, c_1)$ be the solution of 
\[
k_i(y_i,F_1(x,z))-cy_i=0  \quad \text{s.t} \sum_i y_i=1, y_i\ge 0, \; \forall i
\]
now let  $(\{\widehat p^{(2)}_i\}_i)$ the non necessarily normalized solution of 
\[
k_i(y_i,F_2(x,z))-c_1y_i=0  \quad \text{s.t }  y_i\;\ge 0, \; \forall i.
\]
by means of  elementary geometric considerations we can bound: 
\[
|\bar{\pi}^{(1)}_i-\widehat p^{(2)}_i|\le  \frac{ \Delta K(F_1, F_2) }{c_1 - h_1}
\]
where $h_1=\frac{\mathrm dk_i(y, F_1(x))}{\mathrm dy} \mid _{y=\min (\bar{\pi}_i^{(2)}, \widehat p_i^{(2)}) }\le \frac{\mathrm dk_i(y, F_1(x))}{\mathrm dy} \mid _{y=0}$. We recall that by construction (see proof of Theorem \ref{theo1})
we have	$c_1> \max_i k_i(1, F_1(x)))$.

Denoting with $|\widehat p^{(2)}|= \sum_i\widehat p^{(2)}_i$ , we have 
\[
1-\sum_i |\widehat p^{(2)}_i-\bar{\pi}^{(1)}_i| \le | \widehat p^{(2)}|\le 1+\sum_i |\widehat p^{(2)}_i-\bar{\pi}^{(1)}_i|
\]
Now denoted with $(\{\bar{\pi}^{(2)}_i\}_i, c_2)$ the solution of 
\[
k_i(y_i,F_2(x,z))-cy_i=0  \quad \text{s.t} \sum_i y_i=1, y_i\ge 0, \; \forall i
\]
we have, by construction, that: 
\[
\frac{1}{\max(1, |\widehat p^{(2)}| )}<\frac{c_1}{c_2}<\frac{1}{ \min(1, | \widehat p^{(2)}| )}
\]
and therefore, exploiting again elementary geometrical arguments, we can bound: 
\[
|\widehat p^{(2)}_i-\bar{\pi}^{(2)}_i|\le \left| \left(\frac{c_1-h_2}{c_2-h_2}-1\right)\widehat p^{(2)}_i \right|
\]
where $h_2=\frac{d k_i(y, F_2(x,z))}{dy}\mid_{y=\min(\widehat p^{(2)}_i,\bar{\pi}^{(2)}_i)}=\frac{d k_i(y, F_2(x,z))}{dy}\mid_{y=0}$.
Putting everything together, we have proved that:
\[
\max_i| ||\bar{\pi}^{(1)}_i - \bar{\pi}^{(2)}_i  ||=|| \mathcal{G}({F_1}(x,z))-\mathcal{G} ({F_2}(x,z)) ||_{L_\infty}\le M||F_1(x,z)-F_2(x,z) ||_{L_\infty} 
\]
To conclude the proof,  first note that  by properly choosing $\rho(\cdot, \cdot)$ and $\theta(\cdot)$
we can assume  $v_x(x,z)$ and  $\sigma^2_x(x,z)$ to depend sufficiently smoothly on 
${\bar{\pi}}$, i.e.    $ \forall  \varepsilon>0$  we can assume:
\[
\sup_{x}\left|\left| v^{(1)}_x(x,z) -  v^{(2)}_x(x,z) \right |\right|_{L_\infty} \le \varepsilon  ||\bar{\pi}^{(1)} - \bar{\pi}^{(2)} ||_{L_\infty} \qquad  \forall z,
\] 
\[
\sup_{x}\left|\left| \sigma^{2,(1)}_x(x,z) -  \sigma^{2,(2)}_x(x,z) \right|\right|_{L_\infty} \le \varepsilon||\bar{\pi}^{(1)} - \bar{\pi}^{(2)} ||_{L_\infty} \qquad  \forall z,
\]
and
\[
\sup_{x}\left|\left| \frac{\partial  \sigma^{2,(1)}_x(x,z) }{\partial x}-\frac{ \partial \sigma^{2,(2)}_x(x,z)}{\partial x} \right|\right|_{L_\infty} \le \varepsilon ||\bar{\pi}^{(1)} - \bar{\pi}^{(2)} ||_{L_\infty} \qquad  \forall z.
\]

Then observe that  the solution of the Fokker-Planck equation given in \eqref{solution-FP}  on a compact interval (and so also its primitive) 
depends smoothly on function  $v_x(x,z)$, function $\sigma^{2}_x(x,z)$ and its first derivative, as long as
$\inf_{x,z} \sigma^{2}_x(x,z)$ is bounded away from zero. As final remark note that the set  of weakly-increasing functions  $F(x)$, such that $F(a)=0$ and $F(b)=1$ equipped with the $L_\infty$-norm forms a closed set in a complete metric space.
$\square$

\section{Validation of the fluid limit approximation}\label{sec:simulation}
In this section, we compare  predictions of the simplified \textit{fluid limit} against simulation results of  the full  stochastic model described by algorithm \ref{alg:model} (obtained through a Monte-Carlo approach) . We restrict ourselves to a one-dimensional opinion space, as in Section \ref{sec:analysis}, and assume that  all users share the same prejudice~$z$. Again, we consider two \lq \lq competing" influencers. A similar analysis could be performed in scenarios  with any number of influencers at any point in the opinion space, but this would be computationally more challenging since multiple stationary points may exist, each with its own attraction basin.

First, we derive in \ref{sec:preliminary} some preliminary analytical results for the case of two influencers, using the results of the \textit{fluid limit} introduced in Section \ref{sec:stoch_comp}. Then in
\ref{sec:extreme} two extreme instances of the model are solved in closed form. \ref{sec:comparison} is devoted to comparing the analytical results of the \textit{fluid} model with simulations. Finally, we discuss the impact of content personalization.

\subsection{Two competing influencers} \label{sec:preliminary}

Let us specialize the equations presented in Section \ref{sec:stoch_comp} for the mean opinion $\bar{x}(z)$ (Eq. (\ref{eq:o_bar_comp})) and the normalized popularities $\bar{\pi}_i$ (Eq. (\ref{eq:tilda_p})). Note that for $N_i = 2$, $\bar{\pi}_0 = 1 - \bar{\pi}_1$, so it is sufficient to study $\bar{\pi}_1$.

As for the mean user opinion $\bar{x}(z)$, equation (\ref{eq:o_bar_comp}) allows us to write the asymptotic mean directly as a function of $\bar{\pi}_1$ and the opinions of the two influencers $x^{(0)}, x^{(1)}$:
\begin{equation}\label{eq:o_bar_2infl}
	\bar{x}(z) = \frac{\alpha}{1-\beta} z + \frac{\gamma}{1-\beta} \left[ \left( 1-\bar{\pi}_1 \right)x^{(0)} + \bar{\pi}_1 x^{(1)} \right]
\end{equation}

Substituting the functional forms of the \textit{visibility} $\omega$ and \textit{feedback} $\theta$ {(as given in Table \ref{tab:table_param})} into equation (\ref{eq:tilda_p}), we obtain the following expression for the normalized popularity $\bar{\pi}_1$:

\begin{equation}\label{eq:eq_snake}
	\bar{\pi}_1 = \frac{f^{(1)} e^{-\rho \frac{\left(x^{(1)}-\bar{x}\right)^2}{\bar{\pi}_1}} \left(1-|x^{(1)}-\bar{x}|\right)}{\sum_{i \in \{0,1\}}f^{(i)} e^{-\rho \frac{\left(x^i-\bar{x}\right)^2}{\bar{\pi}_1}} \left(1-|x^{(i)}-\bar{x}|\right)}  = f(\bar{\pi}_1, \bar{x})
\end{equation}

Moreover, if we combine the above expression with equation (\ref{eq:o_bar_2infl}) for $\bar{x}$, we get $\bar{\pi}_1=f(\bar{\pi}_1$), which can be solved numerically through a fixed-point approximation (FPA) (a graphical representation is shown on Fig.~\ref{fig:snake}). 
The outcome of this FPA and the corresponding simulation results are compared in Figure \ref{fig:sim_th_comparison}.

\subsection{Closed form computations in extremal cases}\label{sec:extreme}

The combination of equations (\ref{eq:o_bar_2infl}) and (\ref{eq:eq_snake}) cannot be solved in closed form in the general case.
However, there are at least two scenarios in which this is possible, separately considered in the following subsections.

\subsubsection{When an influencer ``wins"}\label{sec:infl_wins}

We consider an influencer a ``winner" if its normalized popularity $\bar{\pi}_i$ approaches 1. Suppose that the influencer whose opinion is $x^{(1)}=1$ \textit{wins}, then $\bar{\pi}_1 \rightarrow 1$. This implies $\bar{\pi}_0 \rightarrow 0$ and thus $\omega \rightarrow 0^+$: the influencer with $x^{(0)}=0$ is seen by a negligible fraction of users and in practice, only influencer $i=1$ remains visible.
Note that in the extreme case in which influencer $1$ wins, users see only $x^{(1)}$, and asymptotically all users move towards it. In this case, the final opinion $\bar{x}(z)$ can be easily calculated with a recursion of the update rule~(\ref{eq:d_model_user_update}):

\begin{equation*}
	x^{(u)}(n) = \sum_{i=0}^n \beta^i\left(\alpha z + \gamma x^{(1)}\right) + \beta^n x(0)
\end{equation*}

For $n \rightarrow \infty$ and considering $\beta < 1$ (the case $\beta=1$ coincides with the trivial case where users remain fixed at their initial opinion) 
we get:

\begin{equation}
	x^{(w)} = \frac{\alpha}{1-\beta} z + \frac{\gamma}{1-\beta}x^{(1)},
\end{equation}

which is in agreement with (\ref{eq:o_bar_2infl}) if one sets $\bar{\pi}_1 = 1$. 
This corresponds to one of the extreme cases that we will use later to examine the model behavior as a function of the personalization parameter $\rho$. 
It should be noted that this construction relies on the knowledge of the \textit{winning} influencer, which is unknown in advance. However in the fluid limit, %it is easy to understand 
we expect that the winning influencer, if any, is the one that has a structural advantage over the others at the beginning (e.g., a higher posting rate $f^{(i)}$, see Figure \ref{fig:metric_rate}).

\subsubsection{Constant personalization function}\label{sec:const_rho}
The other extreme case we consider is the one in which $\rho=0$. In this case, the personalization function $\omega$ no longer depends on $\bar{\pi}_i$, and it is easy to see from Table \ref{tab:table_param} that it returns $\omega \equiv 1$. Moreover, we consider $x^{(1)}=1, x^{(0)}=0$, which further simplifies (\ref{eq:o_bar_2infl}). The above formulas (Eq. \ref{eq:eq_snake} and Eq. \ref{eq:o_bar_2infl}) can then be solved in closed form. 
In particular,  equation (\ref{eq:eq_snake}) for the normalized popularity $\bar{\pi}_1$ becomes:
\vspace{-1mm}
\begin{equation*}
	\bar{\pi}_1 = \frac{f^{(1)} \left(q+m\,\bar{\pi}_1\right)}{f^{(0)}\left( 1-(q+m\, \bar{\pi}_1)\right)+f^{(1)} \left(q+m\,\bar{\pi}_1\right)}
\end{equation*}
where $m \triangleq \frac{\gamma}{1-\beta}$ and $q \triangleq \frac{\alpha}{1-\beta}z$ for compactness. This leads to a second order equation which can be easily solved for $\bar{\pi}_1$:
\vspace{-1mm}
\begin{equation}\label{eq:2nd_eq}
	\bar{\pi}_1^2\, m (f^{(1)}-f^{(0)}) + \bar{\pi}_1 \left[ f^{(0)} (1-q) + f^{(1)} (q-m) \right] - f^{(1)} q = 0
\end{equation}

\subsection{Comparison of analytical results and Monte-Carlo simulations}\label{sec:comparison}
This section is devoted to comparing the analytical results derived in Section \ref{sec:analysis} with simulations of the model. 
%stochastic system described  by algorithm \ref{alg:model}, considering an example scenario specified in \ref{sec:scenario}. 
Numerical and graphical solutions of equation (\ref{eq:eq_snake}) are also provided, shedding light on the impact of the algorithmic personalization performed by the platform.

\subsubsection{Description of the scenario}\label{sec:scenario}

The scenario setting is analogous to that described in Section \ref{sec:sim_scenario} and Table \ref{tab:table_param}. % \ref{tab:table_param}. 
However, here, we consider a  one-dimensional opinion space $[0,1]$ and we assume all  users to have the same prejudice, i.e., $z^{(u)}=z=0.4, \forall u \in \mathcal{U}$ matching their initial opinion $x^{(u)}(0)$. The \lq\lq competing " influencers have opinions at the extremes of the domain, and their posting frequencies are $f^{(1)}=0.7$ and $f^{(0)}=0.3$, i.e., influencer $i=1$ has a \textit{structural advantage} over influencer $i=0$. %The initial absolute popularity is set equal for both influencers, $p_{0,1}(0) =100$.
Note that in a one-dimensional space, the reference direction $r^{(i)}$, and hence the consistency $c^{(i)}$, lose their significance. 
To avoid obtaining trivial results in which influencer 1 obviously wins, regular users are initially placed closer to the disadvantaged influencer $i=0$.

\subsubsection{Simulation, fluid limit and fixed-point approximation}\label{phase-trans}

\begin{figure}[h]
	\centering
	\includegraphics[width=.7\linewidth]{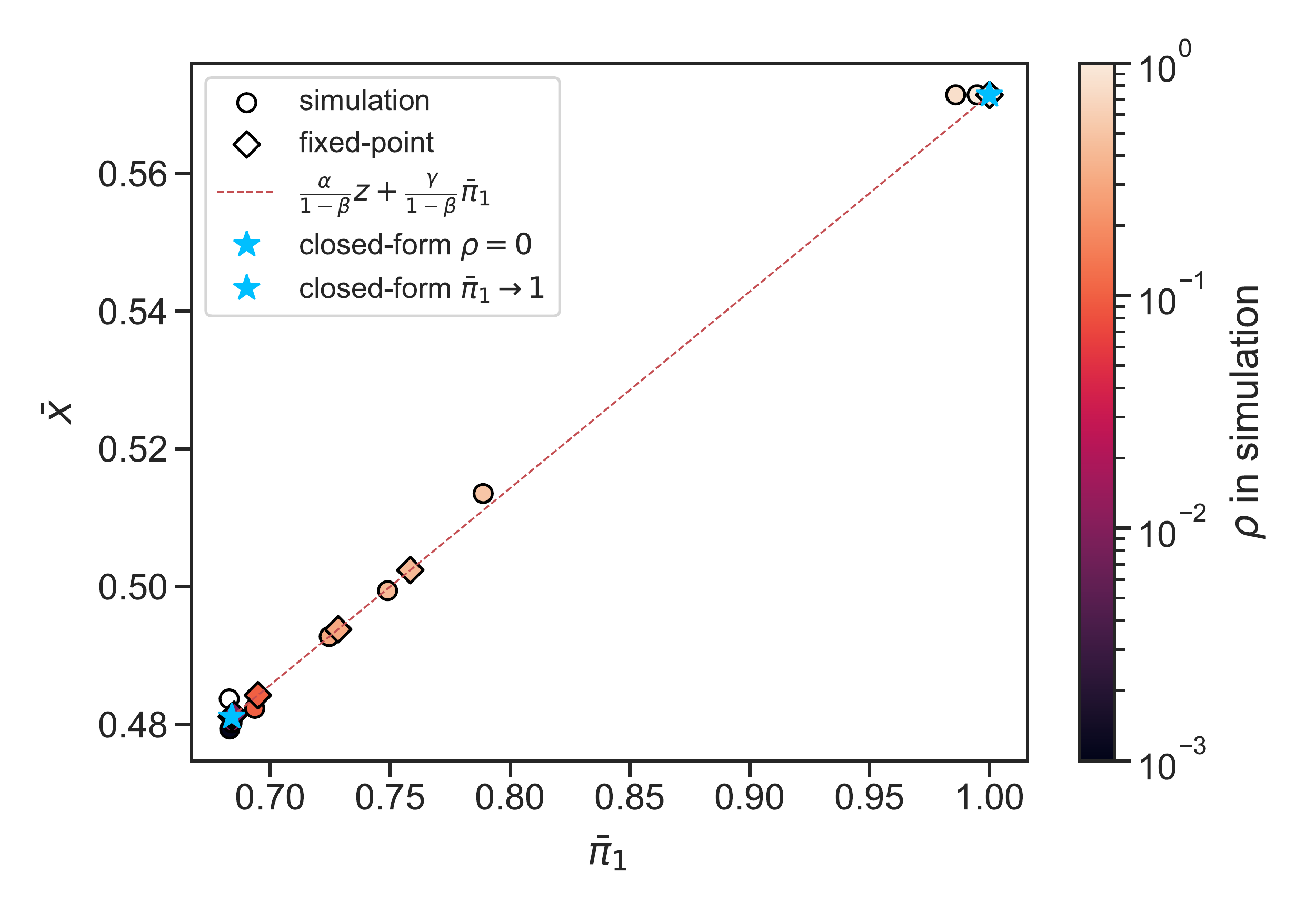}
	\caption{Comparison between analytical results, including the exact extreme points calculated in \ref{sec:const_rho} and \ref{sec:infl_wins}, and the linear relationship 
		between $\bar{x}$ and $\bar{\pi}_1$ according to Equation (\ref{eq:o_bar_2infl}).
		Diamonds represent the fixed-point approximation for the solution of equation (\ref{eq:eq_snake}).
		Simulation results of the stochastic dynamics, represented by circles, 
		were obtained by averaging $100$ realizations of the process as described 
		by Algorithm \ref{alg:model}.
		We consider a scenario in which $\alpha=0.05, \beta=0.93$, with two influencers 
		at the extremes of the domain, with $f^{(0)}=0.3, f^{(1)}=0.7$ and the same 
		initial absolute popularity $p_0=p_1=100$. Numerical values from 
		simulation and fixed-point approximation are reported in the table
		alongside the plot in Fig. \ref{fig:snake}.}
	% , alongside with the graphical solution of \ref{eq:eq_snake}.}
\label{fig:sim_th_comparison}
\end{figure}

Comprehensive validation and comparison of the approaches used to obtain
the system equilibria are shown in Figure \ref{fig:sim_th_comparison}. First, the stochastic model described by Algorithm \ref{alg:model} is \lq\lq simulated"  by obtaining  $100$ different sample whose length is $500000$ elementary steps. 
%{\color{red} ho mediato sugli ultimi 100 samples}
The variables of interest $\bar{x}(z)$ and $\bar{\pi}_1$ are obtained by averaging the process over both discrete times steps $n$ and sample paths and are represented by circle marks. Second, equation (\ref{eq:o_bar_2infl}), which is a specialization of (\ref{eq:o_bar_comp}) obtained from the fluid limit, indicates that the state of the system lies on a line in the plane $\bar{\pi}_1$,$\bar{x}$ (dashed line in Figure \ref{fig:sim_th_comparison}). Third, the extreme cases of the model analyzed in \ref{sec:extreme}, for which we derived a closed-form solution, are represented by star-like marks. Lastly, diamonds are solutions of (\ref{eq:eq_snake}) employing the fixed-point approximation. 

\vspace{3mm}

\begin{figure}[h!]
\centering
\begin{subfigure}{.30\textwidth}
	\vspace{-0.8cm}                  % to allign the table with the figure
	\captionsetup{labelformat=empty} % avoid the (a) in the Table caption
	\caption{Simulation and FPA}
	\centering
	\scalebox{0.80}{
		%\caption{Simulation and fixed-point comparison}
		\label{tab:table1}
		\begin{tabular}{S[table-number-alignment=right]|
				S[table-number-alignment=right]
				S[table-number-alignment=right]}
			\toprule % <-- Toprule here
			\textbf{$\rho$} & \textbf{$\bar{\pi}_1$} & \textbf{$\bar{\pi}_1$}\\
			& $SIM$ & $FPA$ \\
			\midrule % <-- Midrule here
			0.0   & 0.682 & 0.684\\
			0.001 & 0.683 & 0.684\\
			0.01  & 0.684 & 0.685\\
			0.1   & 0.693 & 0.695\\
			0.3   & 0.725 & 0.728\\
			0.4   & 0.749 & 0.758\\
			0.5   & 0.789 & 0.999\\
			0.8   & 0.986 & 0.999\\
			1.0   & 0.995 & 1.0\\
			\bottomrule % <-- Bottomrule here
		\end{tabular}
	}
\end{subfigure}
\begin{subfigure}{.65\textwidth}
	\centering
	\includegraphics[width=0.9\linewidth]{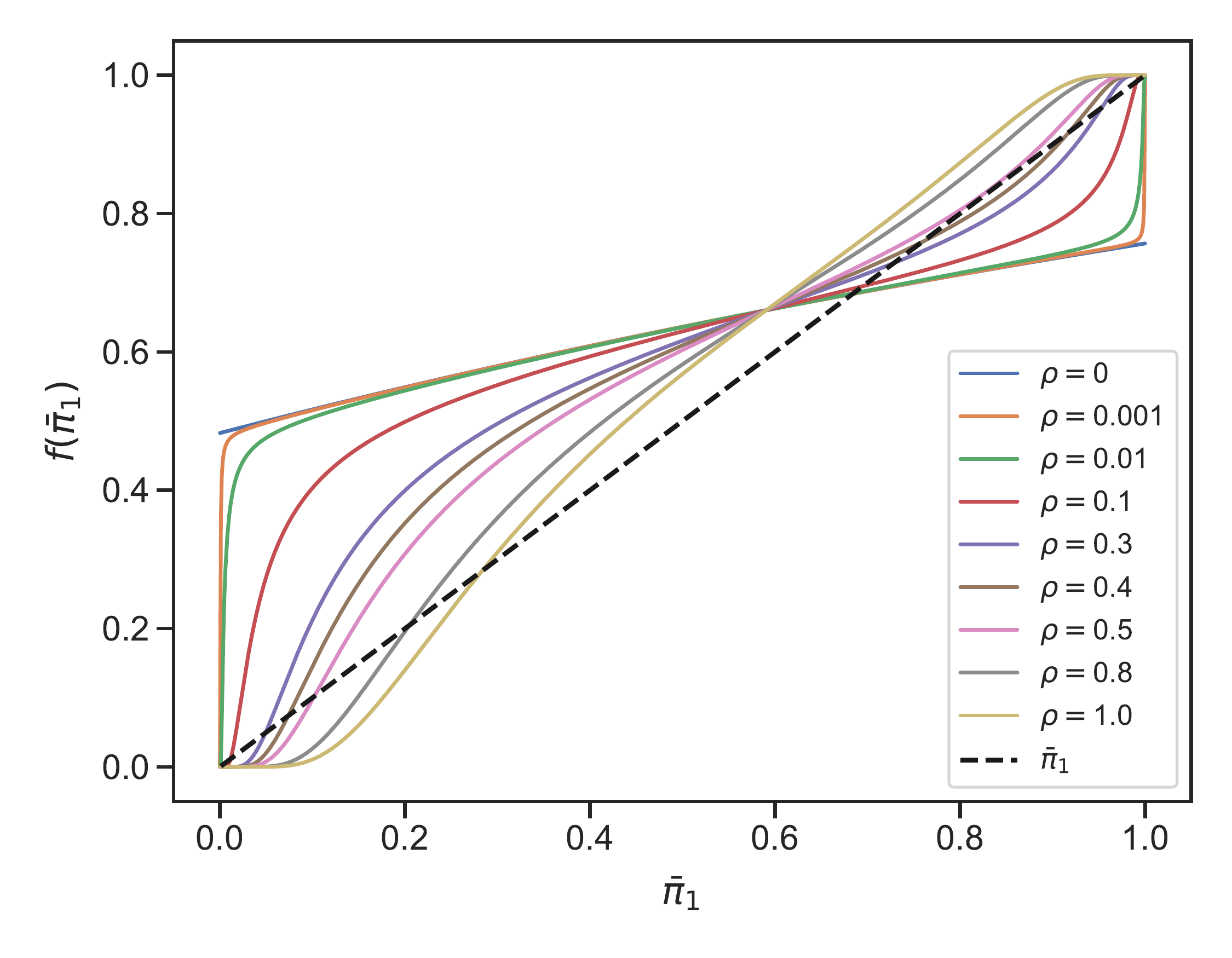}
\end{subfigure}
\caption{Graphical solution of  $\bar{\pi}_i = f(\bar{\pi}_i)$,
	(\ref{eq:eq_snake}). Stable solutions corresponds 
	to intercepts between $f(\bar{\pi}_i)$ and the
	bisector, such that $f'(\bar{\pi}_i) < 1$.
	We observe that non-trivial solutions (i.e., solutions
	in which both influencers survive) exist, roughly in the interval $[0.7,0.8]$, 
	provided that $\rho$ is not too large (i.e., $\rho < 0.5$).
	For $\rho > 0.5$, the only stable solution is $\bar{\pi} = 1$.
	% Interestingly, one can observe a portion of the dashed line that has no intercepts followed by a portion where the intercepts are not stable (i.e. the tangent line has a greater slope than the bisector). 
	This explains the results in Figure \ref{fig:sim_th_comparison}.
	Simulation results reported on the alongside table confirm
	the validity of the analytical predictions. }
\label{fig:snake}
\end{figure}

We observe that, for given $\rho$, simulation marks match well with analytical marks. The only exception is for $\rho = 0.5$, for which simulations provide $\bar{\pi}_1 \approx 0.79$, whereas the analysis provides $\bar{\pi}_1 \approx 1$ (see also the table on Fig. \ref{fig:snake}).
This mismatch is due to the fact that $\rho = 0.5$ is close to a \lq phase transition', at which the system switches from a regime in which two stable solutions exist (in particular, one in which both influencers survive) to a regime in which influencer $i=1$ wins. In such a situation, the population is exposed to the opinions of a single individual, hindering diversity on the social platform.
This behavior is better illustrated in Fig.~\ref{fig:snake}, where the curve corresponding to $\rho=0.5$ is almost tangent to the bisector.
It should be noted that the \lq\lq empty" region in Figure \ref{fig:sim_th_comparison} is directly related to this behavior since no stable solutions can exist for that values of $\bar{\pi}_i$. In fact, there is no \textit{stable} intersection with the bisector in Figure \ref{fig:snake} in the corresponding interval.
\begin{comment}
	% to better allign within the table
	\begin{table}[h!]
		\begin{center}
			\caption{Comparison between simulation and FPA}
			\label{tab:table1}
			\begin{tabular}{S[table-number-alignment=right]|
					S[table-number-alignment=right]
					S[table-number-alignment=right]}
				\toprule % <-- Toprule here
				\textbf{$\rho$} & \textbf{$\tilde{p}_1$} & \textbf{$\tilde{p}_1$}\\
				& $sim$ & $fp$ \\
				\midrule % <-- Midrule here
				0.0   & 0.682 & 0.684\\
				0.001 & 0.683 & 0.684\\
				0.01  & 0.684 & 0.685\\
				0.1   & 0.693 & 0.695\\
				0.3   & 0.725 & 0.728\\
				0.4   & 0.749 & 0.758\\
				0.5   & 0.789 & 0.999\\
				0.8   & 0.986 & 0.999\\
				1.0   & 0.995 & 1.0\\
				\bottomrule % <-- Bottomrule here
			\end{tabular}
		\end{center}
	\end{table}

	\begin{figure}[h]
		\centering
		\includegraphics[width=.9\linewidth]{/vect/sim_vs_equations/snake_num_solution.pdf}
		\caption{Graphical solution of the equation \ref{eq:eq_snake} of the form $\tilde{p_i} = f(\tilde{p_i})$. Interestingly, one can observe a portion of the dashed line that has no intercepts followed by a portion where the intercepts are not stable (i.e. the tangent line has a greater slope than the bisector). This gives an indication of the reason for the area without points in Figure \ref{fig:sim_th_comparison}.}
		\label{fig:snake}
	\end{figure}
\end{comment}


\section{Additional simulation results}\label{sec:appendix_sim}
\begin{figure}[h]
\centering
\includegraphics[width=.5\linewidth]{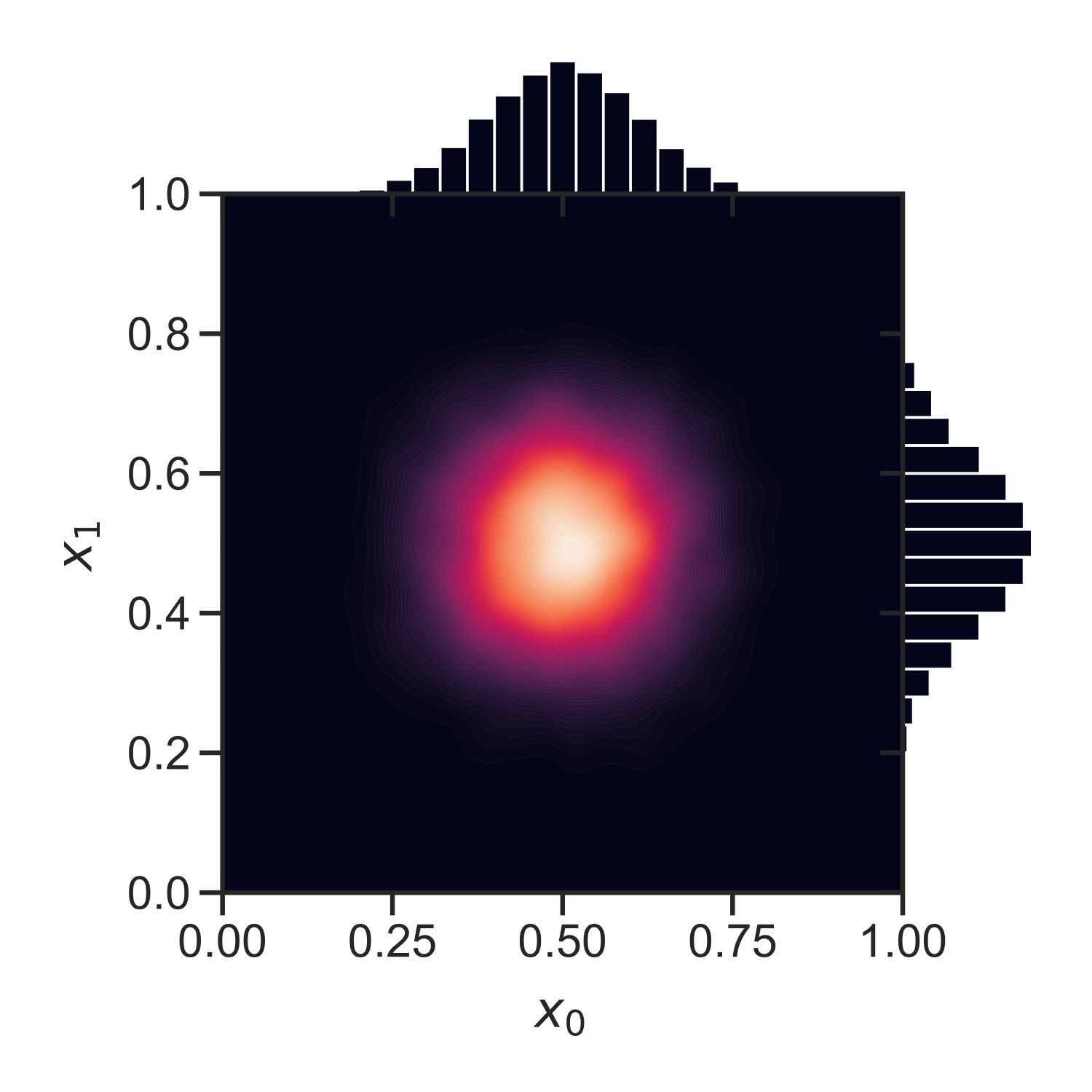} %.6
\caption{Initial opinion (prejudice) distribution of the regular users, recall $\bm{x}^{(u)}(0)=\bm{z}^{(u)}$.}
\label{fig:op_init}
\end{figure}

{
Here we report all the \textit{default} parameters used in the simulations of Section \ref{sec:model_vs_param} (Table \ref{tab:table_param}) and the behavior of the model as a function of the \textit{degree of stubbornness} which is interesting but not the focus of our investigation.
\subsection{Details of the simulation scenario}\label{sec:add_reference_scenario}

In this section, we provide further details on the simulation setting of Section \ref{sec:model_vs_param}. In Figure \ref{fig:op_init} we show the initial distribution of \textit{regular} users' opinions and, since $\bm{x}^{(u)}(0)=z^{(u)}$ also of their prejudice. Table \ref{tab:table_param} contains all the \textit{default} choices of the model's parameters.
}
% TO BE USED IN THE DRAFT VERSION

% TO BE USED IN THE DRAFT VERSION
%\vspace{5mm}
\begin{table}[h!]
\begin{center}
	\footnotesize
	\caption{Parameters and functions shared across experiments}
	\label{tab:table_param}
	\begin{tabular}{l|c|l}
		\toprule % <-- Toprule here
		\textbf{Symbol} & \textbf{Value - Form} & \textbf{Description}\\
		\midrule % <-- Midrule here
		$N_i$            & 2      & Number of influencers\\
		$x_j^{(0)}$      & 0      & Opinion of influencer $0$ on direction $j$\\
		$x_j^{(1)}$      & 1      & Opinion of influencer $1$ on direction $j$\\
		$r^{(0)}$        & 0      & Reference direction of influencer $0$\\
		$r^{(1)}$        & 1      & Reference direction of influencer $1$\\
		$p_{0,1}(0)$     & 100    & Initial absolute popularity of both influencers\\
		$N_u$            & 10000  & Number of regular users\\
		$N_{iter}$       & 100000 & Number of iterations for each simulation\\
		$\alpha$         & 0.05   & First weight in the updating rule in Eq. \ref{eq:d_model_user_update}\\
		$\beta$          & 0.93   & Second weight in the updating rule in Eq. \ref{eq:d_model_user_update}\\
		$\theta(\cdot)$  & $1 - \left|x_j^{(i)} - x_j^{(u)}\right|$                          & Functional form of the \textit{feedback} function\\
		$\omega(\cdot)$  & $e^{-\rho\frac{\left(x_{r}^{(u)}-x_{r}^{(i)}\right)^2}{\bar{\pi}_i}}$ & Functional form of the \textit{visibility} function\\
		$a$              & 10     & First parameter of the initial Beta distribution\\
		$b$              & 10     & Second parameter of the initial Beta distribution\\
		$z_j^{(u)}$      &  $x_j^{(u)}(0)$      & Prejudice coincides with initial opinion\\
		\bottomrule % <-- Bottomrule here
	\end{tabular}
\end{center}
\end{table}

\subsection{Behavior as function of the updating weights}\label{sec:ab}

The behavior of the system depends not only on the characteristics of the influencers and the composition of public opinion, but also on the parameters controlling the opinion update rule in equation (\ref{eq:d_model_user_update}). {The update is a convex combination of the prejudice, the current opinion, and the opinion conveyed by the post. We chose to hold fixed the weight $\beta$ (inertia)
and thet vary the \textit{degree of stubbornness},  $\frac{\alpha}{\alpha+\gamma}$.}

\begin{figure*}[h]
\centering
\includegraphics[width=.5\linewidth]{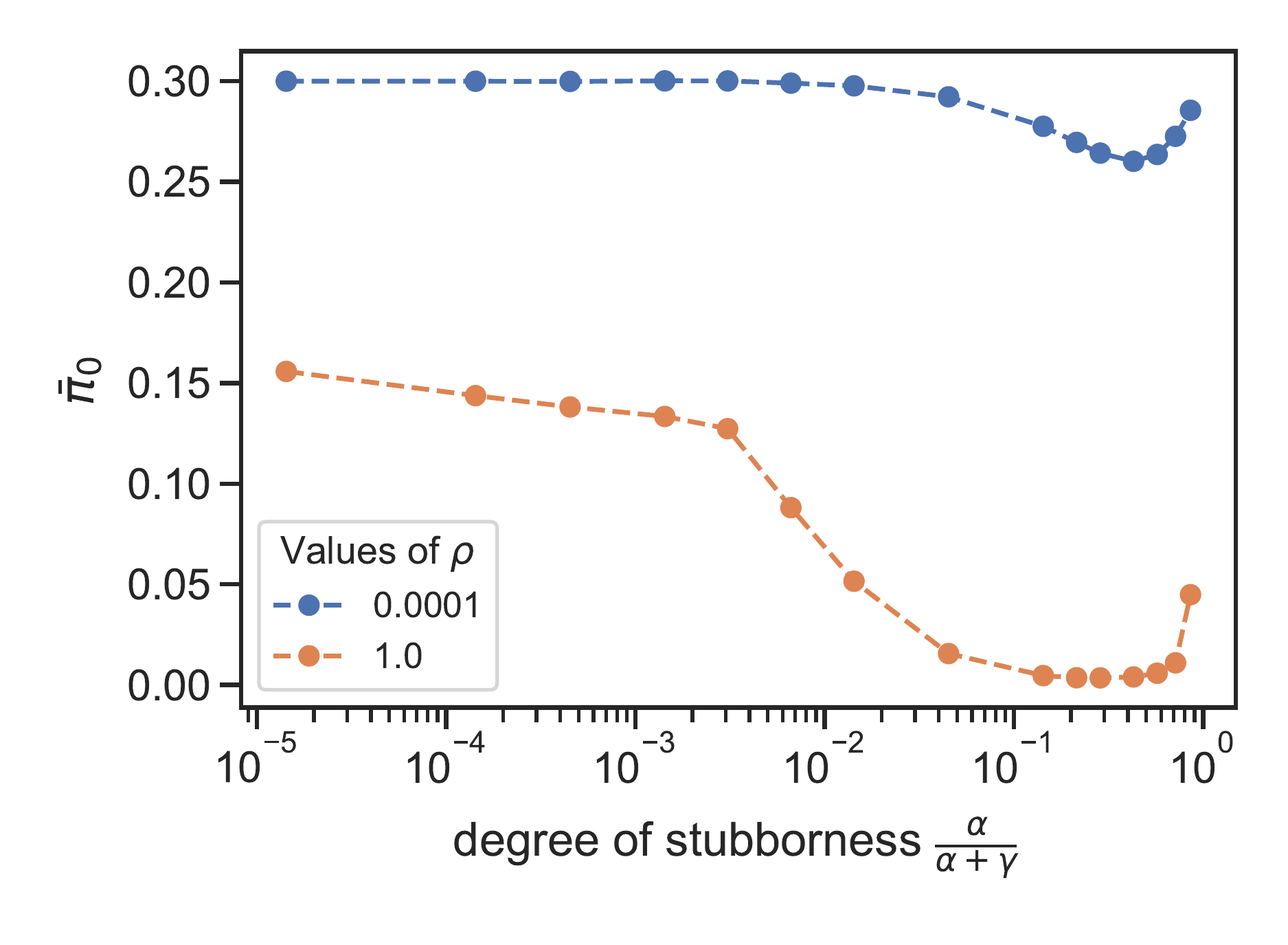}
\caption{Normalized popularity $\bar{\pi}_0$ as a function of the degree of stubbornness {$\frac{\alpha}{\alpha+\gamma}$}, the points are obtained considering 50 realizations of the process and averaging over 100 discrete time instants. Again, smooth ($\rho=0.0001$) and strict ($\rho=1$) personalization are considered. %two levels of algorithmic personalization are considered.
}
\label{fig:ab}
\end{figure*}

We considered an unbalanced scenario in which influencer $i=1$ has a structural advantage, i.e., $f^{(1)}=0.7 > f^{(0)}$. Figure \ref{fig:ab} again shows that personalization favors the structurally advantaged individual (consistent with Section \ref{sec:sim_rate}). Note that the $x$-scale is  logarithmic  to highlight the sudden drop of $\bar{\pi}_0$ {for $\frac{\alpha}{\alpha+ \gamma} \approx 10^{-3}$ (corresponding to small values of $\alpha$)} when \textit{sharp} personalization is applied. The shape of the two curves is quite similar, only the decrease is observed at different values of {$\frac{\alpha}{\alpha+\gamma}$.}  \textit{Smooth} personalization allows the coexistence of influencers on the whole domain, while with \textit{sharp} personalization, for  a wide range of parameters,   influencer $i=1$ \lq\lq wins".

In both cases, there is an initial phase (for low values of {$\frac{\alpha}{\alpha+\gamma}$}) in which the two influencers coexist, and this is followed by a drop of the normalized popularity of the disadvantaged influencer. This can be explained by the fact that small values of {$\frac{\alpha}{\alpha+\gamma}$} imply that a negligible weight
is given to the prejudice, and therefore regular users concentrate around the two influencers' opinions on their reference direction. This can be easily confirmed by looking at the final opinion configuration of users, who concentrate in the upper corners of the opinion space (around $[0,1]$ and $[1,1]$). This is because the influencer $i=1$, whose opinion is $x^{(1)}=[1,1]$ is stronger than the other in terms of popularity and is able to pull users along its non-reference direction as well. We remark that when users are very close in opinion to a particular influencer, it is difficult for the other to persuade them, as the probability of this happening is proportional to the product $\omega \cdot \theta$, both of which are a function of opinion distance. In these scenarios, the distance from the \lq\lq further" influencer is $d_j \approx 1$, which drastically reduces the probability of reaching the users. Thus, as long as {$\frac{\alpha}{\alpha+\gamma}$} is small enough, both influencers can build their user base. These situations represent rather degenerate cases where the population almost disregards their prejudice in favor of the opinion conveyed by the post. It might be interesting to consider users with varying degrees of \lq\lq volatility" who are able to pull along the opinion of their neighborhood.

As the degree of stubborness increases, %so does the inertia of the users. They 
users are more entrenched in their prejudice and therefore no longer concentrate in a small neighborhood of the influencer's opinions. This favors the structurally advantaged influencer, as the other (i.e., $i=0$) is unable to build its user base because users do not get close enough to it (see Figure \ref{fig:ab} for $i=1$ and $\rho=0.0001$ we have $\bar{\pi}_0 \rightarrow 0$). The subsequent rise in $\bar{\pi}_0$ depends on the fact that when {the degree of stubbornness} approaches {one}, users give importance only to their prejudice,  and therefore  they do not deviate too much from their initial position. As a consequence, it can not be triggered the positive feedback  between users' opinion   and influencers' popularity that  leads to the complete victory of one influencer. 

\section{Sensitivity analysis of the case study}\label{sec:sensitivity_cs}

\begin{figure}[h!]
\centering
\includegraphics[width=\linewidth]{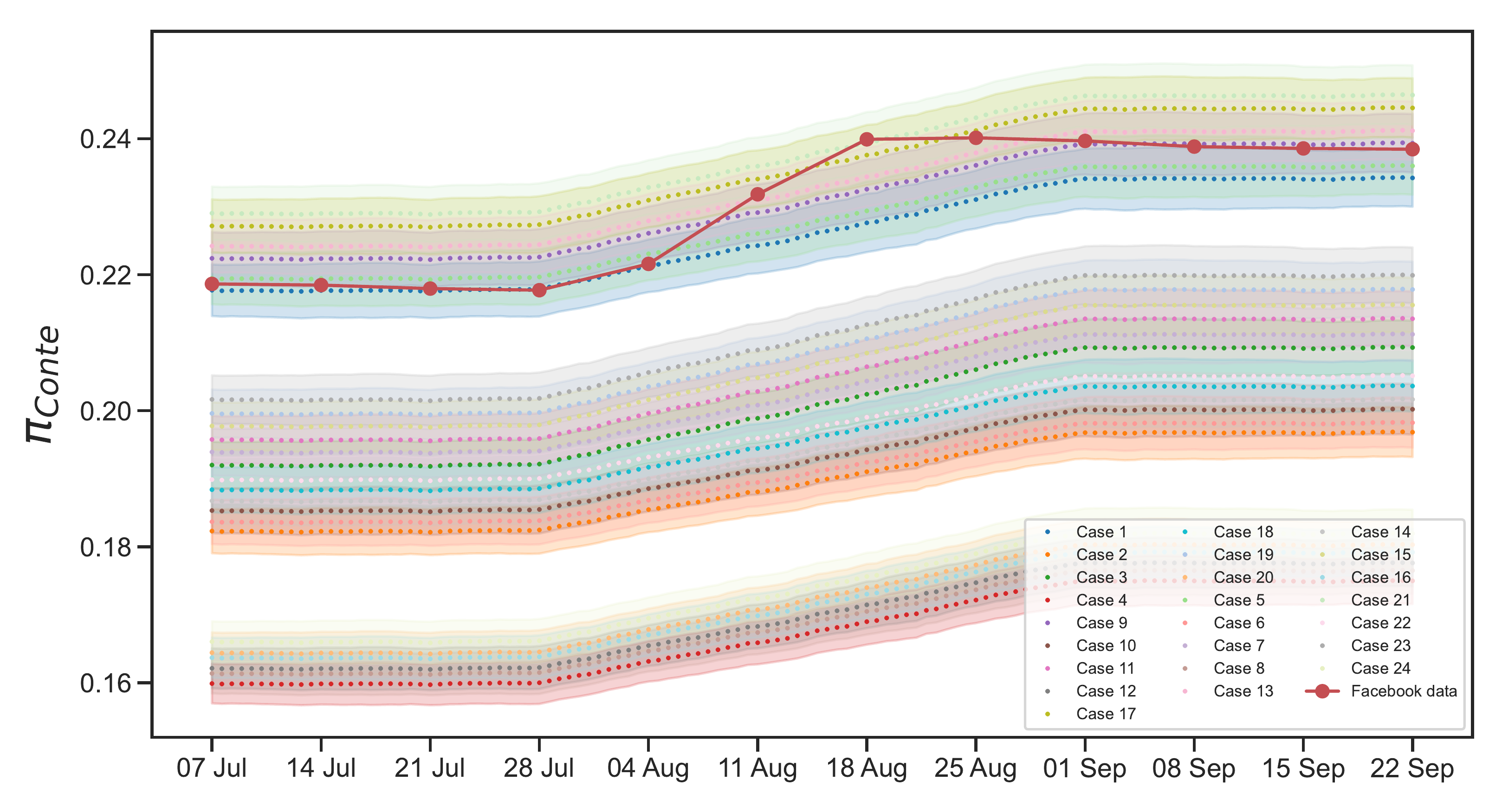}
\caption{Sensitivity analysis for the Figure \ref{fig:conte_salvini_res} considering the 24 different scenarios (in terms of parameters set) defined in Table \ref{tab:sensitivity_scenarios}.}
\label{fig:conte_salvini_sensitivity}
\end{figure}

{For the \textit{Case Study} in Section \ref{sec:conte_salvini}, we were able to measure or define some characteristic parameters (e.g., $f^{(i)}$ and $x_{Endgovernment}$). However, others were unknown, and even if there are methods to measure, for example, user's opinions (see Section \ref{sec:related}), this is far beyond the scope of the present work. We could not measure the updating weights $\alpha, \, \beta$, the \textit{feedback} function $\theta$, and the users' opinion on $Politics$. Nevertheless, it is sufficient to assume that a) Conte and Salvini have \lq\lq opposing views" over \textit{Politics}, i.e., $x^{(Salvini)} < 0.5$ and $x^{(Conte)} > 0.5$ and that b) Salvini has a more extreme viewpoint, i.e., $|x^{(Salvini)}-0.5| > |x^{(Conte)}-0.5|$ c) the population is not too \textit{volatile} in their viewpoints, i.e., the \textit{degree of stubbornness} is not too low c) the \textit{feedback} function $\theta$ decreases quickly with the opinion distance (exponential), motivated by the fact that about \textit{politics} people do not easily like viewpoints too far from theirs, we were able to obtain the results in Figure \ref{fig:conte_salvini_res}. These are obtained with the parameters specified in Table~\ref{tab:table_conte_salvini}.

\begin{table}[h!]\footnotesize % make the font of the table content smaller
	\begin{center}
		%\caption*{Table 3: Parameters and functions for the Case Study} % force the table number
		\caption{Parameters and functions for the Case Study}
		\label{tab:table_conte_salvini}
		\begin{tabular}{l|c|l}
			\toprule % <-- Toprule here
			\textbf{Symbol} & \textbf{Value - Form} & \textbf{Description}\\
			\midrule % <-- Midrule here
			$N_i$                     & 2      & Number of influencers\\
			$x_0^{(Conte)}$           & 0.76   & Opinion of Giuseppe Conte on direction $j$\\
			$x_0^{(Salvini)}$         & 0.0    & Opinion of influencer $1$ on direction $j$\\
			$f^{(Conte)}$             & 0.108  & Opinion of Giuseppe Conte on direction $j$\\
			$f^{(Salvini)}$           & 0.892  & Opinion of influencer $1$ on direction $j$\\
			$r^{(Conte),(Salvini)}$   & 0      & Refrence direction of both influencers\\
			$p_{Conte,Salvini}(0)$    & 20     & Initial absolute popularity of both influencers\\
			$N_u$            & 10000  & Number of regular users\\
			$N_{iter}$       & 15000 & Number of iterations for each simulation\\
			$N_{t}$          & 10000 & Duration of the transient phase\\
			$w$              & 550 & Length of the government crisis\\
			$\alpha$         & 0.3   & First weight in the updating rule in Eq. \ref{eq:d_model_user_update}\\
			$\beta$          & 0.65   & Second weight in the updating rule in Eq. \ref{eq:d_model_user_update}\\
			$\theta(\cdot)$  & $e^{-8.25 (x^{(u)}-x^{(i)})^2}$   & Functional form of the \textit{feedback} function\\
			$\omega(\cdot)$  & $\rho=0 \implies \omega \equiv 1$ & Functional form of the \textit{visibility} function\\
			\bottomrule % <-- Bottomrule here
		\end{tabular}
	\end{center}
\end{table}
%\vspace{-5mm}
To justify our choices, we run a \textit{sensitivity analysis} on the parameters we did not measure, defining the scenarios in Table \ref{tab:sensitivity_scenarios}. Figure \ref{fig:conte_salvini_sensitivity} presents the results. It is evident that the increase in the normalized popularity $\pi_{Conte}$ does not depend on these parameters and is due to the timing (see Fig. \ref{fig:cs_setting}) and the skewness in the population distribution, see Figure \ref{fig:conte_salvini}. Note that in Figure \ref{fig:conte_salvini_sensitivity} there is a negative \textit{offset} that tends to underestimate Conte's normalized popularity. This is because we use the measured rate $f^{(Conte)}=0.108$, which puts him at quite a disadvantage compared to Salvini, who has a much higher publication frequency.}

\begin{table}[h!]
\begin{center}
	\footnotesize
	%\vspace{-1cm}
	\caption{Definition of the scenarios for the sensitivity analysis}
	\label{tab:sensitivity_scenarios}
	\begin{tabular}{c|c|c|c|c|c}
		\toprule % <-- Toprule here
		\textbf{Scenario} & $\alpha$ & $\beta$ & $\theta$ param. & $x_0^{(Conte)}$ &$x_0^{(Salvini)}$ \\
		\midrule % <-- Midrule here
		1 & 0.25 & 0.708 & 8.0 & 0.74 & 0.01 \\
		2 & 0.25 & 0.708 & 8.0 & 0.74 & 0.05 \\
		3 & 0.25 & 0.708 & 8.0 & 0.78 & 0.01 \\
		4 & 0.25 & 0.708 & 8.0 & 0.78 & 0.05 \\
		5 & 0.45 & 0.475 & 8.0 & 0.74 & 0.01 \\
		6 & 0.45 & 0.475 & 8.0 & 0.74 & 0.05 \\
		7 & 0.45 & 0.475 & 8.0 & 0.78 & 0.01 \\
		8 & 0.45 & 0.475 & 8.0 & 0.78 & 0.05 \\
		9 & 0.25 & 0.708 & 8.25 & 0.74 & 0.01 \\
		10 & 0.25 & 0.708 & 8.25 & 0.74 & 0.05 \\
		11 & 0.25 & 0.708 & 8.25 & 0.78 & 0.01 \\
		12 & 0.25 & 0.708 & 8.25 & 0.78 & 0.05 \\
		13 & 0.45 & 0.475 & 8.25 & 0.74 & 0.01 \\
		14 & 0.45 & 0.475 & 8.25 & 0.74 & 0.05 \\
		15 & 0.45 & 0.475 & 8.25 & 0.78 & 0.01 \\
		16 & 0.45 & 0.475 & 8.25 & 0.78 & 0.05 \\
		17 & 0.25 & 0.708 & 8.5 & 0.74 & 0.01 \\
		18 & 0.25 & 0.708 & 8.5 & 0.74 & 0.05 \\
		19 & 0.25 & 0.708 & 8.5 & 0.78 & 0.01 \\
		20 & 0.25 & 0.708 & 8.5 & 0.78 & 0.05 \\
		21 & 0.45 & 0.475 & 8.5 & 0.74 & 0.01 \\
		22 & 0.45 & 0.475 & 8.5 & 0.74 & 0.05 \\
		23 & 0.45 & 0.475 & 8.5 & 0.78 & 0.01 \\
		24 & 0.45 & 0.475 & 8.5 & 0.78 & 0.05 \\
		
		\bottomrule % <-- Bottomrule here
	\end{tabular}
\end{center}
\end{table}
\vspace{-5mm}

\section{Possible model extensions }\label{sec:extensions}

\subsection{Modeling interactions between users}
\label{sec:user_interactions}

In our model we have purposely ignored interactions  among \lq\lq regular\rq\rq  users.
This choice can be justified by the need to keep the model sufficiently simple to be analyzed using the tools introduced so far, and by the hypothesis that opinion
dynamics are primarily driven by the interaction
between influencers and regular users, at least for 
some topics. 

Nevertheless, our model could be extended to incorporate also interactions among regular users, and here we outline one possible way to proceed along this direction. 
Firstly, to preserve the scalability of the model, we do not represent individual interactions among regular users, as done by other authors, but we rather represent the average effect  of such interactions on the generic regular user $u$ through a mean-field approach.
Secondly, we distinguish  between two classes of additional interactions originated by the activity of regular users: {\em direct} (pairwise) interactions between regular users and {\em influencer-triggered} users interactions.

For what concerns the first class, we introduce a monotonic non-increasing function $\zeta(d)$, representing the probability that two \lq\lq  regular \rq\rq users whose current opinion distance 
is $d$ influence each other.  
Note that  $\zeta(d)$ measures the degree of homophily of regular users as function of their
separation in the opinions space: more skewed $\zeta(d)$ shapes  correspond to larger degrees of homophily. 
Moreover, we again assume that user $u$ provides a positive feedback to a message generated by user $u'$ at time $t$ 
on topic $j$ with probability $\theta(| x_j^{(u)(t)}-  x_j^{(u')(t)} | )$. It follows that, to account for the
overall effect on $u$ of pairwise interactions, we need
to compute the \lq center of mass' of
neighbors receiving positive feedback
from $u$ along direction $j$:   
\begin{align*}
%\[
\widehat {{x}}_j^{(u)} (n)  
& = \sum_{ u'\neq u } \bm{x}^{(u')}(n) \theta( x_j^{(u)}(n)-  x_j^{(u')}(n) |) \zeta( || {\bm x}^{(u)}(n)-  
{\bm x}^{(u')}(n)  ||) \\
& \to  \int f({\bm x},t) 
\theta(| x_j^{(u)}-  x_j|)  \zeta( || {\bm x}^{(u)}-  {\bm x} ||) \mathrm{d} {\bm x}
%\]
\end{align*}

Then we extend the user opinion update rule  \eqref{eq:d_model_user_update}  as follows:
\begin{equation}\label{eq:d_model_user_update-extended}
x_j^{(u)}(n+1) =
\begin{cases}
	\alpha z_j^{(u)} + \beta x_j^{(u)}(n) + \gamma x_j^{(i)}+ \delta \widehat x^{(u)}_j  \;\;\; \text{if} \; \Omega \left( \omega(d_r,\pi_i) \right) = 1\,, \Theta \left(\theta(d_j)\right) = 1 \\[5pt]  % \texttt{and}
	\frac{\alpha }{1- \gamma } z_j^{(u)} + \frac{\beta }{1- \gamma }  x_j^{(u)}(n) + \frac{\delta }{1- \gamma }  \widehat x^{(u)}_j  
	\quad\quad\quad\quad\quad\quad\quad\quad\;\, \text{otherwise} %\texttt{o/w}
\end{cases}
\end{equation}

with $\alpha+\beta+\gamma+\delta=1$. 
\begin{comment}
and  
\[
x_{j'}^{(u)}(n+1) =  \frac{\alpha }{1- \gamma }  z_{j'}^{(u)} + \frac{\beta }{1- \gamma }   x_i^{(u)}(n) + 
\frac{\delta}{1-\gamma} \widehat x^{(u)}_{j'}    \qquad j'\neq j
\] 
% \end{equation}
\end{comment}
Different ratios $\delta/\gamma$    correspond to   different relative strengths between {\em direct} interactions between regular users and  {\em influencer-triggered}  users interactions. 

{\em Influencer-triggered} users interactions represent cascading effect of reactions by a user on a post generated by an influencer, appearing on the timeline of its friends. This is another form of (indirect) interaction among regular users which might play a significant role in the overall opinion dynamics. 
It is reasonable to assume that all follow-up reactions to posts of influencer  $i$ mainly circulate among the {\em followers} of $i$ with the resulting effect of amplifying the strength of the original post on the users providing positive feedback to it. As a consequence, one can take into
account the impact the {\em influencer-triggered} users interactions by playing on weights $\alpha$, $\beta$, $\gamma$ and $\delta$, i.e., by increasing the relative  weight of $\gamma$.

\subsection{Strategic influencers}
\label{sec:strategic}
{\
The influence exerted on non-reference directions depends heavily on the size of the user audience reached by the influencer's post, which in turn depends on how popular an influencer is in its main field of expertise. For example, famous public figures (e.g., athletes, models) may express their views on potentially sensitive topics and resonate more than experts due to their popularity in their own field.
This mechanism can be deliberately exploited by an influencer to gain popularity and/or strategically exert influence on a topic other than its own reference topic.

{For example, we have seen in Section \ref{sec:sim_cons} how \textit{low-consistency} values allow an influencer to gain a structural advantage whenever the consistency of the opposing influencer (who has different reference topic) tends to one. We must here remark that a consistency $c^{(i)} < 0.5$ in a bidimensional setting is slightly in contrast with the definition of reference direction itself, if the parameters are static. However, if we consider time-varying consistency, an influencer that is well-known on topic $j$, i.e., $r^{(i)}(n_1)=j$ may modify its consistency at time $n_2$ if it learns to have little competition on its reference direction.}

% We can imagine a setting in which influencers vary their consistency either to increase their popularity or to leverage thei popularity to persuade their users on another topic.

Our model can be rather easily adapted to study such scenarios, 
by  representing  influencers as 
{\em strategic agents},  whose  selected topics and relative opinions  expressed in their generated posts  can 
be dynamically adapted  so to maximize a given utility function.  

}

\bibliography{resources}

\begin{thebibliography}{10}
\expandafter\ifx\csname url\endcsname\relax
  \def\url#1{\texttt{#1}}\fi
\expandafter\ifx\csname urlprefix\endcsname\relax\def\urlprefix{URL }\fi
\expandafter\ifx\csname href\endcsname\relax
  \def\href#1#2{#2} \def\path#1{#1}\fi

\bibitem{found_homophily}
M.~McPherson, L.~Smith-Lovin, J.~M. Cook,
  \href{https://doi.org/10.1146/annurev.soc.27.1.415}{Birds of a feather:
  Homophily in social networks}, Annual Review of Sociology 27~(1) (2001)
  415--444.
\newblock \href {https://doi.org/10.1146/annurev.soc.27.1.415}
  {\path{doi:10.1146/annurev.soc.27.1.415}}.
\newline\urlprefix\url{https://doi.org/10.1146/annurev.soc.27.1.415}

\bibitem{quattrociocchi_echo_chamber}
M.~Cinelli, G.~D.~F. Morales, A.~Galeazzi, W.~Quattrociocchi, M.~Starnini,
  \href{https://doi.org/10.1073/pnas.2023301118}{The echo chamber effect on
  social media}, Proceedings of the National Academy of Sciences 118~(9) (Feb.
  2021).
\newblock \href {https://doi.org/10.1073/pnas.2023301118}
  {\path{doi:10.1073/pnas.2023301118}}.
\newline\urlprefix\url{https://doi.org/10.1073/pnas.2023301118}

\bibitem{das_ash_exp}
A.~Das, S.~Gollapudi, K.~Munagala,
  \href{https://doi.org/10.1145/2556195.2559896}{Modeling opinion dynamics in
  social networks}, in: Proceedings of the 7th ACM International Conference on
  Web Search and Data Mining, WSDM '14, Association for Computing Machinery,
  New York, NY, USA, 2014, p. 403–412.
\newblock \href {https://doi.org/10.1145/2556195.2559896}
  {\path{doi:10.1145/2556195.2559896}}.
\newline\urlprefix\url{https://doi.org/10.1145/2556195.2559896}

\bibitem{Xiong_Liu_2014}
F.~Xiong, Y.~Liu, Opinion formation on social media: An empirical approach,
  Chaos: An Interdisciplinary Journal of Nonlinear Science 24~(1) (2014)
  013130.
\newblock \href {https://doi.org/10.1063/1.4866011}
  {\path{doi:10.1063/1.4866011}}.

\bibitem{scaling_elections}
S.~Fortunato, C.~Castellano,
  \href{https://arxiv.org/abs/physics/0612140}{Scaling and universality in
  proportional elections} (2006).
\newblock \href {https://doi.org/10.48550/ARXIV.PHYSICS/0612140}
  {\path{doi:10.48550/ARXIV.PHYSICS/0612140}}.
\newline\urlprefix\url{https://arxiv.org/abs/physics/0612140}

\bibitem{Ash1955}
S.~E. Asch, \href{http://www.jstor.org/stable/24943779}{Opinions and social
  pressure}, Scientific American 193~(5) (1955) 31--35.
\newline\urlprefix\url{http://www.jstor.org/stable/24943779}

\bibitem{French1956}
J.~R.~P. French, \href{https://doi.org/10.1037/h0046123}{A formal theory of
  social power.}, Psychological Review 63~(3) (1956) 181--194.
\newblock \href {https://doi.org/10.1037/h0046123}
  {\path{doi:10.1037/h0046123}}.
\newline\urlprefix\url{https://doi.org/10.1037/h0046123}

\bibitem{Festinger1954}
L.~Festinger, \href{https://doi.org/10.1177/001872675400700202}{A theory of
  social comparison processes}, Human Relations 7~(2) (1954) 117--140.
\newblock \href {https://doi.org/10.1177/001872675400700202}
  {\path{doi:10.1177/001872675400700202}}.
\newline\urlprefix\url{https://doi.org/10.1177/001872675400700202}

\bibitem{Mastroeni_IEEE_review}
L.~Mastroeni, P.~Vellucci, M.~Naldi,
  \href{https://doi.org/10.1109/access.2019.2913787}{Agent-based models for
  opinion formation: A bibliographic survey}, {IEEE} Access 7 (2019)
  58836--58848.
\newblock \href {https://doi.org/10.1109/access.2019.2913787}
  {\path{doi:10.1109/access.2019.2913787}}.
\newline\urlprefix\url{https://doi.org/10.1109/access.2019.2913787}

\bibitem{Degroot1974}
M.~H. Degroot, \href{https://doi.org/10.1080/01621459.1974.10480137}{Reaching a
  consensus}, Journal of the American Statistical Association 69~(345) (1974)
  118--121.
\newblock \href {https://doi.org/10.1080/01621459.1974.10480137}
  {\path{doi:10.1080/01621459.1974.10480137}}.
\newline\urlprefix\url{https://doi.org/10.1080/01621459.1974.10480137}

\bibitem{Friedkin_Johnsen_1990}
N.~E. Friedkin, E.~C. Johnsen, Social influence and opinions, The Journal of
  Mathematical Sociology 15~(3–4) (1990) 193–206.
\newblock \href {https://doi.org/10.1080/0022250X.1990.9990069}
  {\path{doi:10.1080/0022250X.1990.9990069}}.

\bibitem{HK_model}
R.~Hegselmann, U.~Krause, Opinion dynamics and bounded confidence: Models,
  analysis and simulation, Journal of Artificial Societies and Social
  Simulation 5 (2002) 1--24.

\bibitem{DW_model}
G.~Deffuant, D.~Neau, F.~Amblard, G.~Weisbuch,
  \href{https://doi.org/10.1142/s0219525900000078}{Mixing beliefs among
  interacting agents}, Advances in Complex Systems 03~(01n04) (2000) 87--98.
\newblock \href {https://doi.org/10.1142/s0219525900000078}
  {\path{doi:10.1142/s0219525900000078}}.
\newline\urlprefix\url{https://doi.org/10.1142/s0219525900000078}

\bibitem{Lorenz_2007}
J.~Lorenz, Continuous opinion dynamics under bounded confidence: A survey,
  International Journal of Modern Physics C 18~(12) (2007) 1819–1838,
  arXiv:0707.1762 [physics].
\newblock \href {https://doi.org/10.1142/S0129183107011789}
  {\path{doi:10.1142/S0129183107011789}}.

\bibitem{clifford_and_sudbury_1973}
P.~CLIFFORD, A.~SUDBURY, \href{https://doi.org/10.1093/biomet/60.3.581}{A model
  for spatial conflict}, Biometrika 60~(3) (1973) 581--588.
\newblock \href {https://doi.org/10.1093/biomet/60.3.581}
  {\path{doi:10.1093/biomet/60.3.581}}.
\newline\urlprefix\url{https://doi.org/10.1093/biomet/60.3.581}

\bibitem{Holley_Liggett_1975}
R.~A. Holley, T.~M. Liggett,
  \href{https://doi.org/10.1214/aop/1176996306}{Ergodic theorems for weakly
  interacting infinite systems and the voter model}, The Annals of Probability
  3~(4) (Aug. 1975).
\newblock \href {https://doi.org/10.1214/aop/1176996306}
  {\path{doi:10.1214/aop/1176996306}}.
\newline\urlprefix\url{https://doi.org/10.1214/aop/1176996306}

\bibitem{Holme_Newman_2006}
P.~Holme, M.~E.~J. Newman, Nonequilibrium phase transition in the coevolution
  of networks and opinions, Physical Review E 74~(5) (2006) 056108.
\newblock \href {https://doi.org/10.1103/PhysRevE.74.056108}
  {\path{doi:10.1103/PhysRevE.74.056108}}.

\bibitem{Durrett_evolving_VM}
R.~Durrett, J.~P. Gleeson, A.~L. Lloyd, P.~J. Mucha, F.~Shi, D.~Sivakoff,
  J.~E.~S. Socolar, C.~Varghese,
  \href{https://doi.org/10.1073/pnas.1200709109}{Graph fission in an evolving
  voter model}, Proceedings of the National Academy of Sciences 109~(10) (2012)
  3682--3687.
\newblock \href {https://doi.org/10.1073/pnas.1200709109}
  {\path{doi:10.1073/pnas.1200709109}}.
\newline\urlprefix\url{https://doi.org/10.1073/pnas.1200709109}

\bibitem{Nardini_whos_talking_1st}
C.~Nardini, B.~Kozma, A.~Barrat,
  \href{https://doi.org/10.1103\%2Fphysrevlett.100.158701}{Who's talking first?
  consensus or lack thereof in coevolving opinion formation models}, Physical
  Review Letters 100~(15) (apr 2008).
\newblock \href {https://doi.org/10.1103/physrevlett.100.158701}
  {\path{doi:10.1103/physrevlett.100.158701}}.
\newline\urlprefix\url{https://doi.org/10.1103\%2Fphysrevlett.100.158701}

\bibitem{Granovsky_Madras_1995}
B.~L. Granovsky, N.~Madras, The noisy voter model, Stochastic Processes and
  their Applications 55~(1) (1995) 23–43.
\newblock \href {https://doi.org/10.1016/0304-4149(94)00035-R}
  {\path{doi:10.1016/0304-4149(94)00035-R}}.

\bibitem{Ben-Naim_2005}
E.~Ben-Naim, Opinion dynamics: Rise and fall of political parties, Europhysics
  Letters 69~(55) (2005) 671–677.
\newblock \href {https://doi.org/10.1209/epl/i2004-10421-1}
  {\path{doi:10.1209/epl/i2004-10421-1}}.

\bibitem{toscani_kinetic_model}
G.~Toscani, \href{https://arxiv.org/abs/math-ph/0605052}{Kinetic models of
  opinion formation} (2006).
\newblock \href {https://doi.org/10.48550/ARXIV.MATH-PH/0605052}
  {\path{doi:10.48550/ARXIV.MATH-PH/0605052}}.
\newline\urlprefix\url{https://arxiv.org/abs/math-ph/0605052}

\bibitem{sznajd_weon_model}
K.~SZNAJD-WERON, J.~SZNAJD,
  \href{https://doi.org/10.1142\%2Fs0129183100000936}{{OPINION} {EVOLUTION}
  {IN} {CLOSED} {COMMUNITY}}, International Journal of Modern Physics C 11~(06)
  (2000) 1157--1165.
\newblock \href {https://doi.org/10.1142/s0129183100000936}
  {\path{doi:10.1142/s0129183100000936}}.
\newline\urlprefix\url{https://doi.org/10.1142\%2Fs0129183100000936}

\bibitem{Slanina_sznajd_analytical}
F.~Slanina, H.~Lavicka,
  \href{https://doi.org/10.1140/epjb/e2003-00278-0}{Analytical results for the
  sznajd model of opinion formation}, The European Physical Journal B -
  Condensed Matter 35~(2) (2003) 279--288.
\newblock \href {https://doi.org/10.1140/epjb/e2003-00278-0}
  {\path{doi:10.1140/epjb/e2003-00278-0}}.
\newline\urlprefix\url{https://doi.org/10.1140/epjb/e2003-00278-0}

\bibitem{Castellano_Fortunato_Loreto_2009}
C.~Castellano, S.~Fortunato, V.~Loreto, Statistical physics of social dynamics,
  Reviews of Modern Physics 81~(2) (2009) 591–646, arXiv:0710.3256 [cond-mat,
  physics:physics].
\newblock \href {https://doi.org/10.1103/RevModPhys.81.591}
  {\path{doi:10.1103/RevModPhys.81.591}}.

\bibitem{cox_voter}
J.~T. Cox, \href{http://www.jstor.org/stable/2244439}{Coalescing random walks
  and voter model consensus times on the torus in zd}, The Annals of
  Probability 17~(4) (1989) 1333--1366.
\newline\urlprefix\url{http://www.jstor.org/stable/2244439}

\bibitem{Frachebourg1996}
L.~Frachebourg, P.~L. Krapivsky,
  \href{https://doi.org/10.1103/physreve.53.r3009}{Exact results for kinetics
  of catalytic reactions}, Physical Review E 53~(4) (1996) R3009--R3012.
\newblock \href {https://doi.org/10.1103/physreve.53.r3009}
  {\path{doi:10.1103/physreve.53.r3009}}.
\newline\urlprefix\url{https://doi.org/10.1103/physreve.53.r3009}

\bibitem{Suchecki2005}
K.~Suchecki, V.~M. Egu{\'{\i}}luz, M.~S. Miguel,
  \href{https://doi.org/10.1103/physreve.72.036132}{Voter model dynamics in
  complex networks: Role of dimensionality, disorder, and degree distribution},
  Physical Review E 72~(3) (Sep. 2005).
\newblock \href {https://doi.org/10.1103/physreve.72.036132}
  {\path{doi:10.1103/physreve.72.036132}}.
\newline\urlprefix\url{https://doi.org/10.1103/physreve.72.036132}

\bibitem{Sood2005}
V.~Sood, S.~Redner, \href{https://doi.org/10.1103/physrevlett.94.178701}{Voter
  model on heterogeneous graphs}, Physical Review Letters 94~(17) (May 2005).
\newblock \href {https://doi.org/10.1103/physrevlett.94.178701}
  {\path{doi:10.1103/physrevlett.94.178701}}.
\newline\urlprefix\url{https://doi.org/10.1103/physrevlett.94.178701}

\bibitem{Yildiz2013}
E.~Yildiz, A.~Ozdaglar, D.~Acemoglu, A.~Saberi, A.~Scaglione,
  \href{https://doi.org/10.1145/2538508}{Binary opinion dynamics with stubborn
  agents}, {ACM} Transactions on Economics and Computation 1~(4) (2013) 1--30.
\newblock \href {https://doi.org/10.1145/2538508} {\path{doi:10.1145/2538508}}.
\newline\urlprefix\url{https://doi.org/10.1145/2538508}

\bibitem{valensise_polarization}
C.~M. Valensise, M.~Cinelli, W.~Quattrociocchi,
  \href{https://arxiv.org/abs/2205.15958}{The dynamics of online polarization}
  (2022).
\newblock \href {https://doi.org/10.48550/ARXIV.2205.15958}
  {\path{doi:10.48550/ARXIV.2205.15958}}.
\newline\urlprefix\url{https://arxiv.org/abs/2205.15958}

\bibitem{Peralta_2021_algorithmic_bias}
A.~F. Peralta, M.~Neri, J.~Kert{\'{e}}sz, G.~I{\~{n}}iguez,
  \href{https://doi.org/10.1103\%2Fphysreve.104.044312}{Effect of algorithmic
  bias and network structure on coexistence, consensus, and polarization of
  opinions}, Physical Review E 104~(4) (oct 2021).
\newblock \href {https://doi.org/10.1103/physreve.104.044312}
  {\path{doi:10.1103/physreve.104.044312}}.
\newline\urlprefix\url{https://doi.org/10.1103\%2Fphysreve.104.044312}

\bibitem{Perra2019}
N.~Perra, L.~E.~C. Rocha,
  \href{https://doi.org/10.1038/s41598-019-43830-2}{Modelling opinion dynamics
  in the age of algorithmic personalisation}, Scientific Reports 9~(1) (May
  2019).
\newblock \href {https://doi.org/10.1038/s41598-019-43830-2}
  {\path{doi:10.1038/s41598-019-43830-2}}.
\newline\urlprefix\url{https://doi.org/10.1038/s41598-019-43830-2}

\bibitem{data_in_opdyn_2022}
A.~F. Peralta, J.~Kertész, G.~Iñiguez,
  \href{https://arxiv.org/abs/2201.01322}{Opinion dynamics in social networks:
  From models to data} (2022).
\newblock \href {https://doi.org/10.48550/ARXIV.2201.01322}
  {\path{doi:10.48550/ARXIV.2201.01322}}.
\newline\urlprefix\url{https://arxiv.org/abs/2201.01322}

\bibitem{pnas_opposing_views}
C.~A. Bail, L.~P. Argyle, T.~W. Brown, J.~P. Bumpus, H.~Chen, M.~B.~F.
  Hunzaker, J.~Lee, M.~Mann, F.~Merhout, A.~Volfovsky,
  \href{https://www.pnas.org/doi/abs/10.1073/pnas.1804840115}{Exposure to
  opposing views on social media can increase political polarization},
  Proceedings of the National Academy of Sciences 115~(37) (2018) 9216--9221.
\newblock \href
  {http://arxiv.org/abs/https://www.pnas.org/doi/pdf/10.1073/pnas.1804840115}
  {\path{arXiv:https://www.pnas.org/doi/pdf/10.1073/pnas.1804840115}}, \href
  {https://doi.org/10.1073/pnas.1804840115}
  {\path{doi:10.1073/pnas.1804840115}}.
\newline\urlprefix\url{https://www.pnas.org/doi/abs/10.1073/pnas.1804840115}

\bibitem{Morgan2011}
T.~J.~H. Morgan, L.~E. Rendell, M.~Ehn, W.~Hoppitt, K.~N. Laland,
  \href{https://doi.org/10.1098/rspb.2011.1172}{The evolutionary basis of human
  social learning}, Proceedings of the Royal Society B: Biological Sciences
  279~(1729) (2011) 653--662.
\newblock \href {https://doi.org/10.1098/rspb.2011.1172}
  {\path{doi:10.1098/rspb.2011.1172}}.
\newline\urlprefix\url{https://doi.org/10.1098/rspb.2011.1172}

\bibitem{voter_for_voters}
J.~Fern\'andez-Gracia, K.~Suchecki, J.~J. Ramasco, M.~San~Miguel, V.~M.
  Egu\'{\i}luz,
  \href{https://link.aps.org/doi/10.1103/PhysRevLett.112.158701}{Is the voter
  model a model for voters?}, Phys. Rev. Lett. 112 (2014) 158701.
\newblock \href {https://doi.org/10.1103/PhysRevLett.112.158701}
  {\path{doi:10.1103/PhysRevLett.112.158701}}.
\newline\urlprefix\url{https://link.aps.org/doi/10.1103/PhysRevLett.112.158701}

\bibitem{Barbera_PS2015}
P.~Barberá, J.~T. Jost, J.~Nagler, J.~A. Tucker, R.~Bonneau,
  \href{https://doi.org/10.1177/0956797615594620}{Tweeting from left to right:
  Is online political communication more than an echo chamber?}, Psychological
  Science 26~(10) (2015) 1531--1542, pMID: 26297377.
\newblock \href {http://arxiv.org/abs/https://doi.org/10.1177/0956797615594620}
  {\path{arXiv:https://doi.org/10.1177/0956797615594620}}, \href
  {https://doi.org/10.1177/0956797615594620}
  {\path{doi:10.1177/0956797615594620}}.
\newline\urlprefix\url{https://doi.org/10.1177/0956797615594620}

\bibitem{De_TWEB2019}
A.~De, S.~Bhattacharya, P.~Bhattacharya, N.~Ganguly, S.~Chakrabarti,
  \href{https://doi.org/10.1145/3343483}{Learning linear influence models in
  social networks from transient opinion dynamics}, ACM Trans. Web 13~(3) (nov
  2019).
\newblock \href {https://doi.org/10.1145/3343483} {\path{doi:10.1145/3343483}}.
\newline\urlprefix\url{https://doi.org/10.1145/3343483}

\bibitem{Monti_CIKM2021}
C.~Monti, G.~Manco, C.~Aslay, F.~Bonchi,
  \href{https://doi.org/10.1145/3459637.3482444}{Learning ideological
  embeddings from information cascades}, in: Proceedings of the 30th ACM
  International Conference on Information \& Knowledge Management, CIKM '21,
  Association for Computing Machinery, New York, NY, USA, 2021, p. 1325–1334.
\newblock \href {https://doi.org/10.1145/3459637.3482444}
  {\path{doi:10.1145/3459637.3482444}}.
\newline\urlprefix\url{https://doi.org/10.1145/3459637.3482444}

\bibitem{science_diverse_news}
E.~Bakshy, S.~Messing, L.~A. Adamic,
  \href{https://doi.org/10.1126/science.aaa1160}{Exposure to ideologically
  diverse news and opinion on facebook}, Science 348~(6239) (2015) 1130--1132.
\newblock \href {https://doi.org/10.1126/science.aaa1160}
  {\path{doi:10.1126/science.aaa1160}}.
\newline\urlprefix\url{https://doi.org/10.1126/science.aaa1160}

\bibitem{pnas_homophily}
S.~Aral, L.~Muchnik, A.~Sundararajan,
  \href{https://doi.org/10.1073/pnas.0908800106}{Distinguishing influence-based
  contagion from homophily-driven diffusion in dynamic networks}, Proceedings
  of the National Academy of Sciences 106~(51) (2009) 21544--21549.
\newblock \href {https://doi.org/10.1073/pnas.0908800106}
  {\path{doi:10.1073/pnas.0908800106}}.
\newline\urlprefix\url{https://doi.org/10.1073/pnas.0908800106}

\bibitem{risken1996fokker}
H.~Risken, Fokker-planck equation, in: The Fokker-Planck Equation, Springer,
  1996, pp. 63--95.

\bibitem{cohen_tsang_ONS}
R.~Cohen, A.~Tsang, K.~Vaidyanathan, H.~Zhang, Analyzing opinion dynamics in
  online social networks, Big Data \& Information Analytics 1~(4) (2016)
  279--298.

\bibitem{VassioAsonam}
L.~Vassio, M.~Garetto, C.~Chiasserini, E.~Leonardi,
  \href{https://doi.org/10.1145/3487351.3488340}{Temporal dynamics of posts and
  user engagement of influencers on facebook and instagram}, in: Proceedings of
  the 2021 IEEE/ACM International Conference on Advances in Social Networks
  Analysis and Mining, ASONAM '21, Association for Computing Machinery, New
  York, NY, USA, 2022, p. 129–133.
\newblock \href {https://doi.org/10.1145/3487351.3488340}
  {\path{doi:10.1145/3487351.3488340}}.
\newline\urlprefix\url{https://doi.org/10.1145/3487351.3488340}

\bibitem{SocialCovid}
M.~Trevisan, L.~Vassio, D.~Giordano,
  \href{https://www.sciencedirect.com/science/article/pii/S2468696421000203}{Debate
  on online social networks at the time of covid-19: An italian case study},
  Online Social Networks and Media 23 (2021) 100136.
\newblock \href {https://doi.org/https://doi.org/10.1016/j.osnem.2021.100136}
  {\path{doi:https://doi.org/10.1016/j.osnem.2021.100136}}.
\newline\urlprefix\url{https://www.sciencedirect.com/science/article/pii/S2468696421000203}

\end{thebibliography}

\end{document}